\documentclass[10pt,twocolumn,superscriptaddress,floats,showpacs,nobalancelastpage,longbibliography,prx,aps]{revtex4-1}
\pdfoutput=1

\usepackage{array,longtable}
\newcolumntype{L}{>{\tiny $}p{0.33\columnwidth}<{$}}
\newcolumntype{M}{>{\scriptsize $}p{0.33\columnwidth}<{$}}
\newcolumntype{N}{>{\scriptsize $}p{0.43\columnwidth}<{$}}
\usepackage{amsmath}
\usepackage{amssymb}
\usepackage{amsfonts}
\usepackage{graphicx}
\usepackage{hyperref}
\usepackage{tabularx}

\usepackage{float}
\usepackage{graphics}
\usepackage[caption=false]{subfig}
\usepackage{times}
\usepackage{youngtab}
\usepackage{ifthen}
\usepackage[dvipsnames]{xcolor}
\usepackage{xspace}
\usepackage{braket}
\usepackage{appendix}
\usepackage{multirow}
\usepackage{bbm}
\usepackage{ulem}

\allowdisplaybreaks[1]

\newif\ifhyper
\hypertrue
\ifhyper
\hypersetup{
	citecolor = {green},
	colorlinks = {true}, 
	urlcolor = {blue} 
}
\fi

\newcommand{\overbar}[1]{\mkern 1.5mu\overline{\mkern-1.5mu#1\mkern-1.5mu}\mkern 1.5mu}
\newcommand{\su}[3]{
	\ifthenelse{\equal{#2}{0}}{\ifthenelse{\equal{#1}{0}}{}{\mult{#1}}\overset{\bf #3}{\underset{\phantom{.}}{\bullet}}}{\ifthenelse{\equal{#1}{0}}{}{\mult{#1}}\underset{\phantom{.}}{\overset{\bf #3}{\Yboxdim9pt\yng(#2)}}}
}
\newcommand{\mult}[1]{{ \fcolorbox{gray!70}{gray!70}{\textcolor{white}{\footnotesize \bf #1}}}\;}

\newcommand{\sun}{$\mathfrak{su}(N)$\xspace}
\newcommand{\SUN}{SU($N$)\xspace}
\newcommand{\SUNone}{SU($N$)$_1$\xspace}

  \newcommand{\App}[1]{App.\,\ref{#1}}
  \newcommand{\Sec}[1]{Sec.\,\ref{#1}}
  \newcommand{\Fig}[1]{Fig.\,\ref{#1}}
  \newcommand{\Eq}[1]{Eq.\,\eqref{#1}}

\begin{document}

\title{Abelian SU$(N)_1$ Chiral Spin Liquids on the Square Lattice}

\author{Ji-Yao Chen}
\affiliation{Max-Planck-Institut f\"ur Quantenoptik, Hans-Kopfermann-Stra{\ss}e 1, 85748 Garching, Germany}
\affiliation{Munich Center for Quantum Science and Technology, Schellingstra{\ss}e 4, 80799 M{\"u}nchen, Germany}

\author{Jheng-Wei Li}
\affiliation{Arnold Sommerfeld Center for Theoretical Physics, Center for NanoScience,\looseness=-1\,  and Munich Center for \\ Quantum Science and Technology,\looseness=-2\, Ludwig-Maximilians-Universit\"at M\"unchen, 80333 Munich, Germany}

\author{Pierre Nataf}
\affiliation{Laboratoire de Physique et de Mod\'elisation des Milieux Condens\'es, Maison des Magist\`eres, CNRS,
	25 avenue des Martyrs,
	BP166,
	38042 Grenoble, France}

\author{Sylvain Capponi}
\affiliation{Laboratoire de Physique Th\'eorique, F\'ed\'eration Fermi, Universit\'e de Toulouse, CNRS, UPS, 31062 Toulouse, France}

\author{Matthieu Mambrini}
\affiliation{Laboratoire de Physique Th\'eorique, F\'ed\'eration Fermi, Universit\'e de Toulouse, CNRS, UPS, 31062 Toulouse, France}

\author{Keisuke Totsuka}
\affiliation{Yukawa Institute for Theoretical Physics, Kyoto University, Kitashirakawa Oiwake-Cho, Kyoto 606-8502, Japan}

\author{Hong-Hao Tu}
\affiliation{Institut f\"ur Theoretische Physik, Technische Universit\"at Dresden, 01062 Dresden, Germany}

\author{Andreas Weichselbaum}
\affiliation{Condensed Matter Physics and Materials Science Department, Brookhaven National Laboratory, Upton, New York 11973, USA}

\author{Jan von Delft}
\affiliation{Arnold Sommerfeld Center for Theoretical Physics, Center for NanoScience,\looseness=-1\,  and Munich Center for \\ Quantum Science and Technology,\looseness=-2\, Ludwig-Maximilians-Universit\"at M\"unchen, 80333 Munich, Germany}

\author{Didier Poilblanc}
\affiliation{Laboratoire de Physique Th\'eorique, F\'ed\'eration Fermi, Universit\'e de Toulouse, CNRS, UPS, 31062 Toulouse, France}

\date{\today}

\begin{abstract}
In the physics of the Fractional Quantum Hall (FQH) effect, a zoo of Abelian topological phases can be obtained by varying the magnetic field. Aiming to reach the same phenomenology in spin-like systems, we propose a family of SU($N$)-symmetric models in the fundamental representation, on the square lattice with short-range interactions restricted to triangular units, a natural generalization for arbitrary $N$ of an SU($3$) model studied previously where time-reversal symmetry is broken explicitly. Guided by the recent discovery of SU($2$)$_1$ and SU($3$)$_1$ chiral spin liquids (CSL) on similar models we search for topological SU($N$)$_1$ CSL in some range of the Hamiltonian parameters via a combination of complementary numerical methods such as exact diagonalizations (ED), infinite density matrix renormalization group (iDMRG) and infinite Projected Entangled Pair State (iPEPS). Extensive ED on small (periodic and open) clusters up to $N=10$ and an innovative SU($N$)-symmetric version of iDMRG to compute entanglement spectra on (infinitely-long) cylinders in all topological sectors provide unambiguous signatures of the SU($N$)$_1$ character of the chiral liquids. An SU($4$)-symmetric chiral PEPS, constructed in a manner similar to its SU($2$) and SU($3$) analogs,  is shown to give a good variational ansatz of the $N=4$ ground state, with chiral edge modes originating from the PEPS holographic bulk-edge correspondence. Finally, we discuss the possible observation of such Abelian CSL in ultracold atom setups where the possibility of varying $N$ provides a tuning parameter similar to the magnetic field in the physics of the FQH effect.

\end{abstract}
\maketitle

\section{Introduction}

Quantum spin liquids are states of matter of two-dimensional electronic spin systems not showing any sign of  spontaneous symmetry breaking down to zero temperature~\cite{Misguich2005,Savary2016,Zhou2017}.
Spin liquids with long-range entanglement may also exhibit topological order~\cite{Wen1990} such as the spin-1/2 Resonating Valence Bond (RVB) state on the Kagome lattice~\cite{Poilblanc2012}.
Among the broad family of spin liquids, chiral spin liquids (CSL)~\cite{Kalmeyer1987,Kalmeyer1989,laughlin1989,WWZ1989,Laughlin1990} form a very special and interesting class~\cite{Wen2002} exhibiting  broken time-reversal symmetry and chiral topological order~\cite{Wen1990}. Intimately related to FQH states~\cite{Tsui1982}, CSL are incompressible quantum fluids (i.e. with a bulk gap) and host both (Abelian or non-Abelian) anyonic quasi-particles in the bulk~\cite{Halperin1984} and chiral gapless modes on the edge~\cite{Wen1991a}. After the original  papers, the Kalmeyer-Laughlin CSL lay dormant for
many years until an explicit parent Hamiltonian was constructed~\cite{Schroeter2007,Thomale2009} using Laughlin's idea~\cite{laughlin1989}. Later somewhat simpler Hamiltonians were found using different methods~\cite{Nielsen2012,Greiter2014}. An important step towards the goal of finding a chiral spin liquid in realistic systems was taken by examining a physically motivated model for a Mott insulator (Hubbard model) with broken time-reversal symmetry~\cite{Bauer2014,Nielsen2013}. Then, an Abelian CSL was  identified in the (chiral) spin-1/2 Heisenberg model on the triangular lattice~\cite{Wietek2017,Gong2017}.  Note that CSL hosting non-Abelian excitations (useful for topological quantum computing~\cite{Kitaev2003}) have also been introduced in different contexts~\cite{Kitaev2006,Yao2007,Greiter2009}.

It was early suggested that, in systems with enhanced SU$(N)$ symmetry, realizable with ultracold alkaline earth atoms loaded in optical lattices~\cite{Gorshkov2010}, CSL can naturally appear~\cite{Hermele2009}, although this original proposal on the square lattice remained controversial. 
Later on, an Abelian CSL was indeed identified on  the triangular lattice in  SU($N$) Heisenberg models with $N>2$~\cite{Nataf2016}.  The presence of a chiral spin interaction, achievable experimentally via a synthetic gauge field, seems to be a key feature to stabilize SU($N$) CSL~\cite{Chen2016}. Nevertheless, the T and P violation required for a CSL
could emerge spontaneously in T-invariant models, as found for $N=2$ in a spin-1/2 Kagome Heisenberg model~\cite{He2014a,Gong2014a,Wietek2015} or, for $N=3$, in the Mott phase of a Hubbard model on the triangular lattice~\cite{Boos2020}. 
Note also that, using optical pumping, it is now possible to realize (so far in one dimension) strongly correlated liquids of ultracold fermions with a tunable number $N$ of spin components and SU($N$) symmetry~\cite{Pagano2014}.  This offers the prospect to be able to experimentally tune the system through various topological  liquids, as it is realized in the physics of the FQH effect via a tunable external magnetic field.
Apart from ultracold atom setups, condensed matter systems may also host  SU($N$) CSL. For example, it has been proposed very recently that an SU($4$) CSL could be realized in double-layer moir\'e superlattices~\cite{Zhang2021}.

In recent years, Projected Entangled Pair States (PEPS)~\cite{Verstraete2004b} have progressively emerged as a powerful tool to study quantum spin liquids providing variational ground states competitive with other methods~\cite{Liao2017,Lee2019,Liu2020}. PEPS also offer a powerful framework to encode topological order~\cite{Schuch2010a,Schuch2012,Chen2018a} and construct chiral Abelian~\cite{Poilblanc2015} and non-Abelian~\cite{Chen2018b} SU($2$) spin liquids. Generically, SU($2$) CSL described by PEPS exhibit linearly dispersing chiral branches in the entanglement spectrum (ES) well described by Wess-Zumino-Witten (WZW) SU$(2)_k$ (with the level of the WZW model $k=1$ for Abelian CSL) conformal field theory (CFT) for one-dimensional edges~\cite{Franscesco1997}.

Recently, on a square lattice with three-dimensional spin degrees of freedom which transform as the fundamental representation of  SU($3$) on every site, an Abelian CSL was found as the ground state (GS) of a simple
Hamiltonian involving only nearest-neighbor and next-nearest-neighbor (color) permutations and (imaginary) three-site cyclic permutations~\cite{Chen2020}.
Exact diagonalizations (ED) of open finite-size clusters and infinite-PEPS (iPEPS) in the thermodynamic limit (and encoding the full SU(3) symmetry) unambiguously showed the existence of chiral edge modes following the SU$(3)_1$ WZW CFT. Interestingly, these results can be viewed as extending previous results obtained for an SU($2$) spin-1/2 (i.e. $N=2$) chiral Heisenberg model~\cite{Nielsen2013,Poilblanc2017b}. Exactly the same type of Hamiltonian can be defined for $N$-dimensional spin degrees of freedom transforming according to the fundamental representation of SU($N$), for arbitrary integer $N\ge 2$.
It is then natural to speculate that, if such SU($N$) models also host CSLs for $N>3$, then the later should also be of the SU($N$)$_1$ type. Note however that, although a chiral perturbation necessary induces, from linear response theory, a finite response of the quantum spin system, it, by no means, implies the existence of topological order or the absence of conventional (lattice or magnetic) symmetry breaking, which both characterize a CSL. The emergence of a uniform CSL with protected edge modes is therefore a subtle feature that needs to be investigated on a case by case basis. It is far from clear that the findings for SU($3$) generalize to SU($N>3$) bearing in mind that $N$ may be commensurate or incommensurate with the fixed number of nearest neighbors on the square lattice.
Then, in this work, we have (i) generalized the chiral Hamiltonians of Refs.~\cite{Nielsen2013,Poilblanc2017b,Chen2020} to arbitrary $N$, (ii) defined a subset of these SU($N$) models whose Hamiltonians can be written solely as a sum of $S_3$-symmetric operators acting on all triangles within square plaquettes (as in Ref.~\cite{Chen2020}) and (iii) studied these models up to $N=10$ using a combination of complementary numerical techniques such as ED,  Density Matrix Renormalization Group (DMRG) and iPEPS, supplemented by CFT analytical predictions.

We then start by generalizing the SU($2$) and SU($3$) chiral Hamiltonians  by
placing, on every site of a square lattice, an $N$-dimensional spin degree of freedom, which transforms as the fundamental representation of SU($N$). As for $N=3$, we consider the most general SU($N$)-symmetric short-range three-site interaction:
\begin{eqnarray}
	\label{eq:model}
		H &=& J_1\sum_{\langle i,j\rangle}P_{ij} + J_2\sum_{\langle\langle k,l\rangle\rangle}P_{kl} \\
		& + &J_R\sum_{\triangle ijk}(P_{ijk} + P_{ijk}^{\,\,\,\,\,-1}) + iJ_{\rm I}\sum_{\triangle ijk }(P_{ijk} - P_{ijk}^{\,\,\,\,\,-1}),\nonumber
\end{eqnarray}
where the first (second) term corresponds to two-site permutations over all (next-)nearest-neighbor bonds, and the third and fourth terms are three-site (clockwise) permutations on all triangles of every plaquette. $P_{ij}$ ($P_{ijk}$ ) is defined through its action on the local basis states, $P_{ij}|\alpha\rangle_i |\beta\rangle_j = |\beta\rangle_i |\alpha\rangle_j $ ($P_{ijk}|\alpha\rangle_i |\beta\rangle_j |\gamma\rangle_k= |\gamma\rangle_i |\alpha\rangle_j |\beta\rangle_k$, for a fixed orientation of the triangle $i$, $j$, $k$, let say anticlockwise). To restrict the number of parameters we have chosen $J_2=J_1/2$. In that case, the two-body part ($J_1$ and $J_2$) on the interacting triangular units becomes $S_{3}$ symmetric, hence mimicking the corresponding Hamiltonian on the triangular lattice~\footnote{The chiral spin liquid phase should also exist away from $J_2=J_1/2$, due to its gapped nature.}.  We then use the same parametrization as in Ref.~\cite{Chen2020}:
\begin{eqnarray}
	J_1&=&2J_2=\tfrac{4}{3}\cos \theta \sin \phi, \nonumber\\
	J_R&=&\cos \theta\cos\phi, \label{eq:paras}\\
	J_{\rm I}&=&\sin\theta,\nonumber
\end{eqnarray}
and restrict ourselves to {\it antiferromagnetic} couplings $J_1, J_2 >0$, i.e. $0\le\theta\le\pi/2$ and $0\le\phi\le\pi$. Note however that, for $\phi>\pi/2$, the amplitude of the (real) 3-site permutation $J_R$ becomes ferromagnetic ($J_R<0$).
A detailed analysis of the multiplet structure
of a $2\times 2$ plaquette of the Hamiltonian above
is given in \App{app:plaquette}.

For $N=2$, various forms of the Hamiltonian (\ref{eq:model}) can be found in the literature~\cite{Nielsen2013,Poilblanc2017b}. In the original formulation~\cite{Nielsen2013}, a chiral interaction $4J_3 \,\,{\bf S}_i\cdot ({\bf S}_j\times{\bf S}_k)$ on all triangular units $\triangle (ijk)$ is introduced, corresponding to the 3-site cyclic permutations of (\ref{eq:model}) with amplitudes $J_R=0$ and $J_{\rm I}=J_3$.   Also, the 2-site exchange interactions are introduced here as spin-1/2 Heisenberg couplings, which is equivalent from the identity $2{\bf S}_i\cdot{\bf S}_j= P_{ij}-\frac{1}{2}$~\footnote{This can be extended to all fundamental IRREPs of SU($N$): $P_{ij}={\bf J}_i\cdot{\bf J}_j +\frac{1}{N}$,
	where $J^\alpha$ are the generators defined in Eq.~(\ref{eq:liecom}) of  ~\protect\App{app:WZW}.
Note, the  usual SU($2$) spin operators are given by ${\bf S}=(1/\sqrt{2}) {\bf J}$.}.
A Hamiltonian including a (pure-imaginary) cyclic permutation $i\lambda_c (P_{ijkl}-P_{ijkl}^{\,\,\,\,\,\,\,-1})$ on each plaquette $\square(ijkl)$ was also introduced~\cite{Poilblanc2017b}. In fact, the plaquette cyclic permutation $i(P_{ijkl}-\text{h.c.})$ can be rewritten  as $\frac{i}{2}(P_{ijk}+P_{jkl}+P_{kli}+P_{lij}-\text{h.c.})$~\footnote{This decomposition holds only for $N=2$ (in the fundamental representation)} so that this model corresponds also to $J_R=0$ and we can identify $J_{\rm I}=J_3=\lambda_c/2$. An optimum choice of parameters for the stability of the SU($2$) CSL phase is found to be (in our notations) $J_2/J_1\simeq 0.47$  and $J_{\rm I}/J_1\simeq 0.21$~\cite{Nielsen2013}.
Furthermore, evidence is provided that the CSL survives in a rather extended zone of parameter space around this point. Also, an SU(2)-symmetric PEPS ansatz~\cite{Poilblanc2015} provides an accurate representation of the GS at the optimum values of the parameters~\cite{Poilblanc2017b}, and of its edge modes~\cite{Poilblanc2016} following an SU($2$)$_1$ WZW CFT.

\begin{figure}
	\includegraphics[width=0.9\linewidth]{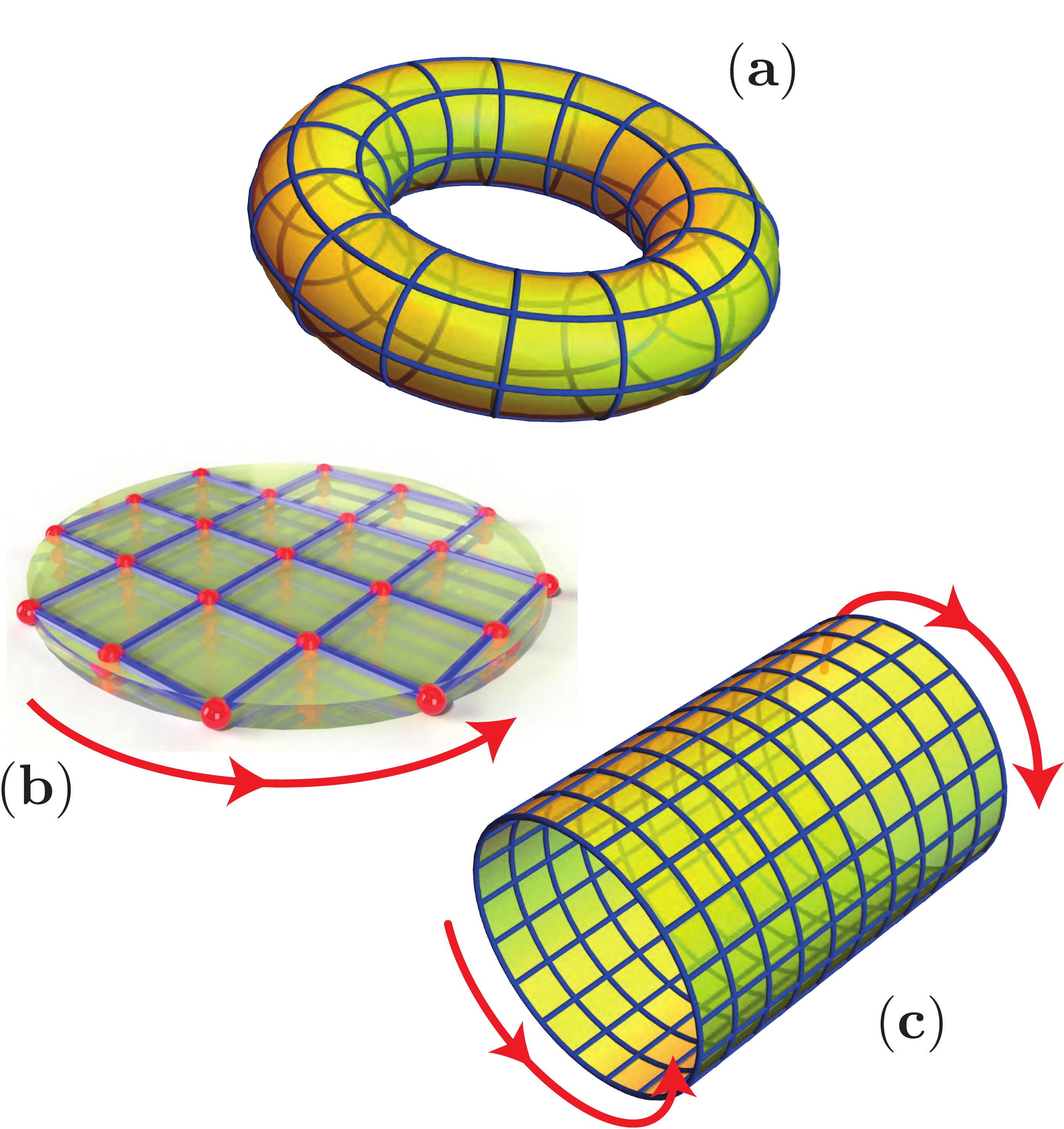} 	
	\caption{
  We considered various system topologies:
  {\bf (a)} periodic cluster topologically equivalent to a torus; {\bf (b)} open cluster topologically equivalent to a disk; {\bf (c)} cylinder with left and right boundaries.
	We used {\bf (a)} and {\bf (b)} in ED and the infinite-length version of {\bf (c)} in DMRG and iPEPS.
      The chiral modes of the CSL are schematically shown on the system edges.
  }
	\label{fig:systems_topo}
\end{figure}

For $N=3$,  from ED, DMRG and iPEPS simulations,
\textit{}clear evidence of a gapped CSL is found  for  $J_2=J_1/2$ and angles like $\theta=\phi=\pi/4$ corresponding to $J_R/J_1= 0.75$ and $J_{\rm I}/J_1\simeq 1.06$~\cite{Chen2020}, and around these values in a rather extended parameter range (see Supplemental Materials of Ref.~\cite{Chen2020}). In addition, edge modes are found to closely follow the predictions of the SU($3$)$_1$ CFT.

\begin{figure*}
		\includegraphics[width=\linewidth]{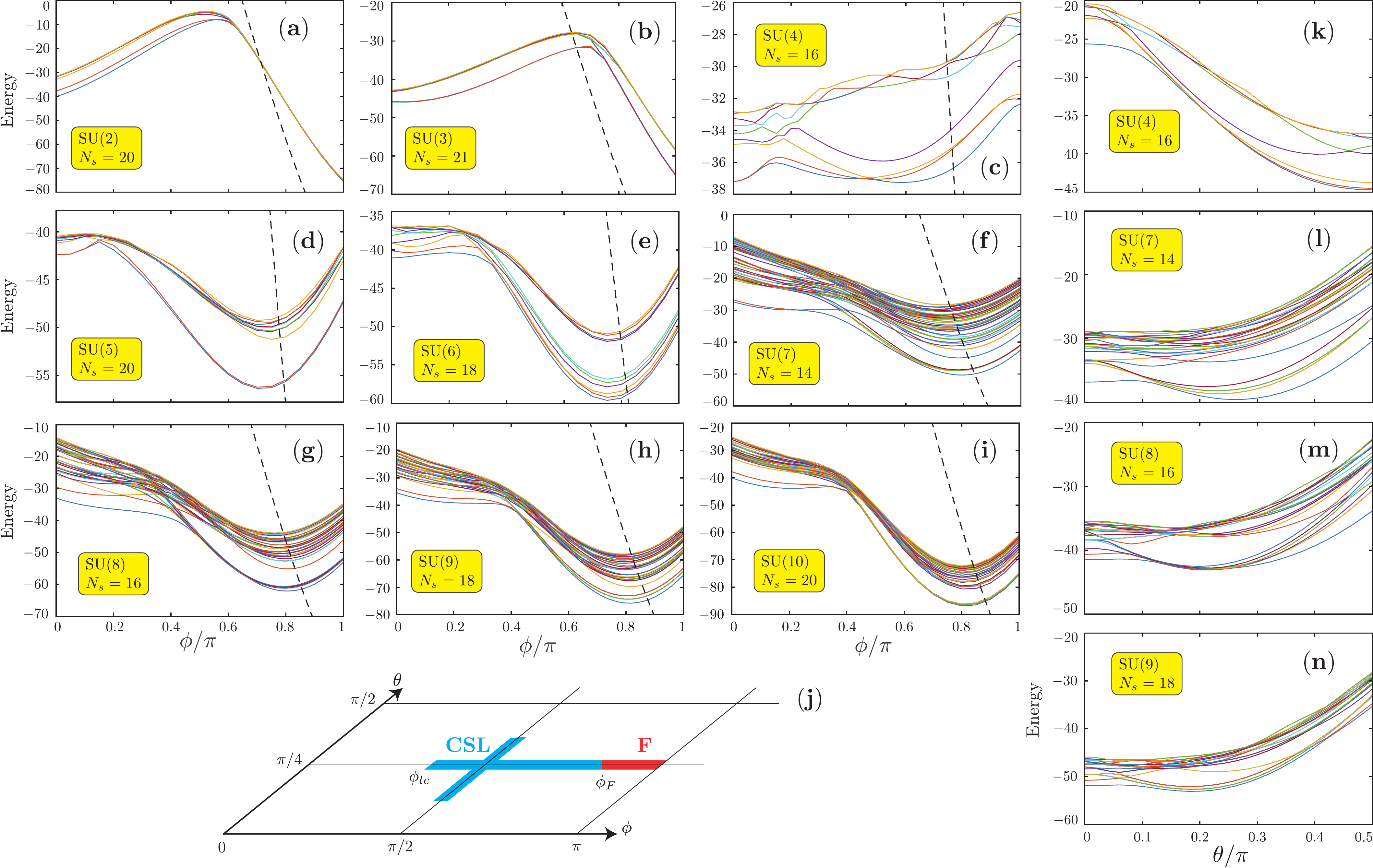} 	
	\caption{Low energy spectra  computed by ED in the SYT singlet basis on periodic clusters of $N_s=kN$ sites, $k\in\mathbb{N}$, and {\bf (a-i)} for a fixed value of $\theta=\pi/4$ as a function of $\phi$, for $N$ ranging from $2$ to $10$, or {\bf (k-n)} for a fixed value of $\phi=\pi/2$ as a function of $\theta\in[0,\pi/2]$, for $N=4,7,8,9$. Only 10 (40) lowest singlet levels are shown at small $N$ in {\bf (a-e)}  and {\bf (k)}  (larger $N$  in {\bf (f-i)} and {\bf (l-n)}).
		The $\phi$ and $\theta$ axis being discretized, lines connecting the data points are used  as guides to the eye (hence, levels crossings around $\phi_{\rm lc}$ may look like anti-crossings).
		$N$ degenerate or quasi-degenerate singlets (see Figs.\protect~\ref{fig:spectrumsinglets},\ref{fig:towers1} and text) are separated from the higher energy states by a gap, in an extended $(\phi,\theta)$  region around  $(\pi/2,\pi/4)$.
		The energy of the (fully polarized) ferromagnetic state ($E_{\rm ferro}=2\sqrt{2}N_s (2\cos{\phi}+\sin{\phi})$), crossing the singlet GS at $\phi=\phi_F$, is shown as a dashed line in {\bf (a-i)}. The location of the CSL and ferromagnetic phases along the cuts {\bf (c-i)} and {\bf(k-n)} are schematized in {\bf (j)}. Note that for $N=2,3$ {\bf (a,b)} the CSL is expected to extend all the way to $\phi=0$.
}
\label{fig:energies_vsPhi}
\end{figure*}

In the following, we will investigate model (\ref{eq:model}) using complementary ED and DMRG techniques, providing overwhelming evidence of  a stable topological CSL phase.
Various systems of different topology, as shown in Fig.~\ref{fig:systems_topo}, will be used. A torus geometry enables to probe bulk properties while a disk or a cylinder geometry, with one or two edges respectively, provides information on the existence and on the nature of edge modes.
More precisely, the topological nature of a CSL phase can be established from (i) the topological GS degeneracy~\cite{Wen1990} on periodic clusters, (ii) the existence of chiral edge modes~\cite{Wen1991a} both in open systems like Figs.~\ref{fig:systems_topo}(b) and in the entanglement spectra of (quasi-)infinite cylinders, and (iii) the content of the edge modes following closely the prediction of some chiral CFT theory.  The Abelian CSL expected here should be revealed by exactly $N$ quasi-degenerate  GS on a closed manifold and by the exact SU($N$)$_1$ WZW CFT content of its edge modes.   The second goal of the paper, beside establishing the existence of  the SU($N$)$_1$ CSL phase itself, is to provide its faithful representation in terms of an  SU($N$)-symmetric PEPS.
Following the prescription for $N=2$ and $N=3$, we shall focus on the $N=4$ case. Common features observed for PEPS with these three values of $N$ allow us to draw heuristic rules and conclusions for  general $N$.

\begin{table}[!h]
			\begin{tabular}{c    c   c  c}
		\hline
		\hline
		$N_s$ & ${\bf t}_1$ & ${\bf t}_2$  & point group  \\
		\hline
		 8     & $(2,2)$     & $(2,-2)$              & $C_{4v}$           \\
		11     & $(1,3)$     & $(3,-2)$              & $C_{2}$           \\
		12     & $(1,3)$     & $(4,0)$          & $C_{2}$           \\
		13      & $(2,-3)$     & $(3,2)$            & $C_{4}$           \\
		14     & $(1,4)$     & $(3,-2)$            & $C_{2}$           \\
		15    & $(1,4)$     & $(4,1)$            & $C_{2v}$           \\
		
		16  & $(4,0)$     & $(0,4)$           & $C_{4v}$           \\
		18    & $(3,3)$     & $(3,-3)$             & $C_{4v}$           \\
		19   & $(1,4)$     & $(4,-3)$           & $C_{2}$            \\
		20    & $(4,2)$     & $(-2,4)$          & $C_{4}$            \\
		21     & $(1,4)$     & $(5,-1)$          & $C_{2}$            \\
		\hline
		\hline
	\end{tabular}
	\caption{\label{tab:clusters}
		List of periodic clusters used here in ED: number of sites $N_s$, cluster size vectors ${\bf t}_1$ and ${\bf t}_2$, and point-group symmetry. Eigenstates can be labeled according to discrete momenta in the BZ.
	At high-symmetry points $\Gamma$, $X$ or $M$ of the BZ, eigenstates can be further labeled by the $C_4$-symmetry ($C_2$-symmetry) IRREP labels, $A$, $B$, $E_a$ and $E_b$ ($A$ and $B$) -- see Fig.~\ref{fig:towers1}.
}
\end{table}

\begin{figure}
	\centering
	\includegraphics[width=\linewidth]{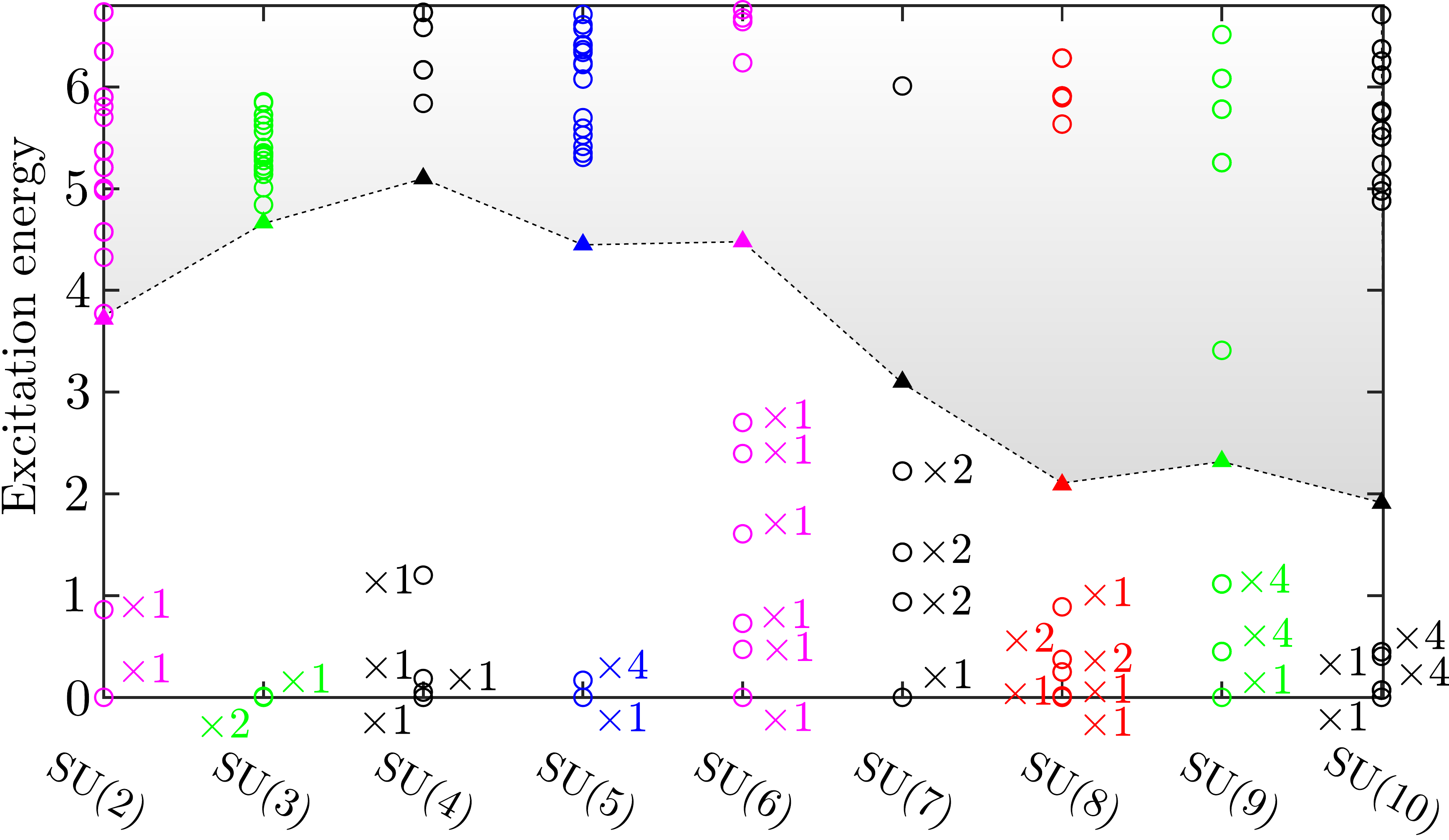}
	\caption{Zoom of the singlet low-energy spectra at $\theta=\pi/4$ and $\phi=\pi/2$, for $N$ ranging from $2$ to $10$, and the same cluster sizes as in Fig.~\ref{fig:energies_vsPhi}. The GS energy is subtracted off for better comparison between the various spectra. The exact degeneracy $g$ of each level is indicated on the plot as $\times g$. The first non-singlet excitation belonging to the adjoint IRREP above the $N$ quasi-degenerate low-energy singlets is shown as a filled triangle (see text). }
	\label{fig:spectrumsinglets}
\end{figure}

\begin{figure*}
	\centering
\includegraphics[width=0.95\linewidth]{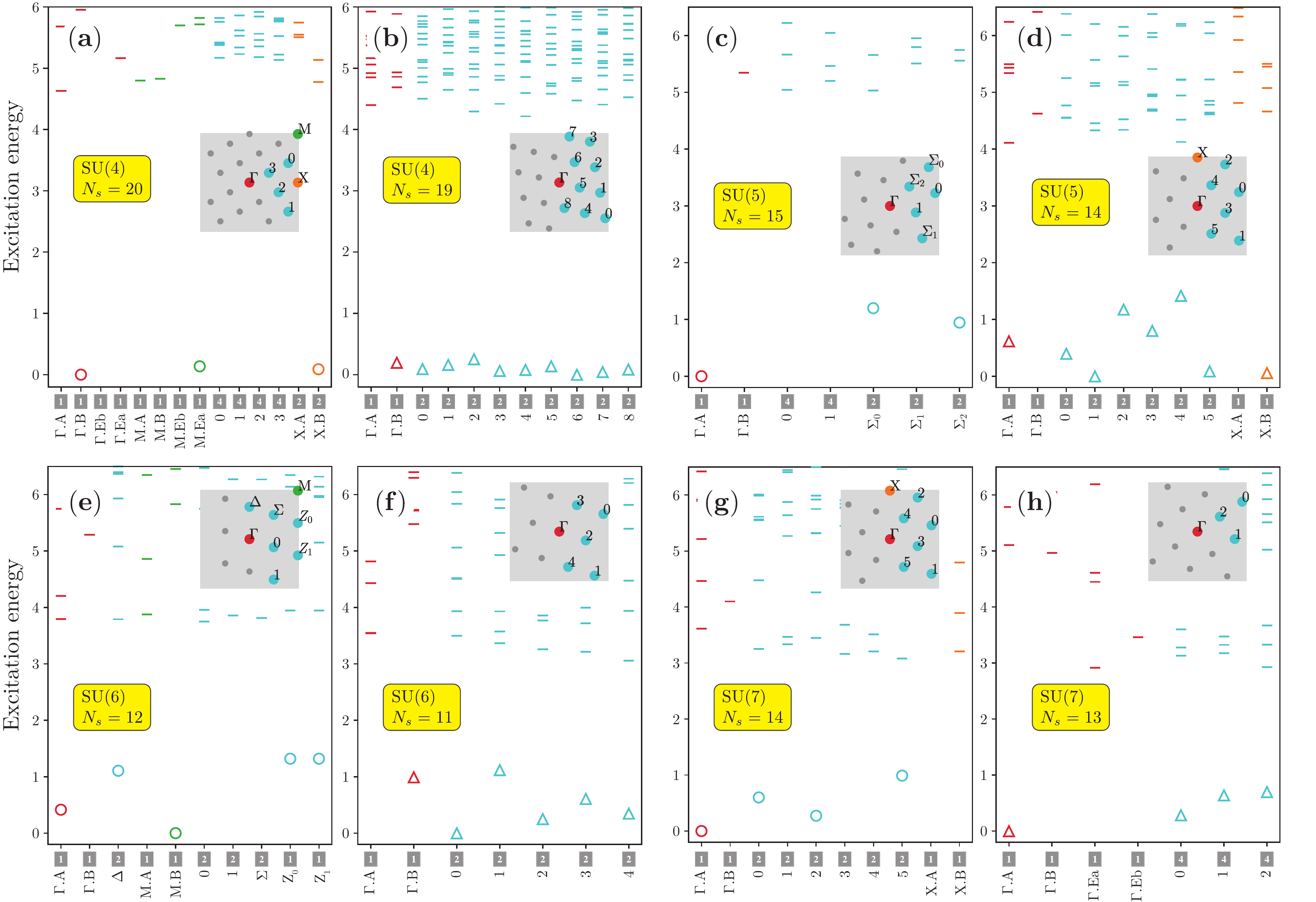}		
	\caption{Low-energy spectra on periodic clusters at fixed $\phi=\pi/2$ and for $\theta=\pi/4$ {\bf (a-d)} or $\theta=\pi/6$ {\bf (e-h)}. Clusters with site numbers $N_s=kN$ (left)  or $N_s=kN-1$ (right), $k\in\mathbb{N}$, are chosen to obtain $0$ and $1$ quasi-hole, respectively, in the putative CSL. The respective BZ with the allowed discrete momenta is shown on each plot as a gray square -- only non-equivalent momenta are labeled --
		and the number of equivalent momenta appearing are listed as grayed squared numbers.
 For  $N_s=kN$ (left), the GS manifold is composed of $N$ singlets (open circles). For $N_s=kN-1$ (right), it is composed of $N_s$ quasi-degenerate levels, one level at each cluster momentum. Each level is comprised of $N$ degenerate states forming a $\bar N$ anti-fundamental IRREP (open triangles). }
	\label{fig:towers1}
\end{figure*}

\section{Exact diagonalizations}
\label{sec:ED}

\subsection{Exact diagonalizations in the U(1) basis and in the  standard Young tableaux (SYT) basis}

We start this section by a brief review of the two distinct and complementary exact diagonalisation methods used in this work.

First, for periodic clusters (see Table~\ref{tab:clusters}), we can implement the spatial symmetries (and in particular the
translations) which allows us to both reduce the size of the matrix to diagonalize by a factor
typically equal to $N_s $ (where $N_s$ is the size of the cluster) and to directly obtain the momenta
associated to each eigenenergy.

However, as $N$ increases, EDs performed this way are severely limited by the size of the available
clusters since the dimension of the Hilbert space increases exponentially with $N_s$. A way to
overcome such limitations is to implement the SU($N$) symmetry and this is the second ED
protocol that we have employed here. In particular, when $N_s$ is a multiple of $N$, the ground state
of Hamiltonian~(\ref{eq:model}) is an SU($N$) singlet state for a wide range of parameters. The singlet sector has a
dimension much smaller than the one of the full Hilbert space. The gain to implement the full SU($N$)
symmetry and to look for the lowest energy states directly in the singlet sector is huge and
increases with $N$. For instance, for $N=10$ and $N_s=20$, the singlet sector has only dimension
16796, while the dimension of the full Hilbert space is $10^{20}$. In addition, to write the matrix
representing the Hamiltonian in the singlet subspace and in the sectors labeled by higher
dimensional SU($N$) irreducible representation (IRREP), we have employed the algorithms detailed in Refs.~\cite{Nataf2014,Wan2017},
which is mainly based on the use of Standard Young Tableaux and on the theory of the representation of the
permutation group.

In particular, it allows one to bypass the need for the Clebsch-Gordan coefficients, which can only
be calculated with an algorithm whose complexity also increases with $N$ (see Ref.~\cite{Alex2011}).
Typically, for the present problem, through this method, we can address clusters with $N_s \sim
20$ sites for $N$ up to 10. Note that contrary to the first ED method based on the implementation of
spatial symmetries, the momenta can only be accessed in a second stage: we first calculate the
eigenvectors and then the effect of translation or rotation on them.

\begin{figure*}
	\centering
\includegraphics[width=0.85\linewidth]{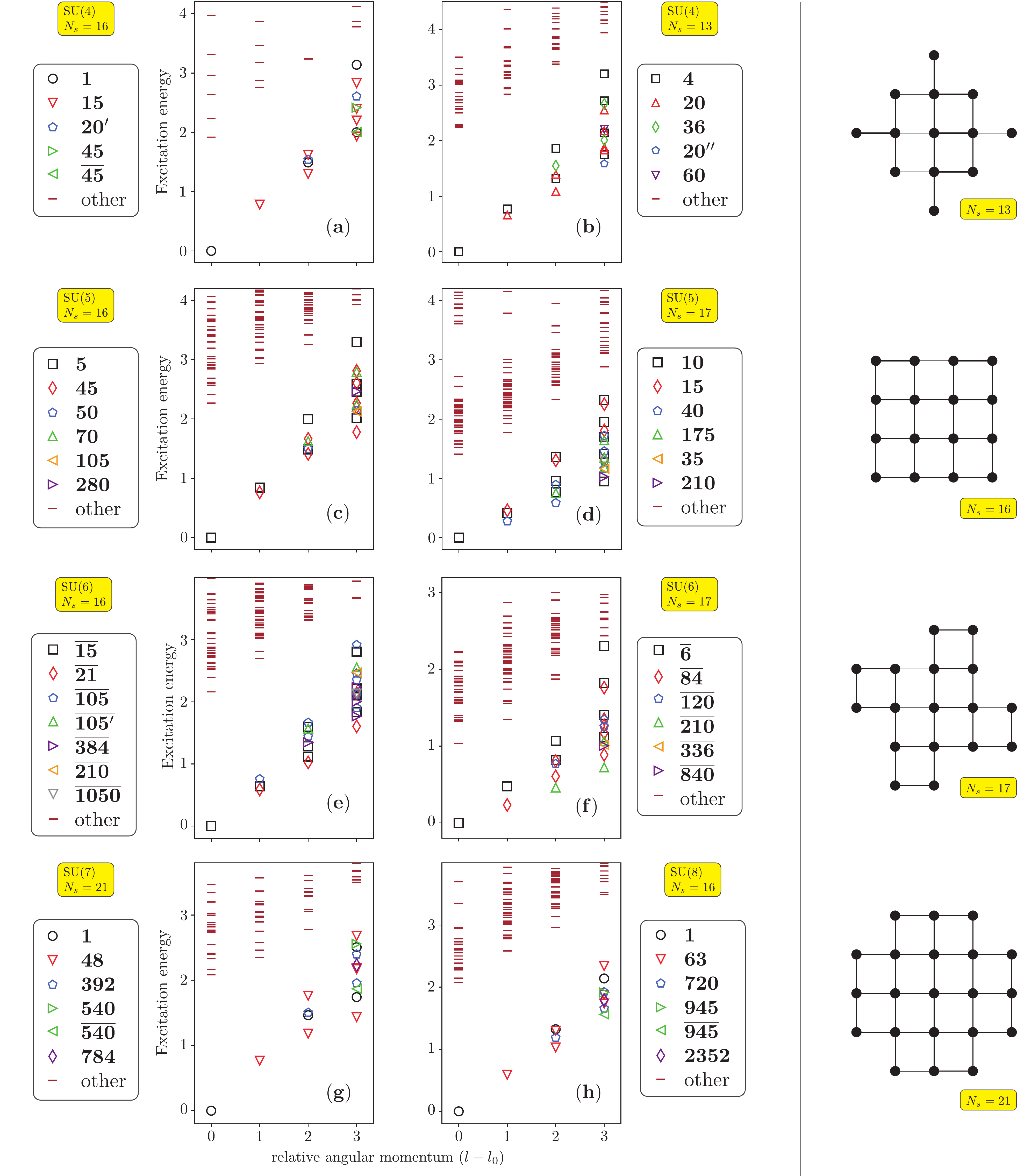}
	\caption{Low-energy spectra on open $C_{4}$-symmetric clusters depicted on the right-hand side of the figure, as a function of the angular momentum $l$ (with respect to   the GS angular momentum $l_0$), at fixed $\phi=\pi/2$ and for $\theta=\pi/4$ {\bf(a-d)} or $\theta=\pi/6$ {\bf(e-h)}.
			Symbols labeling the various SU($N$) IRREPs entering the chiral mode are shown in the legends.
The Young diagrams for the corresponding IRREPs can be identified using the tables in  ~\App{app:WZW}.
	The GS IRREPs are fully antisymmetric, and labeled by Young diagrams consisting of a single column of $r_0={\rm{mod}}(N_s,N)$ boxes, with degeneracy $\frac{N!}{(N-r_0)!r_0!}$.
Identifying $l-l_0$ with the Virasoro level $L_0$, all low-energy ToS in {\bf(a-h)} for $0\le l-l_0\le 3$ follow exactly the WZW CFT predictions  of Tables~\ref{tab:su4_1},\ref{tab:su4_4},\ref{tab:su5_5},\ref{tab:su5_10},\ref{tab:su6_15},\ref{tab:su6_6},\ref{tab:su7_1} and \ref{tab:su8_1}, respectively.
The only exception is the SU($6$)  $\bf\overline{15}$ (SU($8$)  $\bf 1$) tower, for which two multiplets  $\bf\overline{15}$ and  $\bf\overline{21}$ ($\bf  1$ and $\bf\overline{63}$) are missing in the $L_0=3$ Virasoro level.
	}
	\label{fig:towers2}
\end{figure*}

\subsection{Periodic clusters~:  bulk gap and GS manifold}

The results for $N=2$ and $N=3$ described above suggest that the existence of an Abelian CSL may be generic for arbitrary integer $N$. To investigate such an appealing  scenario, we start by examining, for larger $N$, the low-energy spectra obtained on $N_s$-site periodic clusters (see Table~\ref{tab:clusters} for details about clusters used).
For antiferromagnetic and frustrating couplings $J_1>0$, $J_2>0$, we expect the lowest-energy to belong to the  antisymmetric IRREP $\text{aIR}_N(r_0)$ defined by a Young tableau of $r_0$ vertical boxes,  $r_0={\rm{mod}}(N_s,N)$. In particular, in the case where $N_s$ is an integer multiple of $N$ ($r_0=0$), the low-energy states are expected to belong to the singlet subspace.  
However, at e.g. $\theta=\pi/4$, when increasing $\phi$ beyond $\phi=\pi/2$, $J_R$ changes sign and states belonging to the antisymmetric IRREP are gradually destabilized with respect to the completely symmetric (ferromagnetic) state of energy $E_{\rm ferro}/N_s=3J_1+8J_R$. In particular, we clearly see at $\theta=\pi/4$ a {\it macroscopic} energy gain (penalty) of the lowest-energy eigenstate of $\text{aIR}_N(r_0)$ with respect to the ferromagnetic state at $\phi=\pi/2$ ($\phi=\pi$) (see \App{app:FSS}). This fact indicates a transition at $\phi=\phi_F$ (somewhere in the range $\pi/2<\phi_F<\pi$) between one (or several) spin liquid phase(s) and a ferromagnetic phase. Note also that a detailed analysis of the
$2\times 2$ plaquette Hamiltonian in \App{app:plaquette},
shows that the antiferromagnetic states dominate the low energy regime, yet with the ferromagnetic regime in close proximity.

We now focus on the prospective spin liquid region discussed above and consider the case of $N_s=kN$, $k\in\mathbb{N}$, so that no quasiparticle excitations would be populating the GS of a CSL phase. To identify the exact nature(s) of the spin liquid(s), one needs to examin in details the low-energy singlet subspace (gap structure, degeneracies, etc...).
A selection of the singlet energy spectra for fixed $\theta=\pi/4$, plotted versus $\phi$ (for fixed $\phi=\pi/2$, plotted versus $\theta$), is shown in Fig.~\ref{fig:energies_vsPhi} for $N$ ranging from $2$ to $10$ (for $N=4,7,8,9$).
For all the values of $N$ studied here,  in a broad interval of $\phi$ ($\phi<\phi_F$) or $\theta$ values, a clear gap is observed between a group of degenerate and quasi-degenerate states and the rest of the singlet spectrum. 
Interestingly, for $\theta=\pi/4$ and $N>3$, we observe level crossings occuring in the singlet subspace at some value of $\phi_{\rm lc}<\pi/2$, suggesting the existence of two different gapped phases. For $0\le\phi<\phi_{\rm lc}$, we observe a two-fold quasi-degenerate GS manifold  within the singlet subspace which are translationally invariant but which break the lattice point group $\pi/2$-rotation symmetry~\footnote{Both states are translationally invariant and have different $\pm 1$ characters under $\pi/2$-rotation, for $C_4$-symmetric clusters.}. This could correspond to a nematic valence cluster state as also seen in SU($2$) spin-1 models~\cite{Haghshenas2018,Chen2018c}. Note that, as a finite-size effect, the ground state of the total spectrum for small $\phi$ and $\theta$ around $\pi/4$ is not necessarily a singlet state when $N_s<N^2$ (see \App{app:FSS}). A more careful investigation of this  phase, although interesting, is beyond the scope of this work and left for a future study.

We now move to a closer inspection of the gapped spin liquid phase seen for $N=2,3$ and $\phi<\phi_F$, and for $N>3$ and $\phi_{\rm lc}<\phi<\phi_F$, and identify it as a CSL. Interestingly, we note that $\phi=\pi/2$ -- corresponding to a pure imaginary 3-site cyclic permutation --  is always located within this gapped phase (Note, for $N=3$, $\phi=\pi/4$ instead was chosen in Ref.~\cite{Chen2020}).
This gapped phase is also stable within a significant range of the parameter $\theta$, around $\theta=\pi/4$ and $\phi=\pi/2$, e.g. also at $\theta=\pi/6$.
Hence, in the following, we shall mostly report results obtained at fixed $\phi=\pi/2$ (i.e. for a pure imaginary 3-site permutation)  and for $\theta=\pi/4$ or, occasionally, $\theta=\pi/6$.

To identify the type of (singlet) gapped phase, we now investigate the exact degeneracy and the quantum numbers of the singlet GS manifold. Fig.~\ref{fig:spectrumsinglets} shows a zoom of the low-energy spectra at $\theta=\pi/4$ and $\phi=\pi/2$, with the exact degeneracy of each level below the gap. A simple counting shows that there are exactly $N$ states below the gap. Note that the first excitation defining the gap does not belong to the singlet sector but most often belongs to the adjoint IRREP of dimension $N^2-1$, except for some of the largest values of $N$ (like $N=9$) for which finite size effects are the strongest. This is an extension of the SU($2$) case where the first excitation in antiferromagnetic spin liquids are typically spin-1 ``magnons". In the thermodynamic limit, the gap in the singlet sector should be bounded from above by twice the true ``magnetic" gap as two isolated ``magnons" can fuse into a singlet. If  a singlet bound state occurs between two magnons, the singlet gap is then strictly smaller than twice the magnon gap.

The above observation of the $N$-fold degeneracy of the GS space suggests that the gapped phases indeed correspond to Abelian
$\text{SU}(N)_1$ chiral spin liquids. As realized already for $N=3$ in Ref.~\cite{Chen2020}, it is possible to obtain, for arbitrary $N$, the exact momenta  of the various states in the GS manifold expected for an Abelian SU($N$)$_1$ CSL.
This  can be inferred from a simple
generalized exclusion principle (GEP) ~\cite{Estienne2012,Sterdyniak2013}
with clustering rules (see  ~\App{app:Exclude} for details).
As a final check for periodic systems, we then focus on two distinct commensurability relations between the cluster size $N_s$ and $N$; either, (i) $N_s=kN$, $k\in\mathbb{N}$, for which, as above, the GS contains no quasi-particle and (ii) $N_s=kN-1$, $k\in\mathbb{N}$, for which, a single quasi-hole populates the GS.
Note that in case (ii), $r_0=N-1$ so that the IRREP of the GS manifold is the $\bar N$ anti-fundamental IRREP.
The GEP implies a GS (quasi-)degeneracy of  $N$ and $N_s$ for (i) and (ii), respectively.
This is indeed observed as shown in Fig.~\ref{fig:towers1}. The predictions of the GEP are even more precise, providing all GS momenta expected for the (Abelian) CSL on every periodic cluster (see  ~\App{app:Exclude} for details on the way momenta are assigned). We have checked that -- in most cases -- all GS momenta reported in Fig.~\ref{fig:towers1} match the ones predicted by the heuristic rules.  In particular, for $N_s=kN-1$, the GS manifold is made of exactly one $\bar N$ (antifundamental) IRREP at each cluster momentum. Rare failures of the GEP rules (which may be attributed to cluster shapes, etc$\ldots$) to predict the correct momenta will be discussed in  ~\App{app:Exclude}.

Interestingly, the above features predicted and observed in the case of a single quasi-hole can be understood using a simple physical argument. If the single quasi-hole would be static, it could be placed on each of the $N_s$ sites of the cluster, and this, for each of the $N$ topological (singlet) sectors, hence spanning a $N_s N$-dimensional Hilbert space. The effective hopping allows the quasi-hole states to form a weakly dispersing band below the gap, hence with $N$ states at every momentum.  From the SU($N$)-symmetry, these $N$ states should form a single multiplet belonging to the $\bar N$ (antifundamental) IRREP, as predicted by the GEP and found numerically.

\subsection{Open systems~: edge physics and CFT content
\label{sec:open}}

The previous results give strong evidence of the CSL nature of the GS of the model, for the parameters chosen, from its bulk properties on periodic systems (topologically equivalent to tori). We complete the identification of the CSL phase by the investigation by ED of open clusters. The existence of a chiral edge mode fulfilling the SU($N$)$_1$ WZW CFT should be reflected in the precise content of its low-energy spectrum. By choosing finite-size clusters with (i) open boundaries and (ii) $C_4$ point-group symmetry, we can investigate the low-energy spectrum as a function of the angular momentum, $l=0,\pm 1, 2$ (mod[4]) and reveal a single chiral branch linearly dispersing only in one direction, as expected. At a given $N$, changing the cluster size $N_s$ -- whenever such a $C_4$-symmetric cluster is available -- enables to change the topological sector defined by the integer $r_0=\text{mod}(N_s,N)$, $r_0=0,\cdots,N-1$. Indeed, each topological sector is characterized by the SU($N$) IRREP of its GS, corresponding to the antisymmetric IRREP $\text{aIR}_N(r_0)$ (defined by a Young tableau of $r_0$ vertical boxes), and can then be reached whenever $N_s=kN+r_0$. Note that the dimension of $\text{aIR}_N(r_0)$ is given by $\frac{N!}{(N-r_0)! r_0!}$.

The ED investigation of the chiral edge modes has been carried out on two types of open systems, all exhibiting $C_{4}$ symmetry with respect to  the cluster center.  The first type of clusters is build around a central site by adding successive shells of 4 sites at 90-degree angles. The second type of open clusters are built in the same way but from a center $2\times 2$ plaquette.The  $13$-site, $17$-site and $21$-site ($16$-site) clusters  belongs to the first (second) category, as shown on the right-hand side of  Fig.~\ref{fig:towers2}.
Note that the $17$-site cluster is "chiral", i.e. it breaks reflection symmetry (parity), and spectra for $J_I>0$ and $J_I<0$ are expected to be (slightly) different. Here, $J_I>0$ and the $P_{ijk}$ permutation is assumed counterclockwise.
 ED spectra obtained on such clusters for $N=4,5,6,7,8$ are shown in Fig.~\ref{fig:towers2}, for $\phi=\pi/2$ and $\theta=\pi/4$ or $\pi/6$ (as specified in the caption). In all cases, we observed a rather sharply-defined low-energy chiral edge mode, i.e. a group of levels (i) well-separated from higher-energy levels by a gap, (ii) following a linear dispersion with respect to   the angular momentum and (iii) with a very precise and non-trivial content in terms of SU($N$) multiplets. Each edge mode is characterized by its GS given by the antisymmetric IRREP
$\text{aIR}_N(r_0)$. For each pair $(N,r_0)$ occurring in Fig.~\ref{fig:towers2}, we have computed the expected ``tower of states'' (ToS) generated by $\text{aIR}_N(r_0)$, as predicted by the SU($N$)$_1$ WZW CFT -- see  ~\App{app:WZW}.
Numerically, one can use $(N-1)$ $U(1)$ quantum numbers to diagonalize the Hamiltonian and identify the IRREP content for each group of {\it exactly}  degenerate levels.
A careful check shows that, generically, the quantum numbers of the chiral edge mode spectra match exactly the WZW CFT ToS predictions (identifying the angular momentum with the Virasoro level $L_0$), providing a real hallmark of the CSL phase.
For two cases corresponding to the smallest $N_s=16$ cluster, a small number of multiplets in the CFT predictions are missing in Fig.~\ref{fig:towers2}. We have explicitly checked that finite-size effects can indeed lead to incomplete towers.

\section{DMRG}

For characterizing chiral topological states, the correspondence between the entanglement spectrum and the conformal tower of states is a fingerprint evidence. While DMRG is in principle suited for this purpose, a technical difficulty is that the characterization of topological order requires the full set of (quasi-)degenerate ground states and, furthermore, these states should be combined into the so-called minimally entangled state (MES) basis~\cite{ZY2012}. In this section, we use a two-step procedure to accomplish this task: i) build Gutzwiller projected parton wave functions which describe the SU$(N)_1$ CSL, use them to construct the MES basis on the cylinder, and convert them into MPS; ii) initialize DMRG with the parton-constructed MES basis. This strategy allows us to find the full set of $N$ (quasi-)degenerate ground states in the MES basis. The parton picture also helps us to identify the correspondence between the entanglement spectrum and the SU$(N)_1$ conformal towers.

\subsection{Parton wave functions}
In this subsection, we outline the parton approach to construct trial wave functions for the SU$(N)$ CSL model. To construct the minimally entangled states (MESs)~\cite{ZY2012}, we use a fermionic parton representation of the SU$(N)$ generators~\cite{AAA1965,WXG1991,AA1998}, $S^{\mu}_i=\sum_{\sigma\sigma'}c^{\dagger}_{i\sigma}T^{\mu}_{\sigma\sigma'}c_{i\sigma'}$,
where $T^{\mu}_{\sigma\sigma'}$ are matrix representations of the SU$(N)$ generators in the fundamental representation, and $c^{\dagger}_{i\sigma}$ is the creation operator at site $i$. A local constraint $\sum_{\sigma}c_{i\sigma}^{\dagger}c_{i\sigma}=1$ has to be imposed to ensure that singly-occupied fermions represent the $N$ states in the SU$(N)$ fundamental representation, i.e., $\ket{\sigma}=c^{\dagger}_{\sigma}\ket{0}$ (site index suppressed), with $\ket{0}$ being the vacuum of partons.
The SU$(N)$ CSL with SU$(N)_1$ topological order can be constructed by Gutzwiller projecting a fully occupied $C=1$ Chern band of fermionic partons, where $C$ is the Chern number.
To have a systematic construction for all $N$, we design the following quadratic Hamiltonian for partons on a square lattice:

\begin{equation}
	\begin{split}
		H_\mathrm{p} = &-\sum_{m,n,\sigma} \big( t_x c_{m+1,n,\sigma}^{\dagger} c_{m,n,\sigma} + t_y e^{im\varphi} c_{m,n+1,\sigma}^{\dagger} c_{m,n,\sigma} \big) \\
		&- \sum_{m,n,\sigma} \big( t_2 e^{i(m\varphi\pm\pi/N)} c_{m\pm1,n+1,\sigma}^{\dagger} c_{m,n,\sigma} \big) + \mathrm{h.c.} \\
    &- \mu\sum_{m,n,\sigma} c_{m,n,\sigma}^{\dagger} c_{m,n,\sigma}.
	\end{split}
	\label{eq:H_tb}
\end{equation}
The phase $\varphi$ is chosen to be $2\pi/N$, so that the flux through each square plaquette is $2\pi/N$ and each triangular plaquette is $\pi/N$.
To minimize finite-size effects, we maximize the band gap by choosing $t_2=t_y/2$.

\begin{figure}
\includegraphics[width=\linewidth]{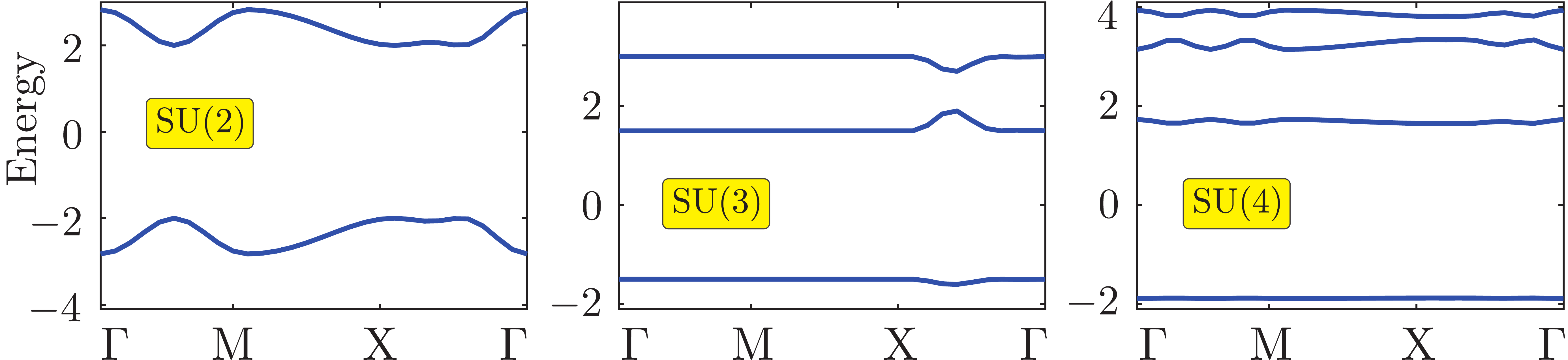}
	\caption{Band structures of the parton Hamiltonian on the torus along high symmetry directions for $N=2,3$ and $4$.
		We set $t_x=t_y$ for $N=2$ and $4$, and $t_x=t_y/2$ for $N=3$.
	}
	\label{fig:torus}
\end{figure}

\begin{figure}
	\includegraphics[width=0.9\linewidth]{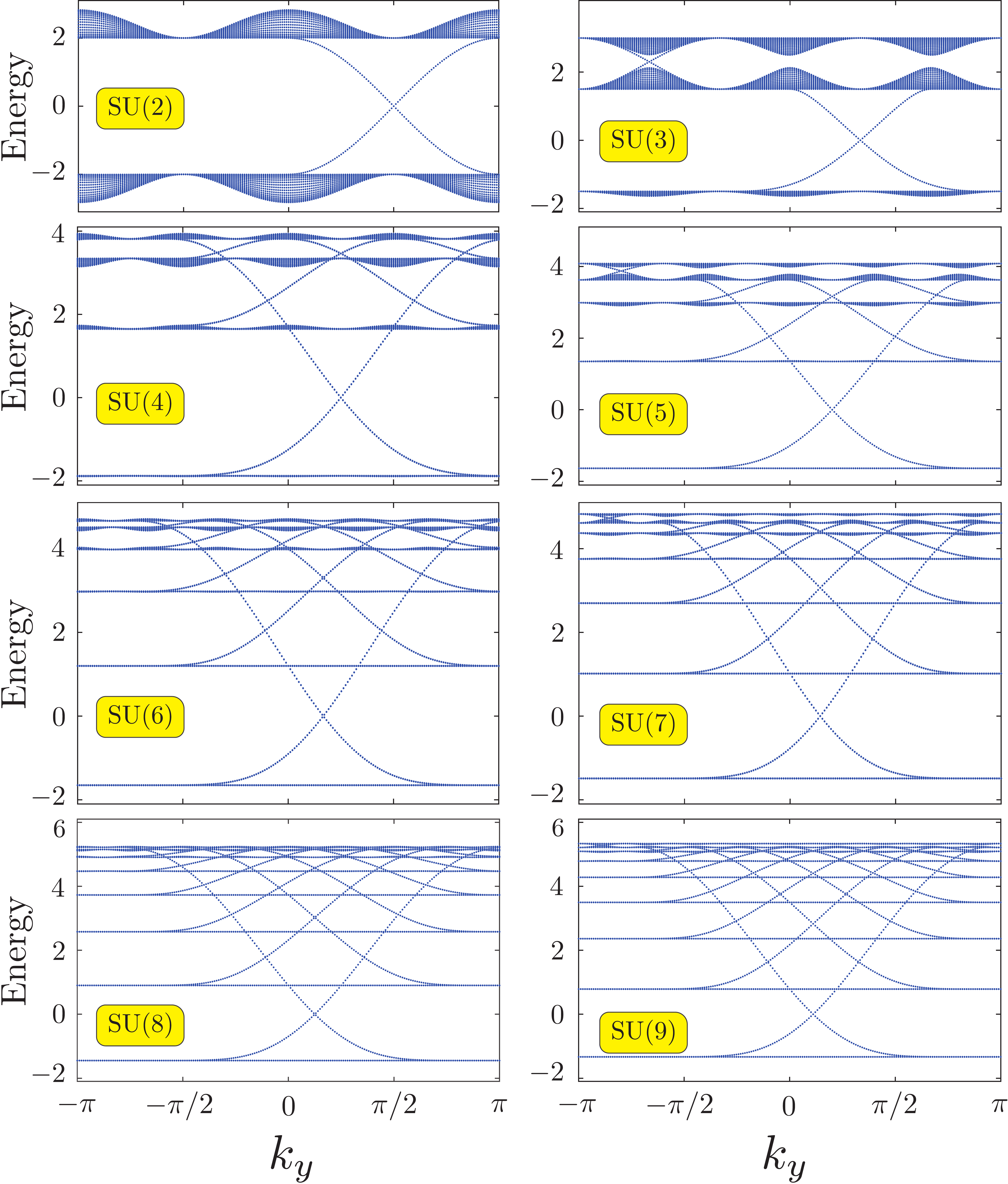}
	\caption{The parton single-particle levels including the  edge states on a wide cylinder for $N=2$ to $9$.
        Filling the Fermi sea up to zero energy corresponds to a filling fraction $1/N$. This fully occupies
        the lowest parton band as well as the edge states up to the
        degenerate zero modes at the single-particle momentum $k_y = \pi/N$.
		These exact zero modes, denoted by $d_{L\sigma}$ and $d_{R\sigma}$, are localized at the left and right boundaries of the cylinder, respectively.
}
\label{fig:parton}
\end{figure}

The design of the parton Hamiltonian~(\ref{eq:H_tb}) follows a lattice discretization of the Landau level problem, i.e., 2D electrons in a strong magnetic field (with the Landau gauge). Under periodic boundary conditions (torus geometry), the fluxes in the square/triangular plaquette are chosen such that there are $N$ bands with the lowest band having Chern number $C = 1$ (see Fig.~\ref{fig:torus}). The $N=2$ case has been considered previously in Refs.~\cite{ZY2012,THH2013,MJW2015,WYH2020}, which was used to constructed Gutzwiller projected wave functions representing the SU$(2)$ CSL of Kalmeyer-Laughlin type. For $N>2$, the lowest band becomes flat and indeed resembles the lowest Landau level. The trial wave functions for describing the SU$(N)_1$ CSL are obtained by (i) tuning the chemical potential $\mu$ such that the lowest band is completely filled and all others empty, yielding a filling of $1/N$ on the lattice when also including the edge mode (see \Fig{fig:parton}) and (ii) Gutzwiller projecting the Fermi sea with fully occupied lowest band. Strictly speaking, this construction does not depend on the flatness of the $C = 1$ band. Here, our extra requirement of a nearly flat band serves another purpose: the single-particle wave functions of a flat band can be made more localized, which helps to suppress the entanglement growth when converting Gutzwiller projected wave functions into MPS~\cite{WYH2020}. Last but not the least, this parton Hamiltonian is also designed to support exact zero modes on the cylinder, which, as we shall see, are important for constructing the MES basis.

\begin{figure}
		\captionsetup[subfigure]{position=top,captionskip=0pt}
	\centering
\includegraphics[width=\linewidth]{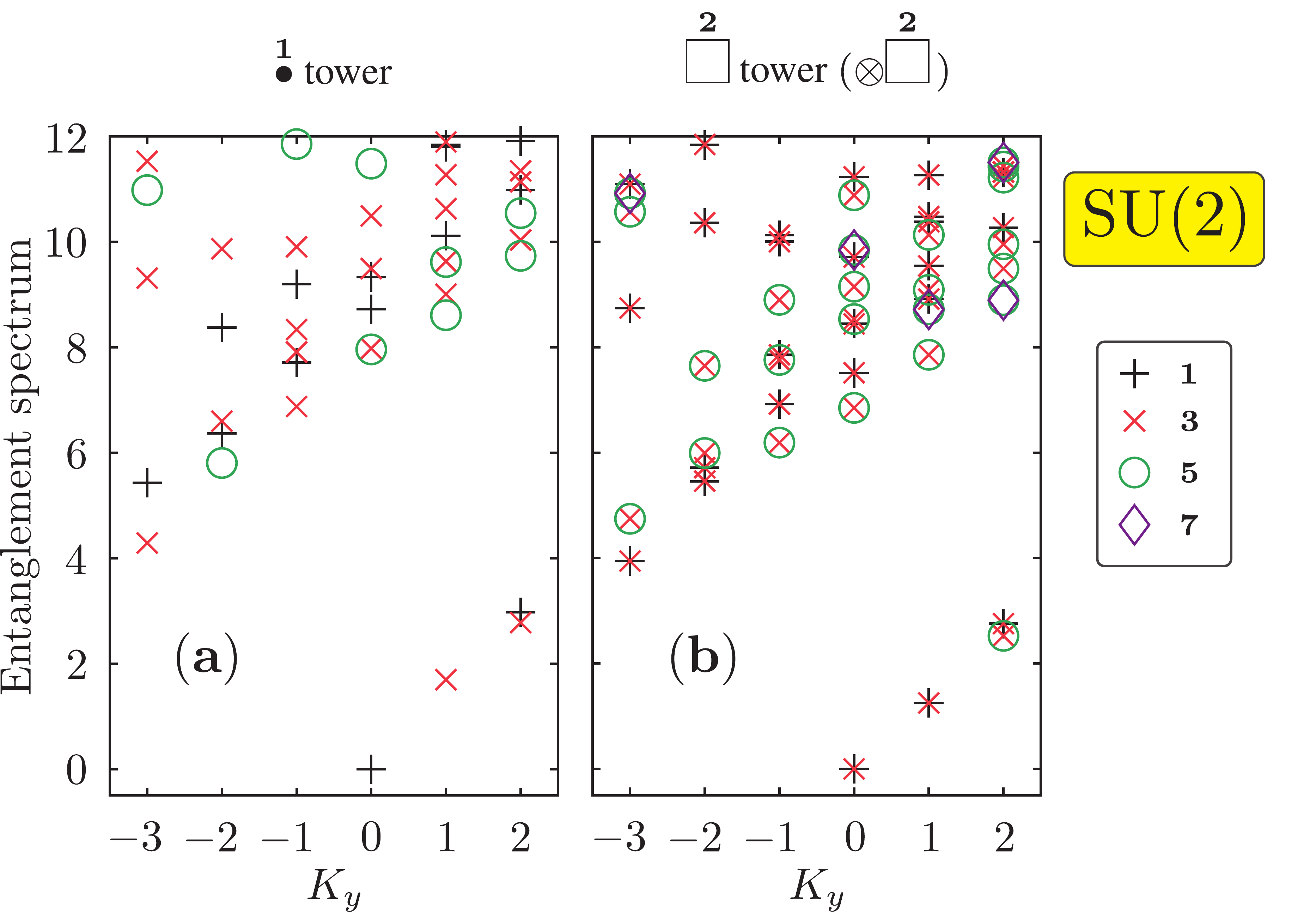}
	\caption{The entanglement spectra on width-6 cylinders for SU$(2)$ CSLs. {\bf(a)} Identity sector. {\bf(b)} Semion sector ($\otimes\frac{1}{2}$).   Identifying $K_y$ with the Virasoro level $L_0$, the content of the chiral branches agrees exactly with the CFT predictions of tables~\ref{tab:su2_1} and \ref{tab:su2_copies} up to $K_y=4$  (mod[6]). }
	\label{fig:es2}
\end{figure}

\begin{table}
	\centering
	\caption{SU$(2)_1$ WZW model -- The direct product of the conformal tower of the spin-$1/2$ primary (left - see Table~\ref{tab:su2_2} in  ~\App{app:WZW}) with a spin-$1/2$ gives a new tower (right) with a doubling of the number of states  in each Virasoro level indexed by $L_0$.}
	\begin{tabular}{|c|l|c|l|}
		\cline{1-2}\cline{4-4}
		$L_0$ & \multicolumn{1}{c|}{ $\protect\su{0}{1}{2}$\text{ tower}} && \multicolumn{1}{c|}{$\protect\su{0}{1}{2}\text{ tower}\otimes \protect\su{0}{1}{2}$} \\ \cline{1-2}\cline{4-4}
		$0$ & \Yvcentermath1$\su{1}{1}{2}$ && \Yvcentermath1$\su{1}{0}{1}\oplus\su{1}{2}{3}$ \\ \cline{1-2}\cline{4-4}
		$1$ & \Yvcentermath1$\su{1}{1}{2}$ &\multirow{3}{*}{$\rightarrow$} & \Yvcentermath1$\su{1}{0}{1}\oplus\su{1}{2}{3}$\\ \cline{1-2}\cline{4-4}
		$2$ & \Yvcentermath1$\su{1}{1}{2}\oplus\su{1}{3}{4}$ && \Yvcentermath1$\su{1}{0}{1}\oplus\su{2}{2}{3}\oplus\su{1}{4}{5}$\\ \cline{1-2}\cline{4-4}
		$3$ & \Yvcentermath1$\su{2}{1}{2}\oplus\su{1}{3}{4}$ && \Yvcentermath1$\su{2}{0}{1}\oplus\su{3}{2}{3}\oplus\su{1}{4}{5}$\\ \cline{1-2}\cline{4-4}
		$4$ & \Yvcentermath1$\su{3}{1}{2}\oplus\su{2}{3}{4}$ && \Yvcentermath1$\su{3}{0}{1}\oplus\su{5}{2}{3}\oplus\su{2}{4}{5}$\\ \cline{1-2}\cline{4-4}
	\end{tabular}
	\label{tab:su2_copies}
\end{table}

For our purpose, we shall consider the cylinder geometry (with circumference $N_y$) rather than the torus geometry, with open boundaries in the $x$ direction and a periodic (or twisted) boundary condition in the $y$ direction. This allows us to characterize the MESs via the entanglement spectrum~\cite{LH2008,THH2013}, and to use these wave functions to initialize our DMRG simulations \cite{JHK2020b}.

By diagonalizing the parton Hamiltonian~(\ref{eq:H_tb}) on the cylinder, we obtain a set of single-particle orbitals composed of local operators, $d_{k\sigma}^{\dagger} = \sum_{m,n} A_{m,n}(k) c_{m,n,\sigma}^{\dagger}$.
For $N=2$, it is known that the exact zero modes play an important role in constructing the MESs~\cite{THH2013,WYH2020}.
These exact zero modes, denoted by $d_{L\sigma}$ and $d_{R\sigma}$, localize at the two boundaries of the cylinder.
Their occurrence at the single-particle momentum $k_y=\pi/2$ requires that for
$\mathrm{mod}(N_y,4)=0$ (2),
the parton Hamiltonian has periodic (antiperiodic) boundary condition in the $y$ direction.
The two MESs with $S_z=0$ are then written as Gutzwiller projected wave functions,
$\ket{\Psi_1} = P_G d_{L\uparrow}^{\dagger}d_{R\downarrow}^{\dagger}\ket{\Phi}$
and $\ket{\Psi_2} = P_G d_{L\uparrow}^{\dagger}d_{L\downarrow}^{\dagger}\ket{\Phi}$,
where $P_G$ imposes the single-occupancy constraint at each site and $\ket{\Phi}$ is the state with all parton modes below the zero modes being fully occupied.
In this representation, it is transparent that the zero mode $d^\dag_{L(R)\sigma}$ creates a semion carrying spin-1/2 (with spin projection $\sigma$) at the left (right) boundary of the cylinder.
It was found~\cite{WYH2020} that the entanglement spectra of $\ket{\Psi_1}$ and $\ket{\Psi_2}$ correspond to the conformal towers of states of the chiral SU(2)$_1$  WZW model in its spin-1/2 (semion) and spin-0 (identity) sectors, respectively.
To qualify as the (quasi-) degenerate ground states of chiral spin liquids, the wave functions should be SU($2$) spin singlets.
While $\ket{\Psi_2}$ is manifestly a spin singlet, $\ket{\Psi_1}$ needs to be combined with $P_G d_{L\downarrow}^{\dagger}d_{R\uparrow}^{\dagger}\ket{\Phi}$ to form a spin singlet
$\ket{\tilde{\Psi}_1} = P_G (d_{L\uparrow}^{\dagger}d_{R\downarrow}^{\dagger}-d_{L\downarrow}^{\dagger}d_{R\uparrow}^{\dagger})\ket{\Phi}$.
However, the entanglement spectrum of $\ket{\tilde{\Psi}_1}$ would then correspond to two copies of spin-1/2 conformal towers due to the entanglement cut of an additional nonlocal singlet formed by a pair of two spin-1/2 semions at the boundaries~\cite{Li13}.

This parton construction of MESs for the SU$(2)$ CSL can be naturally generalized to the SU($N$) CSL.
To allow for exact zero modes, the hopping parameters in Eq.~(\ref{eq:H_tb}) are chosen as $t_x=t_y$ if $N$ is even, and $t_x = t_y \cos(\pi/N)$ otherwise.
This ensures that the exact zero modes, $d_{L\sigma}^{\dagger}$ and $d_{R\sigma}^{\dagger}$, appear at $k_y=\pi/N$ (see Fig.~\ref{fig:parton}), which is always allowed for a suitably chosen boundary condition (i.e., periodic or twisted) in the $y$ direction.
Occupying $N$ of these boundary modes
distributed arbitrarily over left and right boundaries
ensures that the total momentum of the state in
$y$-direction is zero. As such this is then consistent
with a width-$N$ cylinder with plain periodic boundary
conditions around the cylinder.

With that, MESs belonging to $N$ different topological sectors can be written in analogy to the SU($2$) case as
\begin{align}
  \ket{\Psi_p} = P_G \, \Bigl(
  d_{L 1}^{\dagger} \ldots d_{L p}^{\dagger} d_{R,p+1}^{\dagger} \ldots
  d_{R N}^{\dagger} \ket{\Phi} \Bigr)
\label{eq:SUn_wavefunc1}
\end{align}
$p=0,\ldots,N$.
Here $d^\dag_{L(R)\sigma}$ creates an elementary anyon of the
chiral SU($N$)$_1$ theory and also transforms under the SU($N$)
fundamental representation.
Therefore $p=0$ ($N$) corresponds to all $N$ anyons either
located, equivalently and respectively,
at the left or right boundary.
The entanglement spectra of these states $\ket{\Psi_p}$
should be
in one-to-one correspondence with the $N$ Kac-Moody conformal
towers of the chiral SU($N$)$_1$ WZW model, whose $N$ primary
fields are labeled by Young diagrams with $p$
vertical boxes, respectively. However, except for
$p=0$ or $N$
the states above do not yet describe proper
SU($N$)
multiplets. For a more direct comparison with CFT,
the $N$ boundary modes need to be antisymmetrized
over all flavors into an overall SU($N$) singlet.
The corresponding SU($N$) singlets can be written as
\begin{eqnarray}
   \ket{\tilde{\Psi}_p} =P_G \,\Bigl(
   \varepsilon_{\sigma_1{\ldots}\sigma_N} d_{L \sigma_1}^{\dagger}\!\ldots
   d_{L \sigma_{p}}^{\dagger} d_{R \sigma_{p+1}}^{\dagger}  
   \!\ldots d_{R \sigma_N}^{\dagger} \ket{\Phi}\Bigr)
\text{,}\quad
\label{eq:SUn_wavefunc2}
\end{eqnarray}
where $\varepsilon_{\sigma_1\ldots\sigma_N}$ is the totally
antisymmetric Levi-Civita tensor.
\Eq{eq:SUn_wavefunc2} indicates that for non-identity sectors,
multiple branches contribute to the entanglement spectrum. The
number of branches is
$\frac{N!}{(N-p)!p!}$,
where $N!$
comes from the Levi-Civita tensor, and the factors
$(N-p)!$ and $p!$
account for the antisymmetrization of the anyons on
the left or right edge, represented by
  $N-p$ or $p$
vertical boxes in the corresponding Young tableau,
IRREPS $\bar{p}$ and $p$, respectively.
Note that as such this precisely also corresponds to the dimensions
$\mathrm{dim}(\mathrm{aIR}_N(p)) = \mathrm{dim}(\mathrm{aIR}_N(N-p))$
[see \Sec{sec:open} above].

Using the matrix-product-operator matrix-product-state (MPO--MPS) method of Ref.~\cite{WYH2020} to implement the parton construction, we can express the
filled Fermi sea of the above parton wave function
$\ket{\tilde{\Psi}_p}$ as an MPS.
The principal idea for that is as follows:
(i) the vacuum state $\ket{0}$ is an MPS with bond dimension $D=1$.
(ii) the non-local parton operator $d_{k\sigma}^{\dagger}$,
subject to Wannier localization, can be written as an MPO of
bond dimension $D=2$.
(iii) the MPOs $d_{k\sigma}^{\dagger}$ are applied
sequentially onto the MPS with possible
compression after each step, resulting in an MPS with a finite
bond dimension that represents a filled Fermi sea.
(iv) the Gutzwiller projector $P_G=\prod_{\ell=1}^{L}P_{\ell}$
is applied to separately enforce the local constraint,
$\sum_{\sigma}c_{m,n,\sigma}^{\dagger}c_{m,n,\sigma}=1$, on each
site to recover the correct local physical subspace.

\begin{figure*}
	\includegraphics[width=\linewidth]{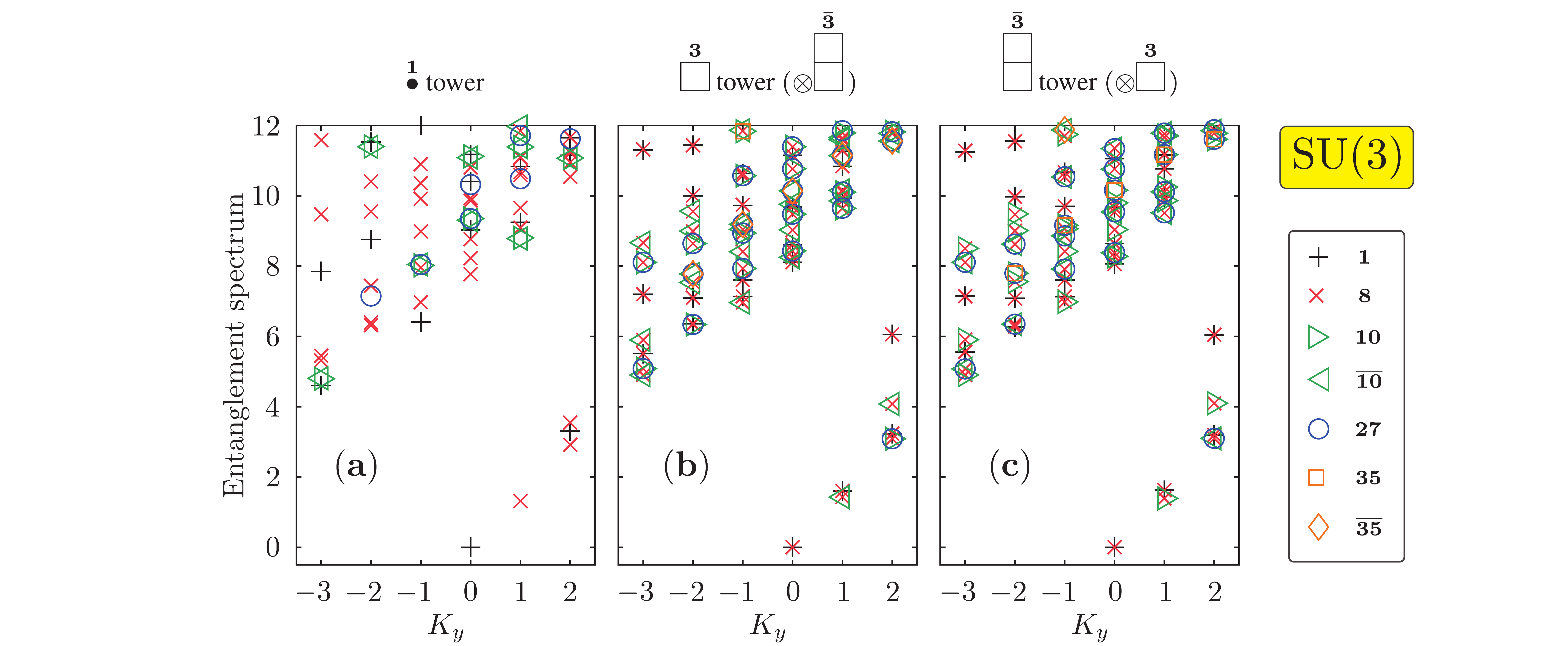}
	\caption{The entanglement spectra on width-6 cylinders for SU$(3)$ CSLs. {\bf(a)} Identity sector. {\bf(b)} $\protect\su{0}{1}{\bf 3}$ sector ($\otimes\protect\su{0}{1,1}{\bar{3}}$). {\bf(c)} $\protect\su{0}{1,1}{\bf\bar 3}$ sector ($\otimes\protect\su{0}{1}{3}$).   Identifying $K_y$ with the Virasoro level $L_0$, the content of the chiral branches  agrees exactly with the CFT predictions of tables~\ref{tab:su3_1} and \ref{tab:su3_copies} up to $K_y=3$ (mod[6]). Note that the towers of the $\bf 3$ and $\bf\bar 3$
		sectors are identical, apart from an overall conjugation of all IRREPs.}
	\label{fig:es3}
\end{figure*}

\begin{figure*}
		\includegraphics[width=\linewidth]{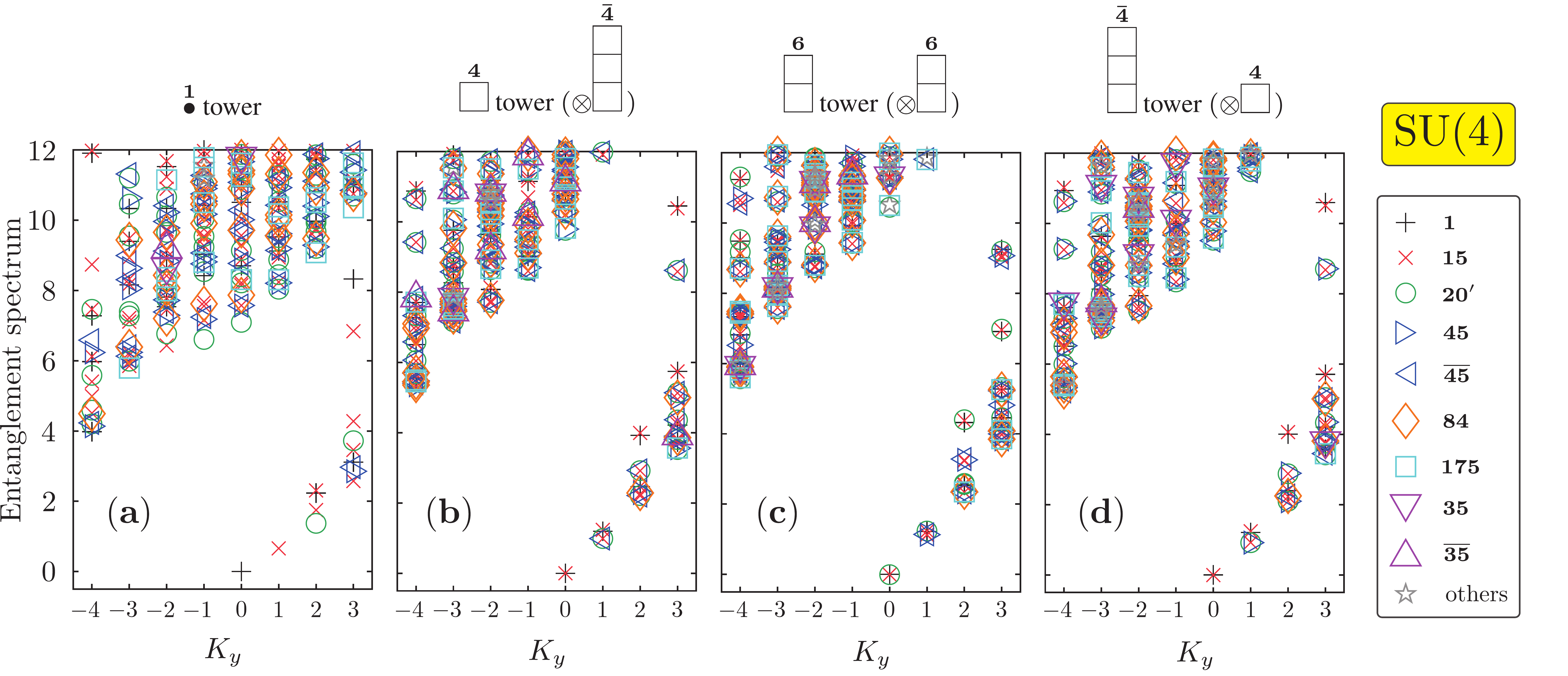}
	\caption{The entanglement spectra on width-8 cylinders for SU$(4)$ CSLs. {\bf(a)} Identity sector. {\bf(b)} $\protect\su{0}{1}{\bf 4}$ sector ($\otimes\protect\su{0}{1,1,1}{\bar{4}}$). {\bf(c)} $\protect\su{0}{1,1}{\bf 6}$ sector ($\otimes\protect\su{0}{1,1}{6}$). {\bf(d)} $\protect\su{0}{1,1,1}{\bf \bar 4}$ sector ($\otimes\protect\su{0}{1}{4}$). Note that the towers of the
		$\bf 4$ and
		$\bf\bar 4$ sectors are identical, apart from an overall conjugation of all IRREPs. Identifying $K_y$ with the Virasoro level $L_0$, the content of the chiral branches  agrees exactly with the CFT predictions of tables~\ref{tab:su4_1}, \ref{tab:su4_copies1} and \ref{tab:su4_copies2} up to $K_y=3$.
	}
	\label{fig:es4}
\end{figure*}

\subsection{Infinite DMRG}

For a cylinder geometry, the $N$ different
minimally entangled states
of the SU$(N)$ CSLs, each carrying distinct anyonic flux threading through the hole in the annulus, form a complete basis for the $N$-fold degenerate ground states.
Finding such a complete basis numerically for the Hamiltonian of Eq.~(\ref{eq:model}) would be a convincing validation for our short-range CSL proposal.

Numerically the finite system width $N_y$ lifts the $N$-fold ground-state degeneracy, with an energy gap which decreases with increasing width.
 If the cylinder is infinitely long, CFT predicts that the energy splittings (with respect to the ground state) are given by $\frac{2\pi v}{N_y} (h_{p}+\bar{h}_p)$, where $v$ is the velocity of the chiral edge states and $h_{p},\bar{h}_p$ are conformal weights of the primary fields (corresponding to the respective anyons at the boundaries). Thus, we expect a power-law  splitting $\mathcal{O}(1/N_y)$ for chiral topological phases
 (rather than exponential, as in the case of  nonchiral topological phases with gapped edges~\cite{AYK2003,Poilblanc2012,CL2013,*CLarXiv}).
This hampers the search for distinct topological sectors via DMRG, a ground-state search algorithm when using cylinders.
Previous DMRG works~\cite{YS2011,CL2013,ZMP2013,He2014a,SSN2016,HS2019,SA2020} have shed some light on this, showing that the presumably higher-energy states can still be examined by adopting tailored boundaries, e.g.  imposing  $\mathbbm{Z}_N$  charges
\footnote{For SU($N$)$_1$ CSL all topological sectors can be obtained in this way. However, for some topological phases, other types of anyon sectors can appear, such as a ``defect line'' cutting along the $x$ direction. This is also very common and appears in, e.g., $\mathbbm{Z}_2$~\protect\cite{Poilblanc2012} and Ising topological phases. Then, adopting tailored boundaries in DMRG is not sufficient to detect such topological sectors.}.
Concretely, DMRG is used to optimize the bulk part of the cylinder, while a small portion of spins at the boundaries are engineered to mitigate finite-width effects, thereby favoring different topological sectors if any exist.
However, how to engineer the boundary spins and choose suitable lattice orientation remains an elusive undertaking.

Our work here is an extension of the above idea, and the parton approach paves a systematic way to construct the boundary spins for different MESs.
For the identity sector, we use typical infinite DMRG (iDMRG) to find the ground state for Eq.~(\ref{eq:model}) \cite{WS1992,McCulloch2008}.
For other sectors that are higher in energy, we use the parton approach outlined above to initialize several possible MESs by occupying edge modes in different ways, then use the infinite DMRG algorithm to minimize the (bulk) ground-state energy with respect to the Hamiltonian of Eq.~(\ref{eq:model}) for each.
The ED calculations in Sec.~\ref{sec:ED} suggest a substantial region of a gapped CSL in the parameter space of $(\theta,\phi) = ({\rm{sin}}^{-1}(J_I),{\rm{tan}}^{-1}(\frac{3}{4}J_1/J_R))$ for each $N$.
Here we focus on only one point within that phase, for $N=2$ up to $4$.
While $N=2$ and $3$ have been investigated by ED and iPEPS previously, a thorough DMRG study for them has not been performed.
We therefore include them here too, to corroborate the consistency of the model as well as the method for different $N$.
We choose $(\theta,\phi)=(\pi/12,\pi/2)$ for $N=2$,  $(\theta,\phi)=(\pi/6,\pi/2)$ for $N=3$, and $(\theta,\phi)=(\pi/4,\pi/2)$ for $N=4$.
The widths of the cylinder are chosen to be a multiple of $N$, so that if $N$ different MESs do exist, all of them they can be found for arbitrary cylinder lengths.

The entanglement spectrum, as the fingerprint of topological order, can be readily extracted from iDMRG wave functions.
To enable a comparison with CFT, we identify the entanglement levels by their SU($N$) irreps
and the momentum $k_y=\frac{2\pi K_y }{N_y}$, $K_y\in \mathbb{N}$ \cite{CL2013,*CLarXiv}, as the converged states should be translationally invariant along the $y$ direction.
They are thus (approximate) eigenstates of the translation operator, with phase factors as eigenvalues, from which we extract the associated momenta $k_y$.
From Fig.~\ref{fig:es2} {\bf (a)}, we see that the identity sector agrees with the SU$(2)_1$ WZW CFT (see Table~\ref{tab:su2_1}) for the first few low-lying states.
For the semion sector, the ES (see Fig.~\ref{fig:es2} {\bf (b)}) consists of a new conformal tower
containing integer spin multiplets, and twice the number of states expected for the semionic conformal tower.
This discrepancy is rooted in the fact that semions carry spin-1/2 quantum numbers and can be best understood from the parton context~\cite{WYH2020}:
the CFT content describes a single edge mode for spin-$1/2$, while the state in our simulation is a spin-singlet, corresponding to an antisymmetric combination of two spin-$1/2$ edge modes.
In other words, neither of the
semion states carrying spin-$1/2$ at the edges, i.e.,
$\ket{\Psi_{1 }} = P_G d_{L\uparrow}^{\dagger}d_{R\downarrow}^{\dagger}\ket{\Phi}$ or
$\ket{\Psi_{1'}} = P_G d_{R\uparrow}^{\dagger}d_{L\downarrow}^{\dagger}\ket{\Phi}$,
does
have a definite total spin.
A spin-singlet can be formed, however, via a linear combination of
$\ket{\Psi_1}$ and $\ket{\Psi_{1'}}$,
which leads to the doubling
of the number of states of  the conformal towers~\footnote{This
is similar to the AKLT state with periodic boundary conditions,
which has four-fold degeneracy in the entanglement spectrum
rather than the two-fold degeneracy suggested by the $D=2$ MPS
representation~\protect\cite{Li13}.}.
This can be easily verified by a direct product of the conformal
towers of the spin-$1/2$ primary of  Table~\ref{tab:su2_2}
( ~\App{app:WZW}) with a spin-$1/2$, as shown in
Table~\ref{tab:su2_copies}.
This observation applies also for cases of $N>2$~:  for
non-identity sectors, the ESs contain, in each Virasoro level,
an integer multiplicity ($\ge N$) of  the number of states of a
single CFT tower.
In general, it is possible to account for such a multiplicity by
taking the direct product of each conformal tower with the
conjugate of its primary spin (see Tables~\ref{tab:su3_copies} ,
\ref{tab:su4_copies1} and \ref{tab:su4_copies2} in
 ~\App{app:WZW2} as examples).
This brings our simulations in overall agreement with CFT as
shown in Figs.~\ref{fig:es3} and \ref{fig:es4} for $N=3$ and
$N=4$, respectively, and a direct comparison with
Tables~\ref{tab:su3_copies}, \ref{tab:su4_copies1} and
\ref{tab:su4_copies2} (see  ~\App{app:WZW2}).
Conversely, one also could have `quenched' the edge spins
$p$ and $\bar{p}$ in the DMRG simulation by coupling them
to an artificial additional physical edge site with spin
$\bar{p}$ and $p$ at the left and right boundary, respectively.
However, we refrained from doing so.

To summarize: in this section we have shown that a DMRG ground-state search for the Hamiltonian of Eq.~(\ref{eq:model}), initialized with an MPS obtained via Gutzwiller-projected parton construction, yields entanglement spectra in excellent agreement with the expectations for SU($N$)$_1$ CSLs.  At a technical level, this required the following innovations: (i) the Gutzwiller projected wave functions for SU($N$)$_1$ CSLs, including the MES basis on the cylinder, are systematically constructed; (ii) the powerful tensor network library incorporating non-Abelian symmetry efficiently converts the projected wave functions into MPSs with high fidelity; (iii) the iDMRG is initialized with the MES basis and preserves the SU($N$) symmetry. The combination of these innovative techniques allows us to obtain all $N$ degenerate ground states of the SU($N$)$_1$ CSL and characterize them from the entanglement spectrum. 

\begin{table}
	\begin{tabular}{ccccccccc} \hline\hline
		& $\{0,0,$ & $\{0,1,$ & $\{1,0,$ & $\{1,3,$ & $\{3,0,$ & $\{0,2,$ & $\{2,1,$ & $\{1,1,$ \\
		& $3,1\}$ & $0,3\}$ & $1,2\}$ & $0,0\}$ & $1,0\}$ & $1,1\}$ & $0,1\}$ & $2,0\}$ \\ \hline
		$A_1$ &               & 1             & 2 & 1             & 2             & 3             & 3             & 4             \\
		$A_2$ & 1             &               & 1 & 2             & 2             & 3             & 3             & 5             \\
		$B_1$ &               & 1             & 2 & 1             & 2             & 3             & 3             & 4             \\
		$B_2$ & 1             &               & 1 & 2             & 2             & 3             & 3             & 5  \\ \hline\hline
	\end{tabular}
	\caption{ Number of  symmetric site-tensors in each class characterized by the IRREP of the $C_{4v}$ point group of the square lattice (rows) and the occupation numbers $\{n_{\bf 6} ,  n_{\bf 4} ,  n_{\bf{\bar 4}} , n_{\bf 1}  \}$ of the  ${\bf  6}$,$ \bf 4$, ${\bf {\bar 4}}$ and ${\bf 1}$ multiplets on the 4 virtual bonds (columns).
		\label{tab:su4tensors}
	}
\end{table}

\begin{figure}
\includegraphics[width=0.95\linewidth]{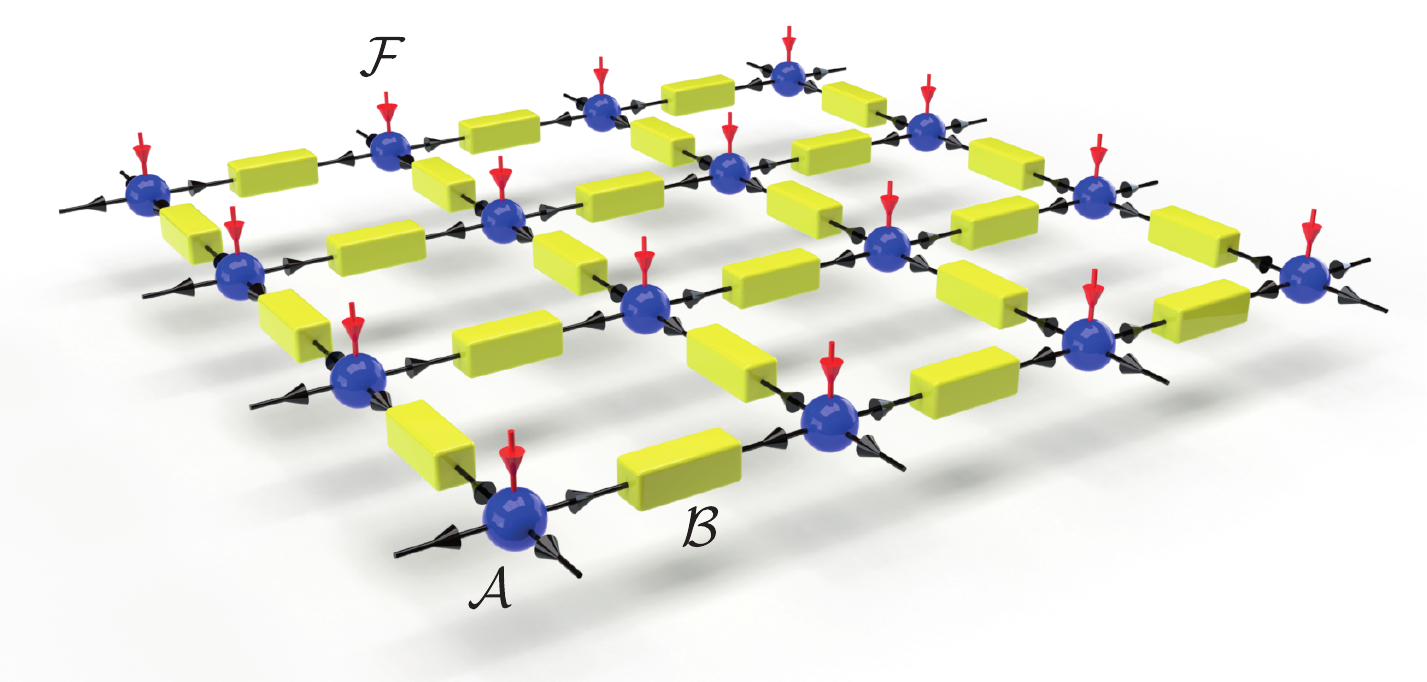}
\caption{
   PEPS on the square lattice involving site $\cal A$ tensors
   and bond $\cal B$ tensors. The bond dimension on the black
   links is $D$, up to $D^\ast=4$ ($D=15$), and the vertical red segments
   correspond to the physical space $\cal F$ spanned by the
   $d=N$ ($d^\ast=1$) physical degrees of freedom. All indices (i.e. legs or lines) carry
   arrows which indicate whether legs enter or leave
   a tensor in terms of state space fusion. This can be
   translated into co- and contravariant tensor index notation,
   respectively \cite{WA2012,WA2020}.
   Note that reverting an arrow also flips all affected
   IRREPS into their dual representations.
}
\label{fig:PEPS}
\end{figure}

\begin{figure*}
	\centering
	\includegraphics[width=0.7\linewidth]{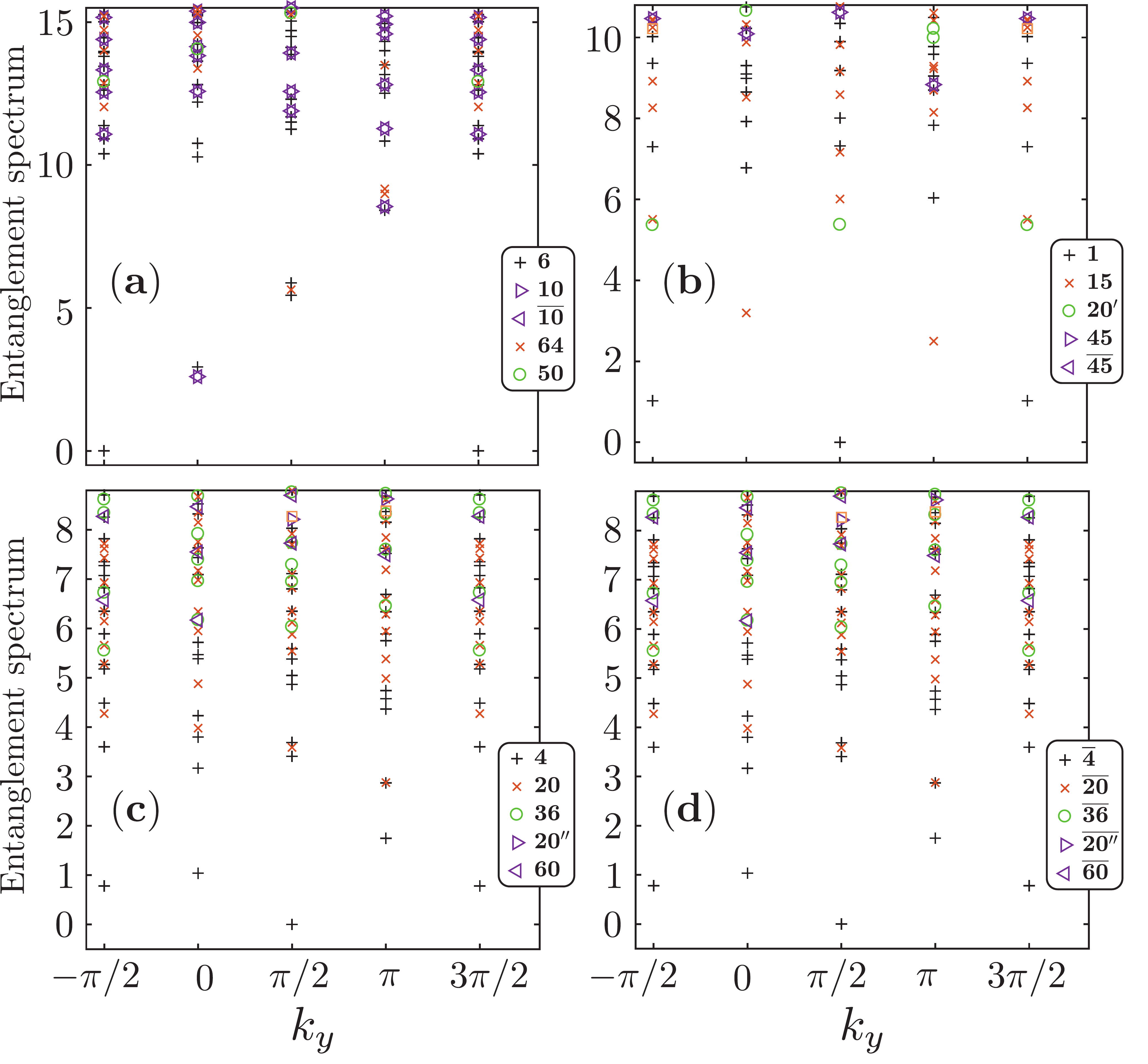}
	\caption{Entanglement spectra on an infinitely-long width-4 cylinder obtained from an SU($4$) ($D=15$) PEPS wave function optimized for $\theta=\pi/4$, $\phi=\pi/2$ and environment dimension $\chi=1350$. Spectra are plotted vs  perimeter momentum $k_y$ and,   to better evidence their chiral nature, the $k_y=-\pi/2$ spectrum is replicated at $k_y=3\pi/2$. Appropriate ${\mathbb Z}_4$ charge boundaries $Q=2$, $0$ and $\pm 1$ are set up to select the $\bf 6$ {\bf (a)}, $\bf 1$ {\bf (b)}, and $\bf 4 / {\bar 4}$ {\bf (c,d)} topological sectors, showing one, two and four branches, respectively. Note that the $\bf 4$ and $\bf\bar 4$ spectra are identical apart from an overall charge conjugation of all IRREPs (and small finite-$\chi$ numerical errors).
		\label{fig:ES_iPEPS}}
\end{figure*}

\section{iPEPS}

The results obtained from ED and iDMRG have shown affirmative evidences for SU$(N)_1$ CSL in a wide range of parameters with arbitrary $N$.
On the other hand, a variational ansatz capturing properties of the CSL phase is also highly desired, especially in terms of symmetric PEPS.
Following the implementation of chiral PEPS for $N=2$ (see Refs.~\cite{Poilblanc2015,Poilblanc2016,Poilblanc2017b}) and $N=3$ (see Ref.~\cite{Chen2020}),
we will first outline the general scheme of the construction, with focus on how the relevant symmetries are realized on the local tensors. We then proceed to a variational optimization of the very few parameters.
Finally, we investigate the entanglement properties and bulk correlations of the optimized chiral PEPS, confronting the results with general considerations.

\subsection{Symmetric PEPS construction}

Let us first extend the construction of  chiral PEPS used for
$N=2$~(see
Refs.~\cite{Poilblanc2015,Poilblanc2016,Poilblanc2017b}) and
$N=3$~(see Ref.~\cite{Chen2020} for more details). The PEPS is
obtained by contracting  the network represented in
Fig.~\ref{fig:PEPS}, i.e., by summing all virtual indices on the
links connecting  rank-($z+1$) site and rank-2 bond tensors, $z$
being the lattice coordination number, $z=4$ for the square
lattice. The physical space $\cal F$ on every lattice site is
spanned by $d=N$ states transforming according to the
fundamental IRREP of SU($N$).
The choice of  the virtual space
on the $z=4$ bonds around each site can be made following
heuristic rules valid for all $N$. In other words, we construct
a SU($N$)-symmetric PEPS from site/bond tensors with virtual (or bond state)
space,

\begin{equation}
	\Yvcentermath1{\cal V}_N= \left.\su{0}{0}{}\oplus \su{0}{1}{}\oplus\cdots \oplus \su{0}{1,1,1,1}{}\right\rbrace N-1
	\label{eq:virtual}
\end{equation}
where the direct sum contains all $N$ IRREPs defined by single column Young diagrams of 0 up to  $N-1$ boxes, consistently with  the $N=2$ and $N=3$ cases, ${\cal V}_2={\bf 1}\oplus{\bf 2}$
	and ${\cal V}_3={\bf  1}\oplus{\bf 3}\oplus {\bf {\bar 3}}$~\footnote{To describe non-Abelian SU($N$)$_k$ CSL, $k>1$, we speculate that one should include all IRREPS in $\cal V$ with up to $k$ columns, consistently with the SU($2$)$_2$ case~\protect\cite{Chen2018b}.}.  For the $N=4$ case we then assume ${\cal V}_4={\bf  1}\oplus{\bf 4}\oplus {\bf 6}\oplus {\bf {\bar 4}}$ (with bond dimension $D=15$).
By construction, the bond state (or virtual) space remains the same
when the direction of arrow in \Fig{fig:PEPS} is reverted,
as $\cal V$ maps into itself
when all IRREPs are flipped into their dual.
Note that the site tensor $\cal{A}$ can be seen as a linear map
(or projection)
$({\cal V}_N)^{\otimes z}\rightarrow {\cal F}$
onto the physical state space,
and the bond tensor $\cal{B}$ as
fusing bond state spaces into a fully entangled
pair singlet state,
$({\cal V}_N)^{\otimes 2}\rightarrow \su{0}{0}{}$.
As such, the tensors $\cal{A}$ and $\cal{B}$ explicitly
correspond to the `P' and `EP' part in the acronym PEPS,
respectively. Up to normalization, the bond tensor
$\cal{B}$ corresponds to an orthogonal matrix inserted
into each bond within the tensor network \cite{WA2012,WA2020}.
It is real and
defined as a weighted sum of three elementary (reflection-symmetric)
tensors representing the three allowed fusion channels
$\su{0}{0}{} \otimes \su{0}{0}{}\rightarrow\su{0}{0}{}$,
${\bf 6}\otimes {\bf 6}\rightarrow\su{0}{0}{}$ and
${\bf 4}\otimes {\bf \bar{4}}\rightarrow\su{0}{0}{}$.
As such, it does not add any variational degrees of freedom.

As for $N=2$ and $3$, we classify the SU(4)-symmetric site-tensors according to (i) the number $n_\alpha$ of $\alpha$-IRREPs appearing on their $z=4$ virtual bonds,
$n_{\rm occ}=\{n_{\bf 6} ,  n_{\bf 4} ,  n_{\bf{\bar 4}} , n_{\bf 1}  \}$
($\sum n_\alpha=z$) and (ii) the (1-dimensional) IRREP of the $C_{4v}$ point group of the square lattice~\cite{Mambrini2016} (see Table~\ref{tab:su4tensors}).
Since the chiral spin liquid only breaks P (parity) and T  (time-reversal) but does not break the product PT, the PEPS complex site tensor $\cal A$ should be invariant (up to a sign) under PT symmetry but acquires a complex conjugation under $P$ or $T$ separately (up to a sign).
The simplest adequate ansatz has the following form:
\begin{equation}
	\label{eq:peps}
	\mathcal{A} = \mathcal{A}_R + i\mathcal{A}_I = \sum_{a=1}^{N_R} \lambda^R_a\mathcal{A}_R^a + i\sum_{b=1}^{N_I} \lambda^I_b\mathcal{A}_I^b,
\end{equation}
where the real elementary tensors $\mathcal{A}_R^a$  and $\mathcal{A}_I^b$
either  transform according to
the $A_1$ and $A_2$  IRREPs, respectively, or  according to
the $B_1$ and $B_2$  IRREPs, respectively, giving rise to two possible families, ${\cal A}_A$ and ${\cal A}_B$.
$N_R=16$ and $N_I=17$
are the numbers of the elementary tensors in each class and
$\lambda^R_{a}$ and $\lambda^I_{a}$ are arbitrary real coefficients of these tensors to be optimized variationally.

To contract the infinite (double layer) tensor network, we have
used the iPEPS method employing a Corner Transfer Matrix
Renormalization Group (CTMRG)
algorithm~\cite{Nishino1996,Orus2009} and obtain the fixed-point
environment tensors used to compute the variational energy (on a
$2\times 2$ plaquette) or the entanglement spectra on infinite
cylinders~\cite{Poilblanc2017b,Chen2020}. In order to cope with
the large bond dimension ($D=15$), the tensor contractions at
each  CTMRG step have been performed using the full SU($N$)-symmetry,
thanks to the QSpace library~\cite{WA2012,WA2020}.
This changes the description of any vector space $\cal V$
from state-based to multiplet-based. For numerical efficiency
then, importantly, the dimensionality is reduced from $D_{\cal V}$
states to an effective dimension of $D^\ast_{\cal V}$ multiplets,
where for SU($N$) it typically holds $D^\ast_{\cal V} \ll D_{\cal V}$.
As an example, the bond dimension $D^2=225$ of  the double layer (rank-4) tensor $\cal AA^\ast$ (used in CTMRG)  can be reduced to
$D^{2\ast}=26$
which represents the number of multiplets in the
product space:
\begin{eqnarray}
	\nu_4^{\otimes 2}  &=& \su{4}{0}{1}\oplus\su{4}{1}{4}\oplus\su{4}{1,1,1}{\bar{4}}\oplus\su{4}{1,1}{6}\oplus\su{1}{2}{10}
	\\
	&\oplus&\su{1}{2,2,2}{\overline{10}}\oplus\su{3}{2,1,1}{15}\oplus\su{2}{2,1}{\overline{20}}\oplus\su{1}{2,2}{20'}\oplus\su{2}{2,2,1}{20} \, .
	\nonumber
\end{eqnarray}
By fully enforcing SU($N$) symmetries on all tensors
and indices, this automatically implies that singular values
within any multiplet are degenerate.
Therefore naturally, state space truncation is also always
performed based on entire multiplets. Degeneracies across
different multiplets, however, can be arbitrarily split
depending on the algorithm and overall convergence.
For SU(4),
we have increased the environment dimension up to
$\chi^\ast=221$ multiplets (corresponding to
$\chi=1350$ states) to control truncation errors.
The optimization of the PEPS (\ref{eq:peps}) with
respect to   its variational parameters is done within a
variational optimization scheme~\cite{Poilblanc2017a}. For
$\theta=\pi/4$, $\phi=\pi/2$, the best variational energy (per
site) $e\simeq -2.105$ (close to the DMRG estimate $-2.14$) is
obtained for the ${\cal A}_B$ ansatz that we shall consider
hereafter.

\subsection{Entanglement spectrum and edge physics}

Both ED and DMRG computations have shown overwhelming evidence of  SU($N$)$_1$ edge modes, both on disk and cylinder geometries, a fingerprint of the Abelian CSL phase.  We note that, apart from the trivial (identity) sector, the conformal towers previously obtained using PEPS on cylinders for $N=2,3$ bear some differences with those obtained in DMRG. For example, the spin-1/2 semionic branch of the SU($2$) spin-1/2  chiral PEPS corresponds exactly to the SU($2$)$_1$ conformal tower  -- consisting of half-integer  spin multiplets  -- associated to the WZW spin-1/2 primary field and  its  descendants, but with an exact two-fold degeneracy~\cite{Poilblanc2015,Poilblanc2016,Poilblanc2017b}.  For the SU($3$)  spin-$\su{0}{1}{}$ chiral PEPS, in the topological sectors defined by imposing $Q=\pm 1$ $\mathbbm{Z}_3$ charges at the boundaries (stricly speaking, infinitely far away),  three chiral branches --  instead of a single one -- separated in momentum by $2\pi/3$ are observed in the ES, whose level contents follow the prediction of the Virasoro levels of the SU($3$)$_1$ WZW CFT~\cite{Chen2020}. Interestingly, both DMRG and PEPS show the same number of states in each Virasoro level, namely $N$ times the WZW CFT content.  These particular features of the PEPS ansatz are now further tested in the case of the SU(4) model in order to draw more general (empirical) statements for SU($N$)  spin-$\su{0}{1}{}$ chiral PEPS.

The ES, revealing the topological properties of the PEPS, is computed by placing the optimized $D=15$ ($D^\ast=4$) PEPS on a width-4 infinite cylinder partitioned in two halves.  The PEPS holographic bulk-edge correspondence~\cite{Cirac2011,Chen2020} enables to compute the ES simply from the (fixed-point) environment tensors.  The four topological sectors are selected by imposing a well-defined total $\mathbb{Z}_4$ charge $Q$ at both ends (strictly speaking at infinity) on the virtual levels. Following the assignment $q_{\bf 1}=0$, $q_{\bf 4}=1$, $q_{\bf \bar{4}}=-1$ and $q_{\bf 6}=2$, we have $Q=\sum q_{\bf\alpha} \,{\rm mod}[4]$, where the sum runs over the virtual open bonds along the circumference at the boundaries. In practice, this is performed by filtering out the components of the environment tensors used to approximate each halves of the cylinder.

A necessary ingredient for identifying the linear dispersing modes in ES is the momentum quantum number associated with each energy level, which originates from the translation invariance along the circumference of the cylinder. For that purpose, we consider the momentum projection operator $P_{k_y}$:
\begin{equation}
	P_{k_y} = \frac{1}{N_y}\sum_{r=0}^{N_y-1}\mathrm{e}^{-ik_yr}T^r,
\end{equation}
where $k_y=\frac{2\pi}{N_y}K_y$, $K_y=0,1,2,\ldots,N_y-1$, and $T$ is the one-site translation operator acting on the virtual degrees of freedom. Since $T$ commutes with $\rho$, we can diagonalize $P_{k_y}\rho P_{k_y}$, whose nonzero eigenvalues are also eigenvalues of $\rho$, and corresponding eigenstates carry momentum quantum number $k_y$, to obtain ES and momentum quantum number simultaneously. In this setup, the action of translation operator on $\rho$ can be implemented as a permutation of indices of $\rho$.

In Fig.~\ref{fig:ES_iPEPS} the ES in the four topological sectors are shown as a function of the momentum $k_y$ along the circumference. For $Q=2$, $0$ and $\pm 1$, we identify one, two or four linearly dispersing chiral branches, respectively. When two or four branches are seen, the later are equally spaced in momentum, i.e. by $2\pi/2=\pi$ and by $2\pi/4=\pi/2$, respectively.  Despite the very small circumference ($N_v=4$), for $Q=2$ and $0$ the expected SU($4$)$_1$ counting of  the first Virasoro levels is satisfied. For $Q=\pm 1$, due to limited resolution in $K$-space, the states of the second Virasoro level of each branch are not clearly separated from the continuum above. Although it is difficult to draw definite conclusions on such a thin cylinder, it seems that the SU($4$) chiral PEPS reveals, as for the SU($2$) and SU($3$) cases, a duplication of the chiral branches for most topological sectors. In the SU($2$) PEPS this was attributed to the so-called ``dressed mirror symmetry" within the virtual degrees of freedom~\cite{Hackenbroich2018}. Note however that there is no exact degeneracy in the $N=3$ and $N=4$ cases, in contrast to $N=2$, so that the duplication of the chiral modes may have a different origin here.
In any case, as for the DMRG wave function, the duplication of the chiral states in the PEPS is linked to the fact that the ansatz is not a MES but, rather, carry an extra entanglement due to its global singlet nature. However, the manifestation in the ES is different in the two cases.

\subsection{Correlation lengths}

\begin{figure}
\centering
\includegraphics[width=0.8\linewidth]{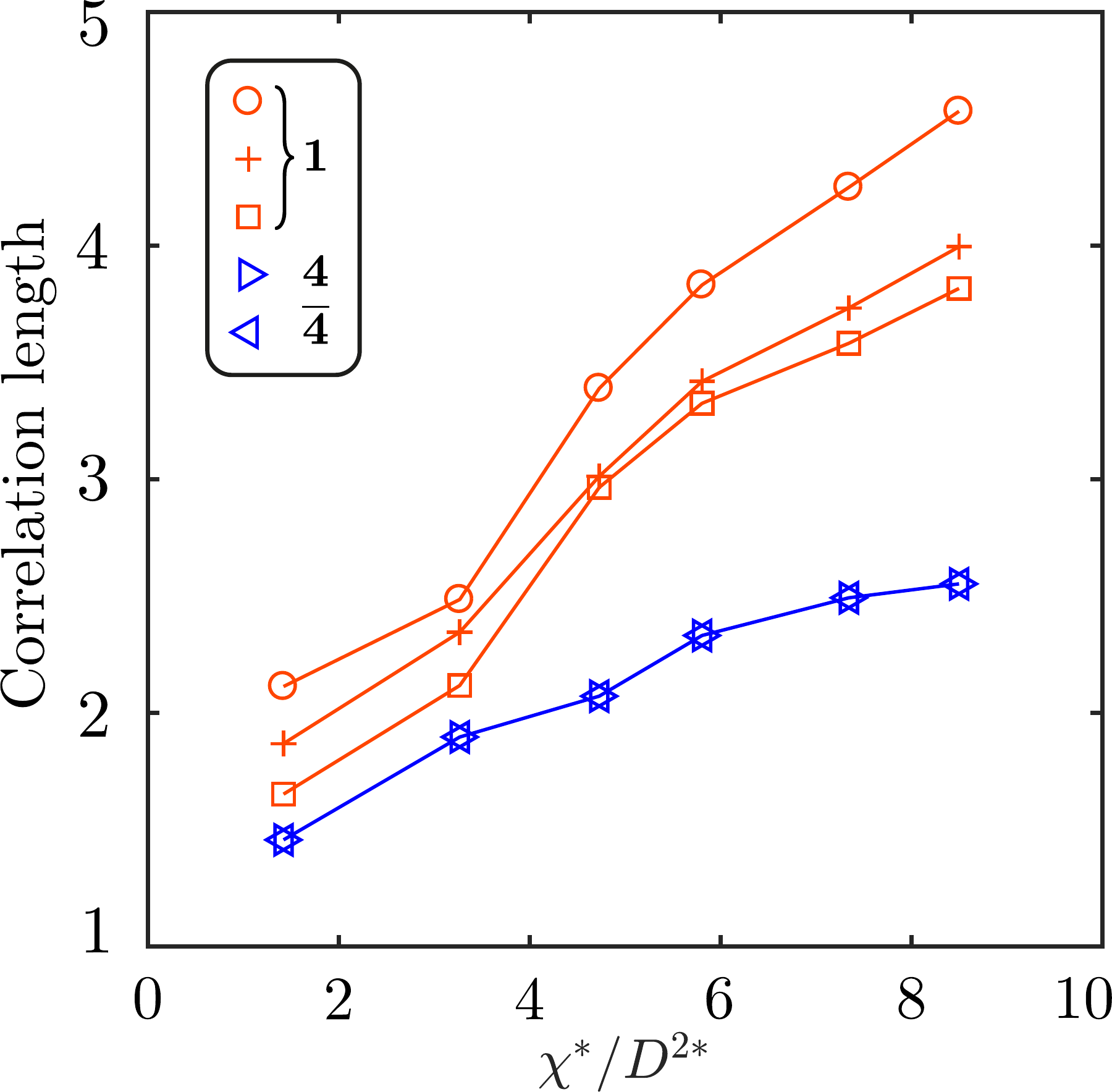}
\caption{
   Leading correlation lengths obtained from the transfer matrix
   (in the absence of  gauge flux) for the case of SU($4$)
   plotted versus
   the number of multiplets $\chi^\ast$ kept in
   the environmental tensors, normalized by $D^{2\ast}=26$
   which represents the number of multiplets in the
   product space of $D\times D$ states with $D$ fixed.
   The SU($4$) IRREPs associated to
   these correlation lengths are indicated.
}
\label{fig:corrLength}
\end{figure}

It was proven that any short-range quadratic parent Hamiltonian for chiral {\it non-interacting}  PEPS is gapless~\cite{Dubail2015}. This suggests that a fundamental  obstruction  or ``no-go theorem" may prevent to  describe  a  gapped CSL  phase  with  a 2D  PEPS (of finite bond dimension $D$).  In fact, the PEPS optimized for the $N=2$ and $N=3$ chiral Heisenberg models~\cite{Poilblanc2017b,Chen2020} reveal rather long-range correlations and growing correlation lengths with environment dimension $\chi$. It is therefore of much interest to also test this important feature in our SU($4$) PEPS. For that purpose, we have computed the leading correlation lengths (associated to the leading correlations in the bulk of the PEPS) from the leading eigenvalues of the transfer matrix (TM)~\cite{Poilblanc2015} (with no gauge ``vison" flux).
These correlation lengths, plotted in Fig.~\ref{fig:corrLength}, show no sign of saturation with $\chi^\ast/D^{2\ast}$, or equivalently with $\chi/D^2$ ($D=15$) -- at least the three largest ones. The latter (shown in orange color) have been obtained from the  singlet eigenvalues of the TM and, probably, correspond to dimer correlations. The next two (shown in blue color) correspond to spinon correlations.
We note that all correlation lengths remain rather short, even for the largest $\chi$ value. However, the data for $N=2$, $3$ and $4$ clearly show that all correlation lengths are comparable at the same value of $\chi/D^2$. For example, the dimer correlation length ranges between $3.5$ and $6$  for  $\chi/D^2=6$, weakly dependent on $N$ and on the model parameters. Since the PEPS bond dimension increases significantly with $N$ ($D=3,7,15$ for $N=2,3,4$, respectively) the maximum achievable value of $\chi/D^2$, and hence of the correlation lengths, decreases strongly with $N$.

Note that in the SU(2) case, the diverging nature of the correlation lengths was shown to be associated, not to a conventional critical behavior but, rather, to the existence of ``long-range tails" (of  very small weight) in most correlation functions~\cite{Poilblanc2017b}. We believe such a property also holds for any SU($N$) CSL, although it could not be established here for $N=4$ due to the large value of the bond dimension $D$.

\section{Conclusion and outlook}

In this work, the previous family of SU($3$) chiral Heisenberg models on the square lattice has been generalized to any SU($N$) fundamental IRREP as physical degrees of freedom. The construction follows two steps: the first one consists in building up the most general fully translational, rotational  and SU($N$)-symmetric model (possibly breaking time-reversal symmetry) whose interactions extend at most to 3-sites within the square plaquettes. In a second step, one restricts to a subset of this model family whose Hamiltonians can be written solely as a sum of $S_3$-symmetric operators defined on all the triangles within the square plaquettes. By doing so, we expect to mimic some of the physics of the triangular lattice with 3-site chiral interactions, although keeping the full $C_{4v}$ point group symmetry of the square lattice.
This procedure defines a sub-family of chiral Heisenberg models spanned by two independent parameters (angles) that we have explored in details.

Extensive ED computations bring overwhelming evidence of extended regions of stability of SU($N$) CSL phases for all $N$, up to $N=10$. The Abelian SU($N$)$_1$ topological nature of these phases has been clearly established from the many-body low-energy spectra of periodic (tori) and open (disks) clusters. When the system size $N_s$ is commensurate with $N$ (so that no anyons is present in the GS) a $N$-fold GS degeneracy is observed on small tori as expected. When the commensurability between $N_s$ and $N$ is such that a single quasi-hole populates the GS, $N_s$ quasi-degenerate GS are found, as expected. Finally, chiral many-body low-energy spectra on open clusters following WZW CFT counting rules provide an even more stringent test of the existence of the SU($N$)$_1$ Abelian CSL.

iDMRG computations by enabling to  access much larger systems -- typically infinitely-long broad cylinders -- provide most valuable and complementary results for $N=2,3,4$. Gutwiller-projected parton wave functions offer a guide to construct iDMRG ansatze in each topological sector. Due to their SU($N$) global singlet nature, the iDMRG wave functions carry larger entanglement than MES (they can be seen as linear combinations of MES, except in the trivial sector) and, hence,
show ES with more structure whose complete understanding has been fully provided.

Following the prescriptions for $N=2$ and $N=3$, we have constructed a family of  chiral SU($4$)-symmetric PEPS and, under optimization, a good variational PEPS ansatz is obtained for the chiral SU($4$) Heisenberg model. The entanglement spectra obtained in the $N=4$ topological sectors of an infinitely-long cylinder reveal chiral modes. The multiplicity of the chiral modes is attributed to the non-MES nature of the singlet PEPS ansatz in most topological sectors. Finally, growing correlation lengths with environment dimension are consistent with the existence of ``long-range tails" (of  very small weight) in correlation functions (evidenced explicitly for $N=2$~\cite{Poilblanc2017b}).
We speculate that these long-range tails would fade away (i.e. their weights would continuously vanish) for increasing $D$, providing a more and more faithful representation of the GS. If correct, this implies that the no-go theorem~\cite{Dubail2015} does not {\it practically} prevent an accurate chiral PEPS representation of the topological gapped CSL phase.

We note that the SU($N$) CSL is stable in some regime where the 3-site interaction is purely imaginary (corresponding to $\phi=\pi/2$), mostly studied here. In fact, this case is relevant in ultracold atom systems which can realize an SU($N$) fermionic Hubbard model~\cite{Gorshkov2010}. In the presence of an artificial gauge field (providing complex amplitudes to the effective hoppings), at $1/N$ filling (one particle per site), the large-$U$ Mott insulating phase~\cite{Bauer2014,Nataf2016,Chen2016} can be approximately described by our Hamiltonian, so that an Abelian SU($N$) phase on the square lattice may be seen experimentally if low-enough temperatures could be reached.
Experimental setups of ultracold atoms at other fractional fillings like $k/N$ ($k\in\mathbb{N}$ particles/per site) could be also of great interest and be described by new types of  SU($N$) spin Hamiltonians, like the two-fermion SU($4$) model~\cite{Gauthe2020} with additional chiral interactions on triangular units, opening the way to observe non-Abelian
CSL.
\section*{Acknowledgments}

J.-Y. C., J.-W. L. and P. N.  contributed equally to this work. D. P. conceptualized the work.
We acknowledge enlightening conversations with Norbert Schuch. We also thank Alexander Wietek for the use of his QuantiPy library and Laurens Vanderstraeten for help on non-abelian symmetries in tensor networks. J.-Y. C. acknowledges support by the European Union's
Horizon 2020 programme through the ERC Starting Grant WASCOSYS (Grant No. 636201) and the ERC Consolidator Grant SEQUAM (Grant No. 863476), and from the Deutsche Forschungsgemeinschaft (DFG, German Research Foundation) under Germany's Excellence Strategy (EXC-2111--390814868).
K.~T. is supported in part by JSPS KAKENHI Grant No.~18K03455 and No.~21K03401.  H.-H. T. is supported by the Deutsche Forschungsgemeinschaft
through project A06 of SFB 1143 (project-id 247310070).
J. v. D. acknowledges support from the Deutsche Forschungsgemeinschaft under Germany's Excellence Strategy EXC-2111390814868, through project No. 409562408.
D. P. acknowledges support by the TNSTRONG ANR-16-CE30-0025  and TNTOP ANR-18-CE30-0026-01 grants awarded by the French Research Council.
J.-W. L. acknowledges support by DFG WE4819/3-1.
A. W. was supported by the U.S. Department of Energy, Office of Science,
Basic Energy Sciences, Materials Sciences and Engineering Division.
This work was granted access to the HPC resources of CALMIP and GENCI supercomputing centers under the allocation 2017-P1231 and A0030500225, respectively, and computations have also been carried out on the TQO cluster of the Max-Planck-Institute of Quantum Optics.

\clearpage
\appendix

\section{Analysis of $2\times 2$ plaquette}
\label{app:plaquette}

The focus of the present paper is on chiral spin liquids
which have the SU($N$) flavor symmetry intact both
locally and globally. In particular, the ground state remains
an SU($N$) singlet in the thermodynamic limit.
This suggests that
also the low-energy regime of smaller clusters should
have a singlet ground state. If that is not possible
by finite size, at least, one may expect to have a
ground state that is closest to a singlet
in the sense that they tend to prefer to fill
up full columns in the corresponding Young tableau (YT).

In this spirit this appendix analyzes the $2\times2$
plaquette as an elementary unit of the Hamiltonian.
The Hamiltonian \eqref{eq:model} on the full 2D square
lattice can be rewritten as
\begin{eqnarray}
  H = \sum_p H_p
\label{eq:H:plaquette}
\end{eqnarray}
where $H_p$ is the Hamiltonian for a single square
plaquette $p$ of $2\times 2$ sites that combines all terms
$i,j,k \in p$ (in order to avoid overcounting along the
edge of the plaquette, we set $J_1 \to \tfrac{1}{2} J_1$
for $H_p$, whereas $J_2$, $J_R$, and $J_I$ remain the same).
Now with $H_p$ the combined set of local operators
that can be used to tile the entire 2D Hamiltonian,
it is natural to analyse its multiplet structure.
Multiplets in $H_p$ that are low in energy are expected
to be important in the low energy physics on the 2D
lattice itself, whereas multiplets of $H_p$ at higher
energies will likely play a minor role.
Clearly, the ground state multiplet of $H_p$ also may
change when tuning the coupling parameters
$\{J_1,J_2,J_R,J_I\}$. This then may signal a
qualitative change of the overall low-energy behavior
of the 2D system, e.g., a low-order phase transition
for similar coupling parameters.

\begin{figure}[tb!]
\begin{center}
\includegraphics[width=\linewidth]{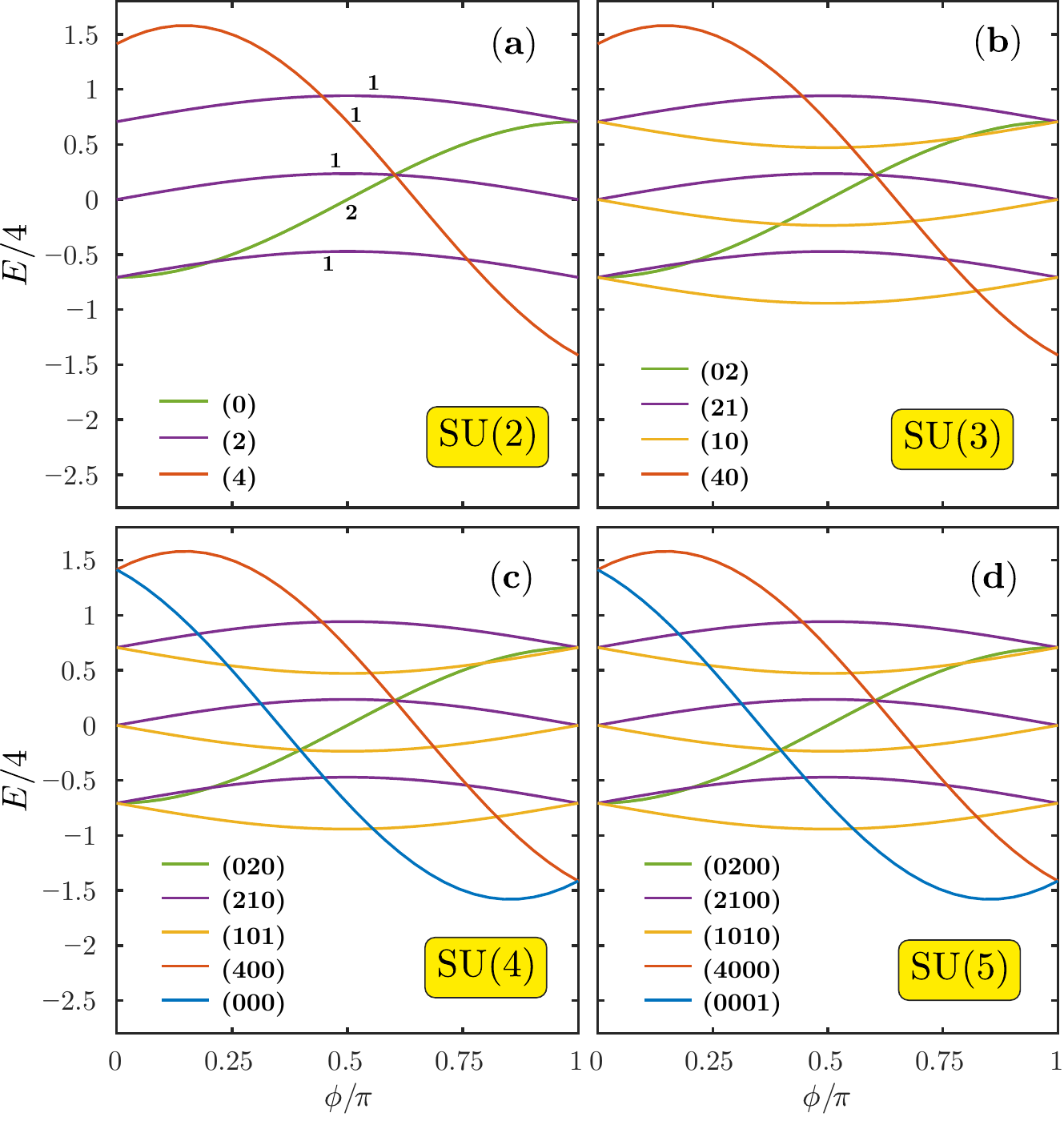}
\end{center}
\caption{Eigenspectrum of a $2\times 2$ cluster
  described by the plaquette Hamiltonian $H_p$ in
  \Eq{eq:H:plaquette}
  vs. $\phi$ using the parametrization in \Eq{eq:paras},
  with $\theta=\pi/4$ fixed as in the main
  text [e.g. see \Fig{fig:energies_vsPhi}].
  To focus on energy per site, the energies are divided
  by the number of sites $N_s=4$ as indicated.
  Panels (a-d) refer to case of $N=2,3,4,5$ symmetric
  flavors, respectively. Colors indicate symmetry sectors
  as indicated with the legend based on Dynkin labels.
  The small numbers on top of each line in panel (a)
  indicate the degeneracy of multiplets which shows
  that the green line only is 2-fold degenerate.
  This also holds for all data in the other panels.
}
\label{fig:4site}
\end{figure}

The eigenspectrum of the $2\times 2$ plaquette
Hamiltonian $H_p$ is analyzed in \Fig{fig:4site} for
$N=2,3,4,5$ in panels (a-d), respectively.
The SU($N$) multiplet structure is fully resolved
as indicated with the legend. For the sake
of the discussion here, we use Dynkin labels
in compact notation to identify symmetry sectors where
$q\equiv (q_1 \ldots q_{N-1})$ directly specifies to
corresponding SU($N$) YT via differential
length offsets of the number of boxes in subsequent
rows of the YT (e.g. see also App. A in \cite{WA2018_SUN}).
For example, $(1 0 \dots 0)_{N-1}$ is the fundamental
or defining representation also labelled as $\bf N$
in the main text,
and $(1 0 \dots 0 1)_{N-1}$ is the adjoint representation.
The reverse order $(q_{N-1} \ldots q_1) \equiv \bar{q}$
specifies the dual IRREP to any $q=(q_1 \ldots q_{N-1})$.
For the case of SU(2), having a single number $(q_1)$
only, the integer $q_1$ simply counts the total number of boxes
in the YT, and thus corresponds to a spin $S\equiv q_1/2$
multiplet. Its adjoint is given by $S=1$, i.e.
multiplet $q=(2)$.

\subsubsection*{%
General aspects of SU($N$) permutation Hamiltonian}

The Hamiltonian \eqref{eq:model} and therefore also
$H_p$ above is defined via simple permutations of
flavors over two or three sites. A direct consequence
of this is, that all eigenenergies appearing for
SU($N$) exactly also must appear for SU($N'>N$),
as can be clearly observed in \Fig{fig:4site}.
The simple reason is that adding additional flavors
$N'-N > 0$ on top of all sites, the Hamiltonian
will not make any reference to these when applying
it to a state that only contains up to the first
$N$ flavors. The multiplet label needs to adapt, though.
By using Dynkin labels, this simply concatenates
additional trailing numbers $q_i$. Considering a
four-site plaquette here, these extra trailing numbers
must all be zero for $N'>4$, as largely already also
observed for SU(4) itself [see legend in \Fig{fig:4site}(d)].
With this clearly also the degeneracy in terms of
states within these multiplets changes as required
by the increased Hilbert space. However
the eigenenergies themselves remain exactly the same.
Therefore given a Hamiltonian that solely consists
of permutations of otherwise symmetric flavors,
the many-body eigenspectrum for a given SU($N$)
is exactly inherited also to all cases SU($N'>N$).
This is made explicit across the panels in
\Fig{fig:4site} by choosing matching color coding.
For example, what was a singlet in SU(2),
i.e., the green line for $q=(0)$,
becomes $q=(02)$ for SU(3),
and then $q=(020\ldots)$ for larger $N$ still.

When increasing the number of flavors $N \to N'>N$,
however, also new eigenenergies can emerge that were
previously absent. For example, in \Fig{fig:4site} this
is seen as additional lines that appear when going from
$N=2 \to 3$ (yellow lines) or
$N=3\to 4$ (blue line).
Given a 4-site plaquette with the fundamental IRREP
on each site, the number of lines will
no longer change for $N'>4$, as seen by going
from $N=4 \to 5$, since all YTs with four boxes
are already present.

\subsubsection*{Low-energy regimes}

Now the analysis in \Fig{fig:4site} tracks the
eigenvalues vs. $\phi$ for fixed $\theta=\pi/4$
similar to  \Fig{fig:energies_vsPhi} in the main text.
The red line in \Fig{fig:4site} corresponds to
the fully symmetric IRREP $q=(4 0 \ldots)_{N-1}$
that is present for all $N\geq 2$. This multiplet
crosses over and becomes the ground state for
$\phi>\pi$ for $N\geq 4$, and already earlier
for $N=2$ and $N=3$. This shows that the
$2\times 2$ plaquette becomes ferromagnetic
around $\phi \gtrsim \pi$
[note that based on \Eq{eq:paras}, $\phi>\pi$
corresponds to negative, and hence ferromagnetic
$J_1$ and $J_2$]. As such, this signals
the onset of ferrogmagnetism on the full 2D system,
also consistent with the analysis of the larger
clusters in \Fig{fig:energies_vsPhi}.

Finally, with focus on a singlet ground state,
on the given four-site plaquette this can only be achieved exactly
for $N=2$ and $N=4$. Interestingly then, the singlet
for SU($2$) [green line in \Fig{fig:4site}(a)]
becomes a non-singlet for $N>2$, i.e.,
(02) for SU(3), and $(020\ldots)$ thereafter.
Instead, an entirely new singlet shows up for SU(4)
in the low-energy regime,
and remains an eigenenergy for $N\geq 4$ (blue line).
Therefore while in the case of SU(2) the singlet
is favored for small $\phi\in[0,\pi]$,
it is favored for larger $\phi\in[0,\pi]$
for SU(4) and onward. What comes closest
to a singlet for SU(3) on the $2\times 2$
plaquette, on the other hand,
is the multiplet (10), i.e. the fundamental
IRREP. Based on the fusion of the four
fundamental IRREPs on the $2\times 2$ plaquette
to start with,
this already fused three of these into a singlet.
As seen by the yellow line in \Fig{fig:4site}(b),
the multiplet (10) is the ground
state for a wide range $\phi\in[0,\pi]$,
including small but excluding large $\phi$
where the system becomes ferromagnetic.
This is perfectly consistent with the analysis
on the larger cluster in \Fig{fig:energies_vsPhi}(a)
in the main text which for $N=3$ also shows the
chiral phase extending all the way down
to $\phi=0$.

The chiral phase was identified in
\Fig{fig:energies_vsPhi} with the gapped phase around
$\phi \gtrsim \pi/2$. However,  when reducing $\phi$,
as seen in \Fig{fig:energies_vsPhi} for $N>3$,
this gapped phase closes at finite $\phi$.
Even more, for certain $N$ it appears to reopen
before approaching $\phi=0$.
Hence based on \Fig{fig:energies_vsPhi}
having the chiral phase identified with the
regime of larger $\phi \gtrsim \pi/2$,
this is entirely consistent with the regime
in the present analysis of the $2\times 2$
plaquette where the system is (or tends
towards becoming) a singlet for $N\geq 4$ in
\Fig{fig:4site}(c,d). Note that
for $N>4$, the blue line in \Fig{fig:4site}(d)
corresponds to the fully antisymmetric
multiplet where four boxes are stacked on top
of each other into a single column
in the corresponding YT.

In the chiral regime $\phi \gtrsim \pi/2$ also
the coupling strength of the real
three-site permutation term $H^R_{ijk} \equiv
J_R (P_{ijk} + P_{ijk}^{\,\,\,\,-\,1})$ turns negative,
i.e., having $J_R<0$.
Its effect is revealed by looking at the
eigenvalues in the 3-site eigenbasis for
given triangle triangle $(ijk)$.
One finds for $N\ge 3$  that the
completely symmetric multiplet $(30\ldots)$ and the
completely antisymmetric multiplet $(001\ldots)$
[equivalent to $(00)$ for SU(3)] are eigenstates
to the same eigenvalue $+2 J_R$, whereas the
2-fold degenerate multiplets $(110\ldots)$
have eigenvalue $-J_R$ (which are eventually
differentiated by the complex term $J_I$).
Hence negative $J_R$ equally favors both, the completely
symmetric multiplet (ferromagnetic) as well as the completely
antisymmetric multiplet (antiferromagnetic)
on any triangle. When considering all triangles
within a $2\times 2$ plaquette as analyzed
in \Fig{fig:4site}, the antiferromagnetic states
dominate the low energy regime, yet with the
ferromagnetic regime in close proximity
(both , the blue and red lines move downward
with increasing $\phi$ for $N\ge 4$).
Eventually, for $\phi>\pi$ when also the two-site
exchange couplings $J_1$ and $J_2$ turn negative,
the ferromagnetic state takes over.

\section{Generalized exclusion principle for Abelian SU($N$) CSL}
\label{app:Exclude}
We provide here complementary details about the heuristics on the content (degeneracy, quantum numbers,
etc$\ldots$)  of the GS manifold within the CSL phase on small periodic clusters (of torus geometry).

As realized already for $N=3$ in Ref.~\cite{Chen2020}, it is possible to obtain, for arbitrary $N$,
the exact momenta  of the various states in the GS manifold expected for an Abelian SU($N$)$_1$ CSL.
This  can be inferred from a simple
generalized exclusion principle (GEP) known for FQH states~\cite{Estienne2012}
or fractional Chern insulators~\cite{Sterdyniak2013}
with clustering properties.

For our SU($N$) model in the fundamental representation, there are $N$ states per site which
can be viewed as a color degree of freedom. The mapping to a bosonic FQH requires to treat
them separately:
one (arbitrarily chosen) color will correspond to a hole while the remaining $C=N-1$ will correspond to spinful
SU($C$) bosons. Hence, Abelian bosonic FQH states can be constructed at a filling
$\nu_{FQH}=C/(C+1)=(N-1)/N$, corresponding to Halperin states~\cite{Halperin1983,THH2014,BQ2014}.
In this terminology, the ground states and quasi-hole states
is given by the number of dressed partitions $(1,2)_C$, see Ref.~\onlinecite{Sterdyniak2013}.
Moreover, the respective momenta can be obtained from the mapping between $N_s$ orbitals obtained when
folding the Brillouin zone~\cite{Regnault2011,Bernevig2012}.

To be more specific, let us consider for instance $N=3$ which maps onto $C=2$ bosons, i.e spin-1/2 particles.
Then, the generalized exclusion principle for the ground-states (for $N_s=kN$)
enforces the occupations $(\downarrow,\uparrow,0,\ldots)$ and its translations, i.e. $3$ states.
This $(1,2)_2$ exclusion rule simply enforces  that identical particles cannot be neighbors
but a $\downarrow$ particle can be followed by
a $\uparrow$ particle.
Such rules can be rephrased in terms of follow-up rules in the string of states, e.g. $0\rightarrow (0,\downarrow,\uparrow)$,
$\uparrow\rightarrow 0$, $\downarrow\rightarrow (0,\uparrow)$, which defines a ``transfer matrix",
\begin{equation}
	T^{(N=3)}=\begin{bmatrix}
			1 & 1 & 1\\
			1 & 0 & 1 \\
			1& 0 & 0
	\end{bmatrix}
	,
\end{equation}
for $N=3$.

The transfer matrix above is easy to generalize to any $N$, with 1's  in the first column and above the diagonal
and zeros otherwise. For example, one gets
\begin{equation}
	T^{(N=5)}=\begin{bmatrix}
			1 & 1 & 1 &1 &1\\
			1 & 0 & 1 &1 &1\\
			1 & 0 & 0 &1 &1\\
			1 & 0 & 0 &0&1\\
			1& 0 & 0 & 0 & 0
	\end{bmatrix}
	,
\end{equation}
for $N=5$.
Note that, in addition to the rules encoded in the transfer matrix (which alone produce a large
number of irrelevant configurations), one should also simultaneously enforce a global property
relating the total appearance of all colors such that the GS belong to the SU($N$) IRREP of smallest
possible dimension compatible with system size. More precisely, defining the integer
$r_0=\text{mod}(N_s,N)$,  the smallest possible IRREP corresponds to the antisymmetric
IRREP with a Young diagram
of $r_0$ vertical boxes (labeled in the text $\text{aIR}_N(r_0)$), and, heuristically, is to
be associated to the GS manifold.
For instance for $N_s=kN$, all colors should appear exactly $k$ times, i.e. $c_1=c_2=\cdots =c_N=k$,
as the singlet character of the GS manifold implies.

For $N_s=k N-1$, $k\in\mathbb{N}$, we expect the low-energy states to represent the quasi-hole
excitations, similar to the quasi-hole Laughlin states when inserting a flux in a fractional
quantum Hall state on a torus. In particular, the quasi-hole counting on a finite cluster should
be the same as in the thermodynamic limit and is given by a generalized Haldane exclusion
principle~\cite{Regnault2011,Bernevig2012}. Moreover, the lattice momenta at which these (quasi)
degenerate states sit can be obtained using a heuristic rule by folding the two-dimensional
Brillouin zone into a one-dimensional lattice of orbitals~\cite{Regnault2011}.
For instance, for all the quasi-hole examples shown in Fig.~\ref{fig:towers1},
since GCD($N$,$N_s$)=1, we expect to find one low-energy SU($N$) multiplet at each momentum
(i.e. a total number of quasi-hole states equal to $NN_s$), which is exactly what is found
numerically.

When $N_s=k N$, we expect $N$-fold quasi-degenerate ground states on a torus. The momenta are
given using a similar heuristic rule and are non-trivial. For completeness, here are the
predictions corresponding to the values shown in Fig.~\ref{fig:towers1} (see the Brillouin
zones as insets for the momenta notations): (i) $N=4$ and $N_s=20$: one state at momentum
$\Gamma$, M and 2-fold degenerate X; (ii) $N=5$ and $N_s=15$: one state at momentum $\Gamma$,
$\pm \Sigma_0$, $\pm\Sigma_2$; (iii) $N=6$ and $N_s=12$: one state at momentum $\Gamma$,
$Z_1$, $\pm \Delta$, $Z_0$, $\Delta$; (iv) $N=7$ and $N_s=14$: one state at momentum $\Gamma$,
$\pm 0$, $\pm 2$, $\pm 5$. All these predictions are verified numerically, and the low-energy
states are always well separated from the higher excited ones as expected in this topological
incompressible gapped phase.

\section{WZW  \SUNone chiral towers of states}
\label{app:WZW}
We provide here an almost self-contained explanation of the Hilbert-space structure of the SU($N$) WZW CFT and
derive the \SUNone WZW towers of states for $N=2$ to $8$, which are to be compared with the ED results for
\SUN open clusters investigated and discussed in the main text.
This appendix is organized as follow. In the first part, we recall some basic facts on \sun Lie algebra and its representation theory
(see Ref.~\cite{Georgi-book-99} for a readable introduction to Lie algebras and their representation).
In a second part, we briefly present the affine extension of \SUN and introduce the primary states on which the Hilbert space
is constructed.
Most of the equations presented in the first two parts are relevant to any (affine) Lie algebras unless otherwise stated.
In the last part, we explain how WZW \SUNone chiral towers of states for open clusters can be computed using this formalism. The appendix closes with the tables showing the explicit form of the towers of states relevant for the present study, up to SU(8).
This appendix in not intended to give a mathematical presentation of the field but rather to introduce, without any mathematical proof, the basic tools needed to identify the expected representations in  WZW \SUNone chiral towers of states.

\subsection{\sun Lie Algebra}

{\it Group, Generators - } The special unitary group \SUN is the Lie group of $N \times N$ unitary matrices with determinant 1. The Lie algebra \sun associated to the Lie group \SUN is determined by a set of $N^2-1$ traceless hermitian generators ${J^\alpha}$ satisfying the commutation relations,
\begin{equation}
	\label{eq:liecom}
	\left [ J^\alpha, J^\beta \right ] = i  f_{\alpha \beta \gamma} J^\gamma,
\end{equation}
where the real fully antisymmetric tensor $f$ encodes the structure constants. Equation (\ref{eq:liecom}) is a direct consequence of the group structure of \SUN and the fact that the Lie group and the Lie algebra are related by  the exponential map which associate to any element $J$ of \sun an element $\exp (i t J)$ of \SUN.

{\it Cartan Weyl basis, Adjoint representation,  roots - }  The maximal subset $\{ H^i \}_{i=1,\ldots,r}$ of \sun composed of commuting generators $[ H^i, H^j ]=0$ forms the Cartan subalgebra of \sun and plays the role of $S^{z}$ in $\mathfrak{su}(2)$.
Obviously, since all $H^i$ can be diagonalized simultaneously, the rank $r$ of \sun is $N-1$, which is equal to
the maximal number of traceless diagonal $N\times N$ matrices.
As $\{ H^i \}$ can be simultaneously diagonalized, we can choose the basis vectors in any irreducible representation to be
the eigenstates $|\boldsymbol{\mu} \rangle$ of $H^i$:
\begin{equation}
	H^i | \boldsymbol{\mu} \rangle = \mu_i | \boldsymbol{\mu} \rangle \; .
\end{equation}
The $(N-1)$-dimensional vector $\boldsymbol{\mu}=(\mu_1,\ldots, \mu_{N-1})$ is called the weight.
The remaining $N(N-1)$ off-diagonal generators will be denoted as $E^{\boldsymbol{\alpha}}$.

To each generator $J^{\alpha}$, we can associate a linear map $\text{ad}_J$ from \sun to itself defined as $\text{ad}_J (X) = [J,X]$ for any $X$ in \sun. This defines the adjoint representation which can be used to classify the generators
$E^{\boldsymbol{\alpha}}$ as eigenvectors of ${\text{ad}}_{H^i}$ :
\begin{equation}
	\label{eq:roots}
	{\text{ad}}_{H^i} (E^{\boldsymbol{\alpha}}) = [H^i , E^{\boldsymbol{\alpha}} ]
	=\alpha_i E^{\boldsymbol{\alpha}} \quad
	(i=1,\ldots, N-1) \; .
\end{equation}
The $(N-1)$-dimensional vectors ${\boldsymbol{\alpha}}=(\alpha_1,\ldots,\alpha_{N-1})$ are called the roots
and $E^{\boldsymbol{\alpha}}$, which play the role of $S^{\pm}$, the ladder operators.
The Cartan-Weyl basis is $\{H^i,E^{\boldsymbol{\alpha}}\}_{i\in\{\ 1, \ldots,r\},{\boldsymbol{\alpha}}\in\Delta}$ where  $\Delta$
denotes the set of all $N(N-1)$ roots. Obviously, only $r=N-1$ roots are linearly independent.
An important remark is the non-degeneracy of roots. Indeed, the existence of a degenerate root would contradict the definition
of the Cartan subalgebra ({\em maximal} set of commuting generators).

It is clear from Eq.~\eqref{eq:roots} that there is some arbitrariness in the determination of  $E^{\boldsymbol{\alpha}}$ and ${\boldsymbol{\alpha}}$ as both depend on the choice of a particular basis for the Cartan subalgebra. Nevertheless,
some general properties can be established.
Once the basis of $r=N-1$ linearly independent roots is fixed, one can expand any root in this basis. Roots with positive coefficients in this expansion are called {\em positive} and form the set $\Delta_{+}$. A root $\boldsymbol{\alpha}^{(i)}$ ($i=1,\ldots,r$)
that cannot be expressed as an integer sum of two positive roots is by definition a {\em simple root}.

The central role of such $r=N-1$ simple roots not only lies in the fact they provide a convenient basis for roots  but also because the
$(N-1)\times(N-1)$ matrix $A$ of the scalar products of simple roots (the {\em Cartan matrix}) completely encode the Lie algebra:
\begin{equation}
	\label{eq:Cartan}
	A^{ij}=	\frac{2 \boldsymbol{\alpha}^{(i)}. \boldsymbol{\alpha}^{(j)}}{\boldsymbol{\alpha}^{(j)}.\boldsymbol{\alpha}^{(j)}}
	= \boldsymbol{\alpha}^{(i)}. \boldsymbol{\alpha}^{(j )\vee},
\end{equation}
with $ \boldsymbol{\alpha}^{(i)\vee}=2  \boldsymbol{\alpha}^{(i)} / \vert \boldsymbol{\alpha}^{(i)} \vert^2$ ({\em coroots}).
The entries of this matrix are always integers and, in the \sun case,  $A$ is symmetric and take the form
$A^{ij}=2\delta_{ij}-\delta_{\vert i-j \vert,1}$.
For \sun in which all the $N(N-1)$ roots have equal length (i.e., {\em simply laced}),
it is convenient to choose $|\boldsymbol{\alpha}^{(i)}|=\sqrt{2}$ so that we do not need to distinguish between the roots and the coroots.
The lattice spanned by the $r=N-1$ basis vectors $\boldsymbol{\alpha}^{(i)}$ ($\boldsymbol{\alpha}^{(i)\vee}$)
is called the {\em root lattice} $\Lambda_{\text{r}}(\text{\sun})$ [the {\em coroot lattice} $\Lambda_{\text{r}}^{\vee}(\text{\sun})$].

\begin{figure}[H]
	\begin{center}
	 \includegraphics[width=0.6\linewidth]{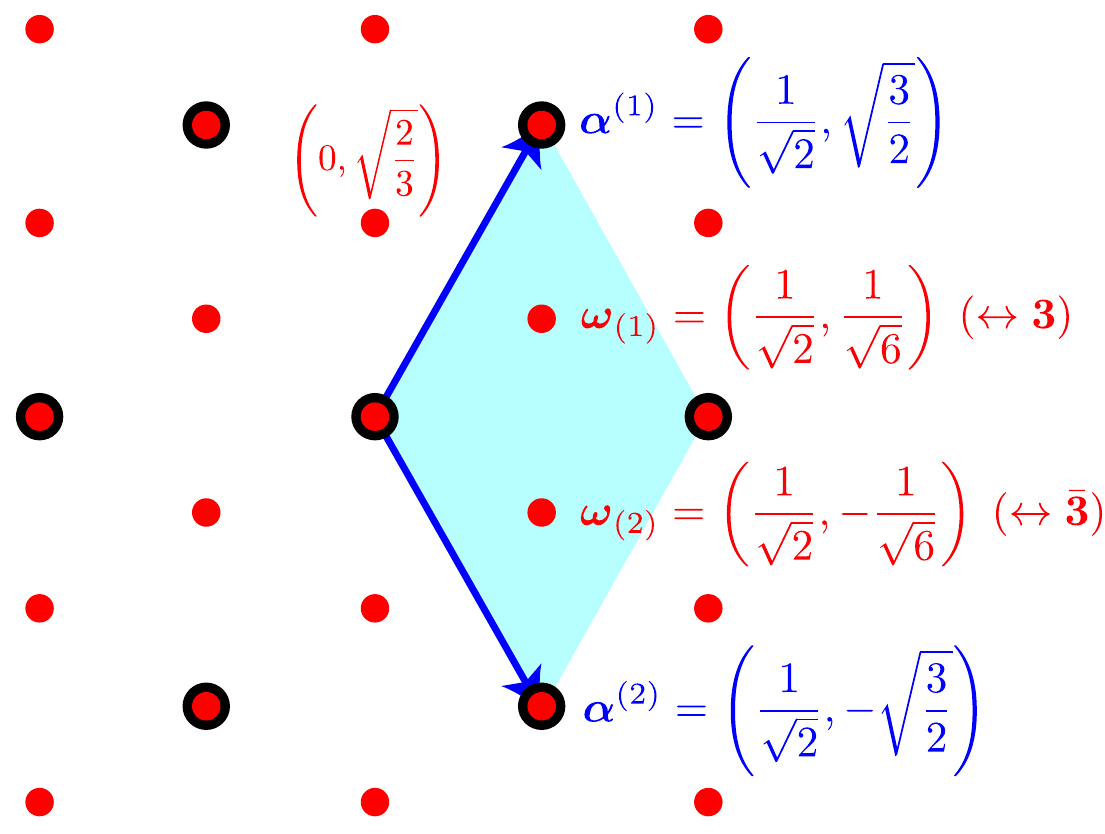}
	\end{center}
	\caption{(color online) The (co)root lattice $\Lambda_{\text{r}}(\mathfrak{g})$ (black circles) and
		and the weight lattice $\Lambda_{\text{w}}(\mathfrak{g})$ (red circles) of $\mathfrak{g}=\mathfrak{su}(3)$.
		The root (weight) lattice is an integer span of two simple (co)roots $\boldsymbol{\alpha}^{(1)}$ and $\boldsymbol{\alpha}^{(2)}$
		(two fundamental weights $\boldsymbol{\omega}_{(1)}$ and $\boldsymbol{\omega}_{(2)}$).
		In $\mathfrak{su}(3)$, $\boldsymbol{\omega}_{(1)}$ and $\boldsymbol{\omega}_{(2)}$ respectively correspond to the highest weights of
		$\mathbf{3}$ and $\bar{\mathbf{3}}$.
		\label{fig:SU3-weight-lattice}}
\end{figure}

\begin{figure}[H]
	\begin{center}
		\includegraphics[scale=0.4]{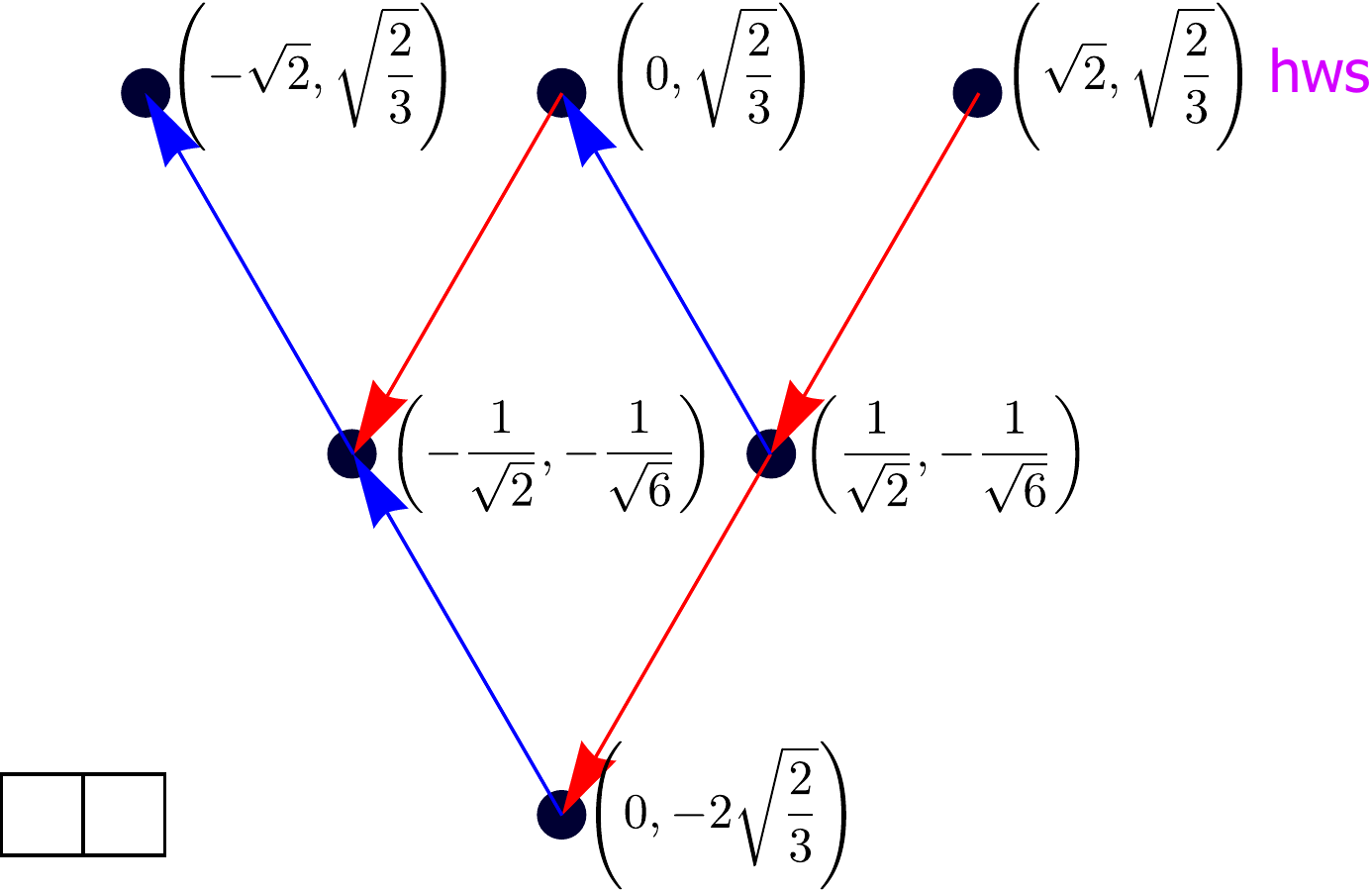}
		
		\vspace{5mm}
		\includegraphics[scale=0.4]{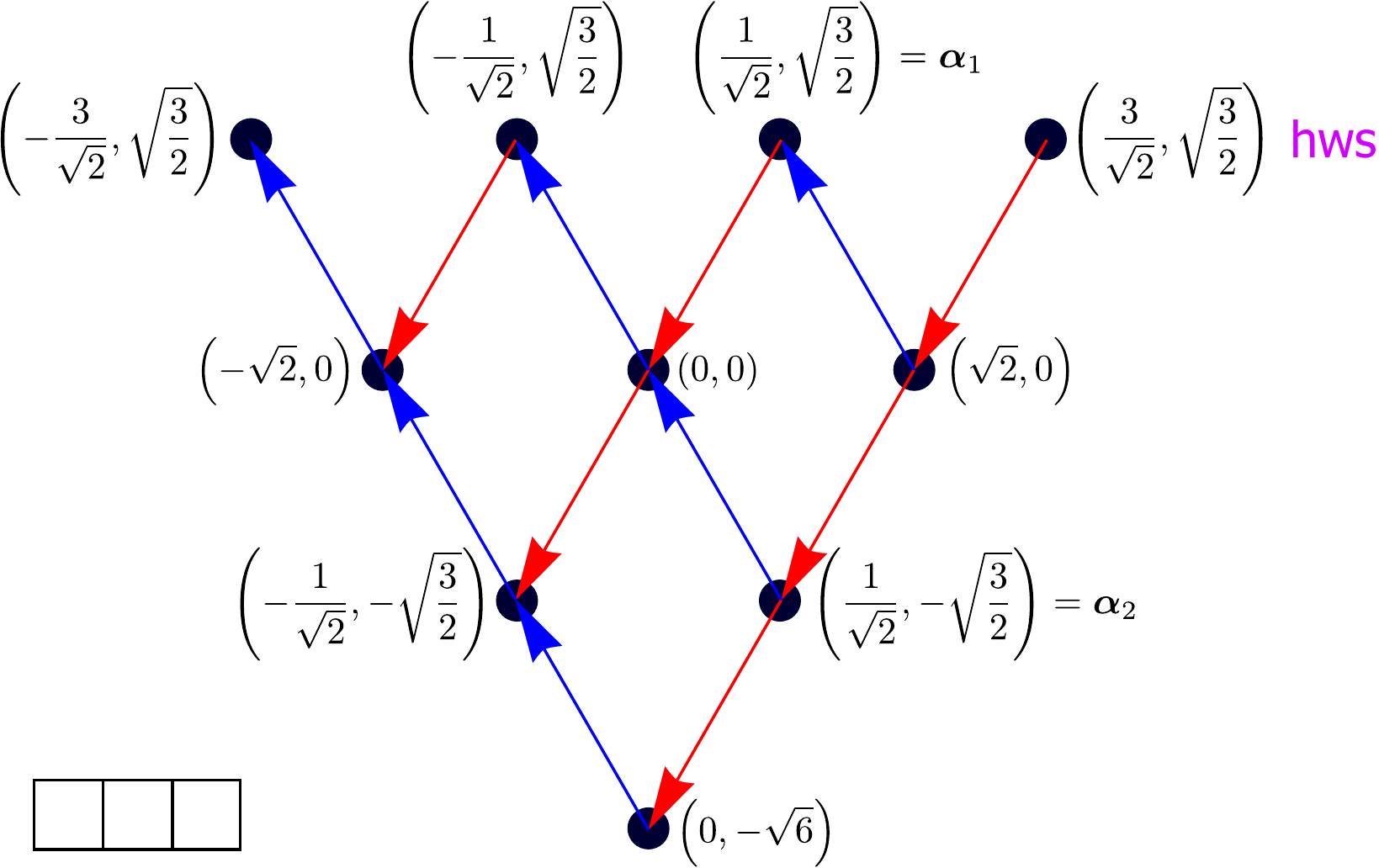}
	\end{center}
	\caption{(color online) Weights of $\mathbf{6}$ and $\mathbf{10}$-dimensional representations of $\mathfrak{su}(3)$.
		The representations $\mathbf{6}$ and $\mathbf{10}$ have highest weights (shown as ``hws'')
		with Dynkin labels $(d^1,d^2)=(2,0)$ and $(3,0)$, respectively.
		Red (blue) arrows show the action of the roots (``lowering operators'') $-\boldsymbol{\alpha}^{(1)}$ ($-\boldsymbol{\alpha}^{(2)}$)
		to the weights (see Fig.~\ref{fig:SU3-weight-lattice} for the definitions of $\boldsymbol{\alpha}^{(1,2)}$).
		\label{fig:weights-SU3-6-rep}}
\end{figure}

{\it Fundamental weights - }  From the set of simple roots $\{ \boldsymbol{\alpha}^{(i)} \}$,
we can introduce its dual, i.e., the fundamental weights
$\boldsymbol{\omega}_{(i)}$ satisfying
\begin{equation}
	\boldsymbol{\alpha}^{(i)\vee}{\cdot} \boldsymbol{\omega}_{(j)} = \delta^{i}_{j} \; ,
\end{equation}
which can be used as the basis of the weights ({\em Dynkin basis}):
\begin{equation}
	\boldsymbol{\mu} = \sum_{i=1}^{N-1} d(\boldsymbol{\mu})^{i} \, \boldsymbol{\omega}_{(j)} \; .
\end{equation}
The coordinates $d(\boldsymbol{\mu})^{i}$ in this basis is called {\em Dynkin labels}.
The lattice spanned by the basis $\{ \boldsymbol{\omega}_{(i)} \}$ is called the {\em weight lattice}
$\Lambda_{\text{w}}(\text{\sun})$ (see Fig.~\ref{fig:SU3-weight-lattice}).
The relation between the coroot lattice $\Lambda_{\text{r}}^{\vee}(\text{\sun})$ and the weight lattice $\Lambda_{\text{w}}(\text{\sun})$ is
analogous to that between the lattices in the real space and the reciprocal space.
Any irreducible representation $\mathcal{R}$ of \sun is specified by its highest weight $\boldsymbol{\lambda}_{\mathcal{R}}$ or its
Dynkin labels $\{  d(\mathcal{R})^{i} \}$
\begin{equation}
	\boldsymbol{\lambda}_{\mathcal{R}} = \sum_{i=1}^{r} d(\mathcal{R})^{i} \boldsymbol{\omega}_{(j)} \quad
	(d^i \in \mathbb{Z}, d^i \geq 0)
	\label{eqn:Dynkin-labels}
\end{equation}
and, by applying the lowering operators $E_{- \boldsymbol{\alpha}}$ ($\boldsymbol{\alpha} \in \Delta_{+}$),
we can construct the corresponding irreducible representation (see Fig.~\ref{fig:weights-SU3-6-rep} for $\mathfrak{su}(3)$ examples).
In \sun, the representation specified by $(d^1,d^2, \cdots, d^{N-1})$ has a Young diagram with
$d^1$ columns with length 1, $d^2$ columns with length 2, $\cdots$, and $d^{N-1}$ columns with length $N-1$.
For example, the fundamental representations are always specified by the Dynkin labels $\{  d(\mathcal{R})^{i} \}$
in which only one of $d^i$ is 1 and the others are zero.
\subsection{Affine Lie Algebras and Wess-Zumino-Witten model}
The affine Lie algebras are characterized by the following commutation relations which generalize \eqref{eq:liecom}:
\begin{equation}
	[ J^{\alpha}_{n} \, , \, J^{\beta}_{m}]
	= i {f_{\alpha\beta\gamma}}J_{n+m}^{\gamma} + \widetilde{k} n \delta_{m+n,0} \delta^{\alpha\beta}  \quad
	\label{eqn:KM-algebra}
\end{equation}
(see, e.g., Refs.~\cite{Goddard-O-86,Franscesco1997} for physicist-friendly reviews of affine Lie algebras).
Physically, \eqref{eqn:KM-algebra} is the algebra of the Fourier modes of the local SU($N$) currents $\{ J^{\alpha}(x) \}$ satisfying:
\begin{equation}
	\begin{split}
		& [\, J^{\alpha}(x)\, , \, J^{\beta} (y)\,]
		= i\, f^{\alpha\beta\gamma} J^{\gamma}(y)\delta(x-y) \\
		&\hskip 2.5cm+ \frac{i}{2\pi} \widetilde{k} \, \delta^{\alpha\beta} \, \partial_{x}\delta(x-y)   \\
		&  J^{\alpha}(x) = \frac{1}{L}\sum_{n\in\mathbb{Z}} e^{-i\frac{2\pi}{L}n x} \, J_{n}^{\alpha}  \;  .
	\end{split}
\end{equation}
The above are anomalous in that the right-hand side contains the central term [which is proportional to $\delta^{\prime}(x)$]
on top of the term expected from the Lie algebra.
Obviously, the coefficient $\widetilde{k}$ of the central term  depends on
the normalization of $J^{\alpha}_{n}$ and it is convenient to introduce the normalization-independent integer
$k$ called the {\em level} of the affine Lie algebra by
\[  \widetilde{k} =\frac{|\boldsymbol{\theta}|^{2}}{2} k  \; , \]
where $|\boldsymbol{\theta}|$ is the length of the highest root (the quantization of $k$ follows, e.g., from the consistency
of the path-integral representation of the WZW model).
The $|\boldsymbol{\theta}|$ depends on the normalization of the generators and, in \sun,
a convenient choice is to normalize the $N$-dimensional hermitian generators $\{ J^{\alpha} \}$
as $\text{Tr}(J^{\alpha} J^{\beta})=\delta^{\alpha\beta}$
which amounts to setting $|\boldsymbol{\theta}| = \sqrt{2}$.  Then, we do not have to distinguish between
the coefficient $\widetilde{k}$ of the central term and the level $k\, (\in \mathbb{Z})$.
The special case $m=n=0$ of \eqref{eqn:KM-algebra} reduces to \eqref{eq:liecom}, which means that the zero modes
$\{ J^{\alpha}_{0} \}$ form the usual \sun Lie algebra (called the {\em horizontal subalgebra}).

As a class of CFT with Lie-algebra symmetry, the WZW CFT has the Virasoro generators $\{ L_n \}$ which are bilinear in $J^{\alpha}_{n}$
({\em Sugawara form}) \cite{Knizhnik-Z-84,Goddard-O-86}:
\begin{equation}
	L_{n} = \frac{1}{|\boldsymbol{\theta}|^{2}(g^{\vee} + k )} \sum_{\alpha} \sum_{m \in \mathbb{Z}}
	: J^{\alpha}_{m} J^{\alpha}_{n-m} :  \; ,
	\label{eqn:L-in-Sugawara}
\end{equation}
where the normal-ordering $:\cdots:$ is defined by
\[
: J^{\alpha}_{m} J^{\alpha}_{n} : \, =
\begin{cases}
	J^{\alpha}_{m} J^{\alpha}_{n}  & \text{when $m<0$} \\
	J^{\alpha}_{n} J^{\alpha}_{m} & \text{when $m \geq 0$} \; .
\end{cases}
\]
The number $g^{\vee}$ (the {\em dual Coxter number}), which is peculiar to a given Lie algebra,
is given by the structure constants as
$- f^{\alpha\beta\mu} f^{\alpha\mu\gamma} = |\boldsymbol{\theta}|^{2} g^{\vee} \delta^{\beta\gamma}$
and is equal to $N$ in \sun.
By a direct calculation, we can show that the above $\{ L_n \}$ satisfy the Virasoro algebra
\begin{equation}
	[ L_m , L_n] = (m-n) L_{m+n} + \frac{1}{12} c(\mathfrak{g},k) m(m^{2}-1) \delta_{m+n,0}
	\label{eqn:Virasoro-WZW-1}
\end{equation}
with the central charge given by
\begin{equation}
	c(\mathfrak{g},k)
	= \frac{k \, \text{dim}(\mathfrak{g})}{k + g^{\vee}}  \quad (k=1,2, \ldots) \; ,
\end{equation}
which, for \sun, reads
\begin{equation}
	c(\text{\sun},k) = \frac{k (N^2-1)}{N+k}  \; .
	\label{eqn:c-SUN-WZW}
\end{equation}
On top of Eq.~\eqref{eqn:Virasoro-WZW-1}, $\{ L_n \}$ satisfy the following commutation relations with the generators
$\{ J^{\alpha}_{n} \}$:
\begin{equation}
	[ L_m , J^{\alpha}_n ] = -n J^{\alpha}_{m+n} \; .
	\label{eqn:Virasoro-WZW-2}
\end{equation}
In particular,
\begin{equation}
	[ L_0 , J^{\alpha}_n ] = -n J^{\alpha}_{n}
	\; , \;\;  [ L_0 , J^{\alpha}_0 ] = 0
	\label{eqn:Virasoro-WZW-3}
\end{equation}
implies that not only $L_{-n}$ ($n>0$) but also $J^{\alpha}_{-n}$ increase the eigenvalue of $L_0$ and that
for each eigenvalue of $L_0$ (i.e., for each level of conformal towers)
we have a {\em reducible} representation of \sun (formed by $\{ J^{\alpha}_{0} \}$).

In CFTs with extended symmetries, it is convenient to define the {\em primary states} $|\phi \rangle$ as those annihilated by
all $J^{\alpha}_{n}$ with positive $n$:
\begin{equation}
	J^{\alpha}_{n} | \phi \rangle = 0  \quad (n>0) \; .
	\label{eqn:KM-primary}
\end{equation}
Then, from \eqref{eqn:L-in-Sugawara}, $|\phi \rangle$ {\em automatically} satisfy the primary condition with respect to
the Virasoro algebra [the converse is not true; in that sense, \eqref{eqn:KM-primary} is stronger than \eqref{eqn:Virasoro-primary}]:
\begin{equation}
	\begin{split}
		L_{n} | \phi \rangle &= 0 \quad (n >0) \\
		L_{0} | \phi \rangle &= \frac{1}{|\boldsymbol{\theta}|^{2}(g^{\vee} + k)} \sum_{\alpha}
		J^{\alpha}_{0} J^{\alpha}_{0} | \phi \rangle \\
		&=  \frac{1}{|\boldsymbol{\theta}|^{2}(g^{\vee} + k)} \mathcal{C}_{2} | \phi \rangle
		 = h_{\phi} | \phi \rangle  \; ,
	\end{split}
	\label{eqn:Virasoro-primary}
\end{equation}
where $J^{\alpha}_{\phi}$ is a matrix representation of $J^{\alpha}$ and $\mathcal{C}_{2}$ is the quadratic Casimir of \sun.
All these mean that the primary states of the WZW model transform under
the irreducible representations $\mathcal{R}$ of the ordinary \sun spanned by the subset $\{ J^{\alpha}_{0} \}$:
\begin{equation}
	\begin{split}
		& |\phi \rangle = | \mathcal{R}; \boldsymbol{\mu}(\mathcal{R}) \rangle
		\quad (\boldsymbol{\mu}(\mathcal{R}): \; \text{weights of $\mathcal{R}$})   \\
		& J^{\alpha}_{0} | \mathcal{R}; \boldsymbol{\mu}(\mathcal{R}) \rangle
		= - J^{\alpha}(\mathcal{R}) | \mathcal{R}; \boldsymbol{\mu}(\mathcal{R}) \rangle   \\
		& [  J^{\alpha}(\mathcal{R}): \text{$J^{\alpha}$ in representation $\mathcal{R}$} ]
	\end{split}
\end{equation}
and that the conformal weights $h_{\phi}$ are given essentially by the quadratic Casimir $\mathcal{C}_{2}$ of $\mathcal{R}$:
\begin{equation}
	h(\mathcal{R}) = \frac{\mathcal{C}_{2}(\mathcal{R})}{|\boldsymbol{\theta}|^{2}(g^{\vee} + k)}
	\quad (g^{\vee}=N \text{ for \sun})  \; .
\end{equation}
As in other CFTs, these are the lowest states in a given $\mathcal{R}$-sector and the higher-lying states are
generated by applying $J^{\alpha}_{-n}$ ($n>0$).

There is a selection rule about the allowed $\mathcal{R}$ for a given level $k$, which,
in terms of the Dynkin labels $(d^1,\ldots, d^{r})$ [see Eq.~\eqref{eqn:Dynkin-labels}], reads for \sun
\begin{equation}
	\sum_{i=1}^{N-1} d (\mathcal{R})^{i} \leq k \; .
	\label{eqn:selection-by-k}
\end{equation}
In the level-1 ($k=1$) SU($N$) WZW model which is relevant in this paper,
only the vacuum [$\mathbf{1}$; SU($N$)-singlet with $\boldsymbol{d}=(0,\ldots,0)$] and the $N-1$ antisymmetric representations
$\text{aIR}_N(r_0)$ [rank-$r_0$ antisymmetric tensor with $\boldsymbol{d}=(0,\ldots,0,\underbrace{1}_{r_0},0,\ldots,0)$; $r_0=1,\ldots, N-1$]
in Sec.~\ref{sec:open} are allowed for primary states:
\begin{equation}
	\begin{split}
		& \mathcal{C}_{2}\left( \text{\scriptsize $r_0$} \Bigl\{
		\raisebox{-1.4ex}{
			{\tiny \yng(1,1,1) }
		}
		\right)
		= \frac{N+1}{2N} r_0 (N-r_0) |\boldsymbol{\theta}|^{2}    \\
		&  h \left( \text{\scriptsize $r_0$} \Bigl\{
		\raisebox{-1.4ex}{
			{\tiny \yng(1,1,1) }
		}
		\right)
		= \frac{1}{2N}r_0 (N - r_0)   \quad  (r_0 =0,\ldots, N-1) \; .
	\end{split}
\end{equation}
These $N$ different primary states (fields) correspond to $N$ topologically degenerate ground states of
SU($N$)$_1$ CSL on a torus.
For the selection rule for general $\mathfrak{g}$, see, e.g.,  Sec.~3.4 of Ref.~\cite{Goddard-O-86}.
\subsection{Finite-size spectrum}
For the clarity of the explanation, we assume $\mathfrak{g}=\mathfrak{su}(N)$ and normalize
the generator as $|\boldsymbol{\theta}| = \sqrt{2}$ in this section.
In this normalization, the coefficient $\widetilde{k}$ of the central term is equal to the level $k$, and $\mathcal{C}_2$ is given simply by
\begin{equation}
	\begin{split}
		&  \mathcal{C}_{2}(\mathcal{R}) =
		\sum_{i,j=1}^{N-1}(\boldsymbol{d}(\mathcal{R}) + \boldsymbol{e})_{i} (A^{-1})^{ij}
		(\boldsymbol{d}(\mathcal{R}) + \boldsymbol{e})_{j}  \\
	&\hskip 0.9cm  	-  \frac{1}{12}N(N^{2}-1)  \, , \\
		& \boldsymbol{e} \equiv \underbrace{%
			(1,1,\ldots, 1)
		}_{N-1} \; ,
	\end{split}
\end{equation}
where the matrix $A$ is the Cartan matrix defined in \eqref{eq:Cartan} and
$\boldsymbol{d}(\mathcal{R})$ is the Dynkin labels that characterizes the highest weight $\boldsymbol{\lambda}_{\mathcal{R}}$
of the representation $\mathcal{R}$ by Eq.~\eqref{eqn:Dynkin-labels}.
When we normalize the $N$-dimensional generators as $\text{Tr}(J^{\alpha} J^{\beta})= \kappa \delta^{\alpha\beta}$,
we need to multiply the right-hand side by $\kappa$.

The Hamiltonian of the chiral CFT is given by \cite{WZWreview1988,Franscesco1997}:
\begin{equation}
	H_{\text{chiral}} = \frac{2\pi}{l} v \left( L_{0} - \frac{c}{24} \right)  \quad (l : \text{ system size})  \; ,
\end{equation}
where $v$ is the velocity parameter of the system.
As $L_0$ and $c$ in the level-$k$ SU($N$) WZW CFT are given respectively by \eqref{eqn:L-in-Sugawara}
and \eqref{eqn:c-SUN-WZW}, we obtain:
\begin{equation}
	\begin{split}
		H_{\text{chiral}}^{\text{\sun}} = &  \frac{2\pi v}{l}
		\frac{1}{2(N+k)} \sum_{\alpha \in\text{SU($N$)}}
		\left\{
		J_{0}^{\alpha}J_{0}^{\alpha}
		+
		2 \sum_{n=1}^{\infty}   J_{-n}^{\alpha}J_{n}^{\alpha}
		\right\}  \\
		& - \frac{\pi v}{12 l} \frac{(N^2 - 1) k }{N + k}   \; .
	\end{split}
\end{equation}
The results in the previous section show that the Hilbert space in the sector specified by an irreducible
representation $\mathcal{R}$ of \sun [$\mathcal{R}$ obeys the selection rule \eqref{eqn:selection-by-k}]
consists of the ground (lowest) states with energy
\[
\frac{2\pi v}{l} \frac{\mathcal{C}_{2}(\mathcal{R})}{2(N + k)}
- \frac{\pi v}{12 l} \frac{(N^2 - 1) k }{N + k}
\]
and the equally-spaced excited states (with the level spacing $2\pi/l$).  All these states are labeled by the eigenvalues of
$L_{0}$ (energy) and $\{ H_{0}^1,\ldots, H_{0}^{N-1} \}$ (weight $\boldsymbol{\mu}$ of horizontal subalgebra
$\{ J^{\alpha}_{0} \}$).
As the action of the \sun-generators $J_0^{\alpha}$ does not change
the value of $L_0$ (i.e., energy) [see Eq.~\eqref{eqn:Virasoro-WZW-3}],
each excited level decomposes into a direct sum of several irreducible representations of \sun
(TABLES \ref{tab:su2_1}--\ref{tab:su8_70} shown below give such decompositions).

There is a compact way of encoding the information on the structure (i..e., energy, degeneracy, and the Lie-algebraic structure) of
the Hilbert space of the WZW CFT.
Consider the finite-temperature ($T$) partition function of the system:
\begin{equation}
	\begin{split}
	& Z = \text{Tr}_{\mathcal{R}} \, e^{- \frac{2\pi}{Tl}v \left( L_0 - \frac{c}{24} \right)}
	= q^{- \frac{c}{24}} \text{Tr}_{\mathcal{R}} \, q^{L_0}  \equiv Z_{\mathcal{R}}(q)  \\
	& (q \equiv e^{- \frac{2\pi}{Tl}v } ) \; ,
	\end{split}
\end{equation}
where the subscript $\mathcal{R}$ means that the trace is taken over all the excited states within
the $\mathcal{R}$-sector.
Since $L_0$ takes values $h(\mathcal{R})+N$ (with $N$ being non-negative integers), if we expand
$ Z_{\mathcal{R}}(q)$ in a power-series
\begin{equation}
	Z_{\mathcal{R}}(q) =  q^{h(\mathcal{R}) - \frac{c}{24}} \sum_{N =0}^{\infty} D(N) q^{N} \; ,
\end{equation}
it immediately gives the degeneracy $D(N)$ of the $N$-th excited state.

In order to know the Lie-algebraic structure, it is convenient to introduce the ``fugacities'' $\{ z_i \}$ for the weight and consider
the following generalized partition function:
\begin{equation}
	\widetilde{Z}_{\mathcal{R}}(q;\{z_i \})
	= q^{- \frac{c}{24}} \text{Tr}_{\mathcal{R}} \left\{ q^{L_0} \prod_{i=1}^{N-1} z_{i}^{H_{0}^{i}} \right\} \; ,
\end{equation}
where $\prod_{i}$ is over all the $N-1$ Cartan generators $\{ H_{0}^i \}$ of the \sun subalgebra $\{ J_{0}^{\alpha} \}$.
Now the coefficient of $q^{N+h(\mathcal{R}) - \frac{c}{24}}$ is a polynomial of $z_1^{\mu_1}z_2^{\mu_2}\cdots z_{N-1}^{\mu_{N-1}}$
that gives the multiplicity of the weight $\boldsymbol{\mu}$ in the $N$-th excited level.
In fact, the generalized partition function $\widetilde{Z}_{\mathcal{R}}(T,L)$ is nothing but the character of
the affine Lie algebra and its expression using the generalized theta function is known explicitly
(see, e.g., section 14.4 of Ref.~\cite{Franscesco1997} for more details).
TABLES \ref{tab:su2_1}--\ref{tab:su8_70}, which show the contents of irreducible representations appearing
at the excited levels of a given $\mathcal{R}$-sector, are obtained in this manner.
For example, TABLE~\ref{tab:su2_2} shows the structure of the Hilbert space of the level-1 SU(2) WZW CFT in
the sector of spin-1/2 representation [$h(j=1/2)=1/4$] and ``Order'' denotes $q^{L_0}$.
The degeneracy 2 of the first entry ($q^{1/4}$) is a direct consequence of the doublet level (primary states) constitutes
the $j=1/2$ representation of $\mathfrak{su}(2)$.
The third entry from the top implies that the second excited level ($q^{9/4}=q^{1/4+2}$) is six-fold degenerate and decomposes into
one $j=1/2$ (${\tiny \yng(1)}$) and one $j=3/2$ (${\tiny \yng(3)}$) representations:
\[   \mathbf{2} ({\tiny \yng(1)}) \oplus \mathbf{4} ({\tiny \yng(3)})  \; .\]

For level-1 \sun WZW CFT (for level-1 simply-laced $\mathfrak{g}$, in general), there is a simple way of
constructing the Hilbert space in terms of $N-1$ (i.e., rank-$\mathfrak{g}$) free bosons ({\em Frenkel-Kac construction}).
First we note that the central charge \eqref{eqn:c-SUN-WZW} of level-1 ($k=1$) \sun WZW CFT is $c=N-1$,
which clearly suggests its close relation to a system of $N-1$ free bosons.  Below, we quickly sketch
how we derive the partition function of the SU($N$)$_1$ WZW CFT.
To begin with, we prepare a set of $N-1$ bosons $\phi_{i}(z)$ ($i=1,\ldots,N-1$) which are normalized as:
\begin{equation}
	\langle \phi_{i}(z) \phi_{j}(w) \rangle \sim - \delta_{ij} \log(z-w) \; .
\end{equation}
The key properties of these bosons are the following operator-product expansions (OPE)
\cite{WZWreview1988,Franscesco1997}:
\begin{equation}
	\begin{split}
		&	 \partial_z \phi_{i}(z) \partial_w \phi_{j}(w) \sim \frac{- \delta_{ij}}{(z-w)^{2}} \\
		&	 \partial_z \phi_{i}(z) : e^{i \boldsymbol{v}{\cdot} \boldsymbol{\phi} (w) } : \,
		= \partial_z \phi_{i}(z) : e^{i \sum_{j} v_j \phi_j (w) } : \\ 
		& \phantom{\partial_z \phi_{i}(z) : e^{i \boldsymbol{v}{\cdot} \boldsymbol{\phi} (w) } :  }
		\sim \frac{-i v_i }{z-w}   : e^{i \boldsymbol{v}{\cdot} \boldsymbol{\phi} (w) } :    \\
		& T (z)  : e^{i \boldsymbol{v}{\cdot} \boldsymbol{\phi} (w) } : \\
		& = - \frac{1}{2} \sum_{i=1}^{N-1}:  (\partial_z \phi_{i}(z) )^{2} :  e^{i \boldsymbol{v}{\cdot} \boldsymbol{\phi} (w) } :   \\
		&\sim
		\frac{\boldsymbol{v}^{2}/2}{(z-w)^{2}}  : e^{i \boldsymbol{v}{\cdot} \boldsymbol{\phi} (w) } :
		+ \frac{1}{z-w} \partial_{w} : e^{i \boldsymbol{v}{\cdot} \boldsymbol{\phi} (w) } :  + \cdots \; ,
	\end{split}
\end{equation}
where $\boldsymbol{v}=(v_1,\ldots,v_{N-1})$ and $\boldsymbol{\phi} =(\phi_1,\ldots,\phi_{N-1})$.
Therefore, if we identify
\begin{equation}
	H^{i} (z) = i \partial_z \phi_{i}(z)  \; , \;\;
	E^{\boldsymbol{\alpha}}(z) = \, : e^{i \boldsymbol{\alpha}{\cdot} \boldsymbol{\phi} (w) } :
\end{equation}
(all the roots $\boldsymbol{\alpha}$ have the length $|\boldsymbol{\alpha} | = \sqrt{2}$),
they satisfy the OPEs expected for the generators of $k=1$ \sun (with scaling dimension 1) \cite{Goddard-O-86,Franscesco1997}:
\begin{equation}
	\begin{split}
		& H^{i} (z)  H^{j} (w) \sim \frac{\delta_{ij}}{(z-w)^{2}}  \\
		& H^{i} (z) E^{\boldsymbol{\alpha}}(w)  \sim \frac{\alpha_i }{z-w} E_{\boldsymbol{\alpha}}(w)  \\
		& E^{\boldsymbol{\alpha}}(z) E^{\boldsymbol{\beta}}(w)   \\
		& \sim (z-w)^{\boldsymbol{\alpha}{\cdot}\boldsymbol{\beta}} E^{\boldsymbol{\alpha}+\boldsymbol{\beta}}(w) \\
		& \phantom{ \sim}
		+ i (z-w)^{\boldsymbol{\alpha}{\cdot}\boldsymbol{\beta}+1} \boldsymbol{\alpha}{\cdot} \partial_{w} \boldsymbol{\phi} (w)
		E^{\boldsymbol{\alpha}+\boldsymbol{\beta}}(w)
	\end{split}
\end{equation}
[in \sun with $|\boldsymbol{\alpha} | = \sqrt{2}$,
$\boldsymbol{\alpha}{\cdot}\boldsymbol{\beta}=-1$ when $\boldsymbol{\alpha}+\boldsymbol{\beta}$ is a root
and $\boldsymbol{\alpha} \neq - \boldsymbol{\beta}$, and $\boldsymbol{\alpha}{\cdot}\boldsymbol{\beta}=-2$
when $\boldsymbol{\alpha} = - \boldsymbol{\beta}$].
This suggests that we can construct the Hilbert space of the $k=1$ SU($N$) WZW CFT by applying
$H^{i}(z) = i \partial_z \phi_{i}(z)$ ($i=1,\ldots,N-1$) repeatedly to the bosonic primary states
$|\boldsymbol{\mu}\rangle \equiv |\mu_1,\ldots, \mu_{N-1}\rangle= : e^{i \boldsymbol{\mu}{\cdot} \boldsymbol{\phi} (0) } : | 0\rangle$
[with $\boldsymbol{\mu}$ being the weights of \sun], that has the eigenvalue
$L_0 |\boldsymbol{\mu}\rangle = \boldsymbol{\mu}^{2}/2 |\boldsymbol{\mu}\rangle$.
The summation over all the possible excited states (with the mode $E_n=(2\pi/l)n$ being occupied with $N_n$ bosons)
of the $i$-th linearly-dispersive boson above the primary state $|\boldsymbol{\mu} \rangle$ yields the partial partition function
\[
e^{-\frac{2\pi v}{Tl}\frac{1}{2}\mu_i^{2}} z_{i}^{\mu_i} \prod_{n=1}^{\infty} \left\{ \sum_{N_n=0}^{\infty} e^{-\frac{2\pi v}{Tl}n N_n}  \right\}
=q^{\frac{1}{2}\mu_i^{2}} z_{i}^{\mu_i} /\prod_{n=1}^{\infty}(1-q^{n}) \; ,
\]
which is to be combined together for all $N-1$ bosons
yielding $q^{\frac{1}{2}\boldsymbol{\mu}^{2}} \prod_{i=1}^{N-1}z_{i}^{\mu_i} \prod_{n=1}^{\infty}(1-q^{n})^{N-1}$.
As the application of the other generators $E^{\boldsymbol{\alpha}}(z)$ changes the ``weight'' of
the primary states as $|\boldsymbol{\mu}\rangle \to |\boldsymbol{\mu}+ \boldsymbol{\alpha} \rangle$,
all these bosonic conformal towers specified by weights $\boldsymbol{\mu}$ that
are related to each other by translation by $\boldsymbol{\alpha}$ must be regarded as belonging to the same WZW conformal tower.
In $\mathfrak{su}(3)$, for instance, the weights $\boldsymbol{\mu}$ on the root lattice all together constitute a {\em single} WZW tower of
the identity representation $\mathbf{1}$ (see Fig.~\ref{fig:SU3-weight-lattice}).
Summing up the partial partition functions for those ``equivalent'' $\boldsymbol{\mu}$, we obtain the partition function of $k=1$ SU($N$) WZW CFT
(see section 15.6 of Ref.~\cite{Franscesco1997} for more details):
\begin{equation}
	\begin{split}
	& \widetilde{Z}_{\mathcal{R}}(q;\{z_i \})   \\
	& \equiv
	\frac{q^{-\frac{N-1}{24}}}{\prod_{n=1}^{\infty}(1-q^{n})^{N-1}}
	\left\{
	\sum_{\boldsymbol{\mu}\in \boldsymbol{\lambda}_{\mathcal{R}}
		+\Lambda_{\text{r}} }
	q^{\frac{1}{2}\boldsymbol{\mu}^{2}}\left(\prod_{i=1}^{N-1}z_{i}^{\mu_{i}}\right)
	\right\}  \; ,
	\end{split}
	\label{eqn:level-1-SUN-character}
\end{equation}
where $\boldsymbol{\lambda}_{\mathcal{R}}$ is the highest weight of the representation $\mathcal{R}$ and
the summation is taken over all the points $\boldsymbol{\mu}$ of the weight lattice $\Lambda_{\text{w}}$ which are equivalent to
$\boldsymbol{\lambda}_{\mathcal{R}}$ modulo the root lattice $\Lambda_{\text{r}}$ spanned by the simple roots
$\{ \boldsymbol{\alpha}^{(i)} \}$.
Since such $\boldsymbol{\mu}$ are given explicitly as
\begin{equation}
	\boldsymbol{\mu}
	= \boldsymbol{\lambda}_{\mathcal{R}} + \sum_{i=1}^{N-1} n_i \boldsymbol{\alpha}^{(i)}  \; ,
\end{equation}
we can trade the sum over $\boldsymbol{\mu}\in \boldsymbol{\lambda}_{\mathcal{R}} +\Lambda_{\text{r}}$ with
that over the $N-1$ integers $\{n_i \}$.   By construction, the representations $\mathcal{R}$ allowed as primary
in the SU($N$)$_1$ WZW CFT, which is relevant in this paper,
are restricted to the points of $\Lambda_{\text{w}}$ within the unit cell of $\Lambda_{\text{r}}$.
As all those $\mathcal{R}$ have the Dynkin labels $\sum_{i=1}^{N-1}d (\mathcal{R})^{i} =0,1$, this selection rule
is consistent with the general one \eqref{eqn:selection-by-k}.
For instance, in order to obtain the partition function for $\mathcal{R}=\mathbf{3}$ (${\tiny \yng(1)}$) of $\mathfrak{su}(3)$,
we sum over all the red points in
Fig.~\ref{fig:SU3-weight-lattice} connected to the point $\boldsymbol{\omega}_{(1)}$ (i.e., the highest weight of $\mathbf{3}$)
by the translation generated by two simple roots $\boldsymbol{\alpha}^{(1)}$ and $\boldsymbol{\alpha}^{(2)}$
(red and blue arrows, respectively); the three inequivalent points in the hatched ``unit cell'' correspond to the three primary fields
$\phi_\mathbf{1}$ (singlet vacuum), $\phi_{\mathbf{3}}$, and $\phi_{\overline{\mathbf{3}}}$ allowed in level-1 $\mathfrak{su}(3)$.


\begin{widetext}

\begin{table*}[!ht]
	\centering

	\caption{	\label{tab:su2_1}  SU(2)$_1$ WZW model  -- Tower of states starting from $\protect\su{0}{0}{1}$.}
	\begin{tabular}{|c|c|l|}
		\hline \hline
		\rotatebox[origin=c]{90}{$\;\;L_0\;\;$} & \rotatebox[origin=c]{90}{Order} & Irreps / Multiplicities	\\	\hline
		$0$	& $q^{0}$& \Yvcentermath1$\su{1}{0}{1}$				\\ \hline				
		$1$ & $q^{1}$& \Yvcentermath1$\su{1}{2}{3}$ 		\\ \hline	
		$2$	& $q^{2}$& \Yvcentermath1$\su{1}{0}{1}\oplus\su{1}{2}{3}$ \\ \hline	
		$3$	& $q^{3}$& \Yvcentermath1$\su{1}{0}{1}\oplus\su{2}{2}{3}$ 								\\ \hline	
		$4$	& $q^{4}$& \Yvcentermath1$\su{2}{0}{1}\oplus\su{2}{2}{3}\oplus\su{1}{4}{5}$ \\ \hline
		$5$	& $q^5{5}$& \Yvcentermath1$\su{2}{0}{1}\oplus\su{4}{2}{3}\oplus\su{1}{4}{5}$ \\ \hline
		$6$	& $q^{6}$& \Yvcentermath1$\su{4}{0}{1}\oplus\su{5}{2}{3}\oplus\su{2}{4}{5}$ \\ \hline
		$7$	& $q^{7}$& \Yvcentermath1$\su{4}{0}{1}\oplus\su{8}{2}{3}\oplus\su{3}{4}{5}$ \\ \hline
		\hline
	\end{tabular}
\end{table*}

\begin{table*}[!ht]
	\centering
	\caption{	\label{tab:su2_2}
		SU(2)$_1$ WZW model  -- Tower of states starting from $\protect\su{0}{1}{2}$.}
	\begin{tabular}{|c|c|l|}
		\hline \hline
		\rotatebox[origin=c]{90}{$\;\;L_0\;\;$} & \rotatebox[origin=c]{90}{Order} & Irreps / Multiplicities	\\	\hline
		$0$	& $q^{1/4}$& \Yvcentermath1$\su{1}{1}{2}$				\\ \hline				
		$1$ & $q^{5/4}$& \Yvcentermath1$\su{1}{1}{2}$ 		\\ \hline	
		$2$	& $q^{9/4}$& \Yvcentermath1$\su{1}{1}{2}\oplus\su{1}{3}{4}$ \\ \hline	
		$3$	& $q^{13/4}$& \Yvcentermath1$\su{2}{1}{2}\oplus\su{1}{3}{4}$ 								\\ \hline	
		$4$	& $q^{17/4}$& \Yvcentermath1$\su{3}{1}{2}\oplus\su{2}{3}{4}$ \\ \hline
		$5$	& $q^{21/4}$& \Yvcentermath1$\su{4}{1}{2}\oplus\su{3}{3}{4}$ \\ \hline
		$6$	& $q^{25/4}$& \Yvcentermath1$\su{6}{1}{2}\oplus\su{4}{3}{4}\oplus\su{1}{5}{6}$ \\ \hline
		$7$	& $q^{29/4}$& \Yvcentermath1$\su{8}{1}{2}\oplus\su{6}{3}{4}\oplus\su{1}{5}{6}$ \\ \hline
		\hline
	\end{tabular}
\end{table*}


\begin{table*}[!ht]
	\centering
	
	\caption{\label{tab:su3_1}
		SU(3)$_1$ WZW model  -- Tower of states starting from $\protect\su{0}{0}{1}$.}
	\begin{tabular}{|c|c|l|}
		\hline \hline
		\rotatebox[origin=c]{90}{$\;\;L_0\;\;$} & \rotatebox[origin=c]{90}{Order} & Irreps / Multiplicities	\\	\hline
		$0$	& $q^{0}$& \Yvcentermath1$\su{1}{0}{1}$				\\ \hline				
		$1$ & $q^{1}$& \Yvcentermath1$\su{1}{2,1}{8}$ 		\\ \hline	
		$2$	& $q^{2}$& \Yvcentermath1$\su{1}{0}{1}\oplus\su{2}{2,1}{8}$ \\ \hline	
		$3$	& $q^{3}$& \Yvcentermath1$\su{2}{0}{1}\oplus\su{3}{2,1}{8}\oplus\su{1}{3}{10}\oplus\su{1}{3,3}{\overline{10}}$ 								\\ \hline	
		$4$	& $q^{4}$& \Yvcentermath1$\su{3}{0}{1}\oplus\su{6}{2,1}{8}\oplus\su{1}{3}{10}\oplus\su{1}{3,3}{\overline{10}}\oplus\su{1}{4,2}{27}$ \\ \hline
		$5$	& $q^{5}$& \Yvcentermath1$\su{4}{0}{1}\oplus\su{10}{2,1}{8}\oplus\su{3}{3}{10}\oplus\su{3}{3,3}{\overline{10}}\oplus\su{2}{4,2}{27}$ \\ \hline
		$6$	& $q^{6}$& \Yvcentermath1$\su{8}{0}{1}\oplus\su{16}{2,1}{8}\oplus\su{5}{3}{10}\oplus\su{5}{3,3}{\overline{10}}\oplus\su{5}{4,2}{27}$ \\ \hline
		$7$	& $q^{7}$& \Yvcentermath1$\su{10}{0}{1}\oplus\su{27}{2,1}{8}\oplus\su{9}{3}{10}\oplus\su{9}{3,3}{\overline{10}}\oplus\su{8}{4,2}{27}\oplus\su{1}{5,4}{\overline{35}}\oplus\su{1}{5,1}{35}$ \\ \hline
		\hline
	\end{tabular}
\end{table*}

\begin{table*}[!ht]
	\centering
	\caption{\label{tab:su3_3}
		SU(3)$_1$ WZW model  -- Tower of states starting from $\protect\su{0}{1}{3}$ (resp. $\protect\su{0}{1,1}{\bar{3}}$ by conjugation of all IRREPs).}
	\begin{tabular}{|c|c|l|}
		\hline \hline
		\rotatebox[origin=c]{90}{$\;\;L_0\;\;$} & \rotatebox[origin=c]{90}{Order} & Irreps / Multiplicities	\\	\hline
		$0$	& $q^{1/3}$& \Yvcentermath1$\su{1}{1}{3}$				\\ \hline				
		$1$ & $q^{4/3}$& \Yvcentermath1$\su{1}{1}{3}\oplus\su{1}{2,2}{\bar{6}}$ 		\\ \hline	
		$2$	& $q^{7/3}$& \Yvcentermath1$\su{2}{1}{3}\oplus\su{1}{2,2}{\bar{6}}\oplus\su{1}{3,1}{15}$ \\ \hline	
		$3$	& $q^{10/3}$& \Yvcentermath1$\su{3}{1}{3}\oplus\su{3}{2,2}{\bar{6}}\oplus\su{2}{3,1}{15}$ 								\\ \hline	
		$4$	& $q^{13/3}$& \Yvcentermath1$\su{6}{1}{3}\oplus\su{4}{2,2}{\bar{6}}\oplus\su{4}{3,1}{15}\oplus\su{1}{4,3}{24}$ \\ \hline
		$5$	& $q^{16/3}$& \Yvcentermath1$\su{9}{1}{3}\oplus\su{8}{2,2}{\bar{6}}\oplus\su{1}{4}{15'}\oplus\su{7}{3,1}{15}\oplus\su{2}{4,3}{24}$ \\ \hline
		$6$	& $q^{19/3}$& \Yvcentermath1$\su{15}{1}{3}\oplus\su{12}{2,2}{\bar{6}}\oplus\su{1}{4}{15'}\oplus\su{13}{3,1}{15}\oplus\su{4}{4,3}{24}\oplus\su{1}{5,2}{42}$ \\ \hline
		$7$	& $q^{22/3}$& \Yvcentermath1$\su{22}{1}{3}\oplus\su{21}{2,2}{\bar{6}}\oplus\su{3}{4}{15'}\oplus\su{21}{3,1}{15}\oplus\su{8}{4,3}{24}\oplus\su{2}{5,2}{42}$ \\ \hline
		\hline
	\end{tabular}
\end{table*}

\begin{table*}[!ht]
	\centering
	\caption{	\label{tab:su4_1}
		SU$(4)_1$ WZW model -- Tower of states starting from $\protect\su{0}{0}{1}$.}
	\begin{tabular}{|c|c|l|}
		\hline \hline
		\rotatebox[origin=c]{90}{$\;\;L_0\;\;$} & \rotatebox[origin=c]{90}{Order} & Irreps / Multiplicities	\\	\hline
		$0$	& $q^0$& \Yvcentermath1$\su{1}{0}{1}$				\\ \hline				
		$1$ & $q^1$& \Yvcentermath1$\su{1}{2,1,1}{15}$ 		\\ \hline	
		$2$	& $q^2$& \Yvcentermath1$\su{1}{0}{1}	\oplus\su{2}{2,1,1}{15}	\oplus\su{1}{2,2}{20'}$ \\ \hline	
		$3$	& $q^3$& \Yvcentermath1$\su{2}{0}{1}	\oplus\su{4}{2,1,1}{15}	\oplus\su{1}{2,2}{20'} \oplus\su{1}{3,1}{45}
		\oplus\su{1}{3,3,2}{\overline{45}}$ 								\\ \hline	
		$4$	& $q^4$& \Yvcentermath1$\su{4}{0}{1}	\oplus\su{7}{2,1,1}{15}	\oplus\su{4}{2,2}{20'} \oplus\su{2}{3,1}{45}
		\oplus\su{2}{3,3,2}{\overbar{45}}	\oplus\su{1}{4,2,2}{84}$ \\ \hline
		\hline
	\end{tabular}
\end{table*}

\begin{table*}[!ht]
	\centering
	\caption{\label{tab:su4_4}
		SU$(4)_1$ WZW model -- Tower of states starting from $\protect\su{0}{1}{4}$ (resp. $\protect\su{0}{1,1,1}{\overline{4}}$ by conjugation of all IRREPs).}
	\begin{tabular}{|c|c|l|}
		\hline \hline
		\rotatebox[origin=c]{90}{$\;\;L_0\;\;$} & \rotatebox[origin=c]{90}{Order} & Irreps / Multiplicities	\\	\hline
		$0$	& $q^{3/8}$& \Yvcentermath1$\su{1}{1}{4}$				\\ \hline				
		$1$ & $q^{11/8}$& \Yvcentermath1$\su{1}{1}{4}\oplus\su{1}{2,2,1}{20}$			\\ \hline	
		$2$	& $q^{19/8}$& \Yvcentermath1$\su{2}{1}{4}\oplus\su{2}{2,2,1}{20}\oplus\su{1}{3,1,1}{36}$ 			\\ \hline	
		$3$	& $q^{27/8}$& \Yvcentermath1$\su{4}{1}{4}\oplus\su{1}{3,3,3}{20''}\oplus\su{4}{2,2,1}{20}\oplus\su{2}{3,1,1}{36}\oplus\su{1}{3,2}{60}$			\\ \hline	
		$4$	& $q^{35/8}$& \Yvcentermath1$\su{7}{1}{4}\oplus\su{1}{3,3,3}{20''}\oplus\su{8}{2,2,1}{20}\oplus\su{5}{3,1,1}{36}\oplus\su{2}{3,2}{60}\oplus\su{1}{4,3,2}{140}$			 \\ \hline
		\hline
	\end{tabular}
\end{table*}

\begin{table*}[!ht]
	\centering
	\caption{	\label{tab:su4_6}
		SU$(4)_1$ WZW model -- Tower of states starting from $\protect\su{0}{1,1}{6}$.}
	\begin{tabular}{|c|c|l|}
		\hline \hline
		\rotatebox[origin=c]{90}{$\;\;L_0\;\;$} & \rotatebox[origin=c]{90}{Order} & Irreps / Multiplicities	\\	\hline
		$0$	& $q^{1/2}$& \Yvcentermath1$\su{1}{1,1}{6}$				\\ \hline				
		$1$ & $q^{3/2}$& \Yvcentermath1$\su{1}{1,1}{6}\oplus\su{1}{2}{10}\oplus\su{1}{2,2,2}{\overline{10}}$			\\ \hline	
		$2$	& $q^{5/2}$& \Yvcentermath1$\su{3}{1,1}{6}\oplus\su{1}{2}{10}\oplus\su{1}{2,2,2}{\overline{10}}\oplus\su{1}{3,2,1}{64}$ 			\\ \hline	
		$3$	& $q^{7/2}$& \Yvcentermath1$\su{4}{1,1}{6}\oplus\su{3}{2}{10}\oplus\su{3}{2,2,2}{\overline{10}}\oplus\su{3}{3,2,1}{64}$			\\ \hline	
		$4$	& $q^{9/2}$& \Yvcentermath1$\su{9}{1,1}{6}\oplus\su{5}{2}{10}\oplus\su{5}{2,2,2}{\overline{10}}\oplus\su{1}{3,3}{50}\oplus\su{6}{3,2,1}{64}\oplus\su{1}{4,3,3}{\overline{70}}\oplus\su{1}{4,1,1}{70}$			 \\ \hline
		\hline
	\end{tabular}
\end{table*}

\begin{table*}[!ht]
	\centering
	\label{tab:su5_1}
	\caption{SU$(5)_1$ WZW model -- Tower of states starting from $\protect\su{0}{0}{1}$.}
	\begin{tabular}{|c|c|l|}
		\hline \hline
		\rotatebox[origin=c]{90}{$\;\;L_0\;\;$} & \rotatebox[origin=c]{90}{Order} & Irreps / Multiplicities	\\	\hline
		$0$	& $q^{0}$& \Yvcentermath1$\su{1}{0}{1}$				\\ \hline				
		$1$ & $q^{1}$& \Yvcentermath1$\su{1}{2,1,1,1}{24}$			\\ \hline	
		$2$	& $q^{2}$& \Yvcentermath1$\su{1}{0}{1}\oplus\su{2}{2,1,1,1}{24}\oplus\su{1}{2,2,1}{75}$ 			\\ \hline	
		$3$	& $q^{3}$& \Yvcentermath1$\su{2}{0}{1}\oplus\su{4}{2,1,1,1}{24}\oplus\su{2}{2,2,1}{75}\oplus\su{1}{3,1,1}{126}\oplus\su{1}{3,3,2,2}{\overline{126}}$			\\ \hline	
		\hline
	\end{tabular}
\end{table*}

\begin{table*}[!ht]
	\centering
	\caption{\label{tab:su5_5}
		SU$(5)_1$ WZW model -- Tower of states starting from $\protect\su{0}{1}{5}$ (resp. $\protect\su{0}{1,1,1,1}{\overline{5}}$ by conjugation of all IRREPs).}
	\begin{tabular}{|c|c|l|}
		\hline \hline
		\rotatebox[origin=c]{90}{$\;\;L_0\;\;$} & \rotatebox[origin=c]{90}{Order} & Irreps / Multiplicities	\\	\hline
		$0$	& $q^{2/5}$& \Yvcentermath1$\su{1}{1}{5}$				\\ \hline				
		$1$ & $q^{7/5}$& \Yvcentermath1$\su{1}{1}{5}\oplus\su{1}{2,2,1,1}{45}$			\\ \hline	
		$2$	& $q^{12/5}$& \Yvcentermath1$\su{2}{1}{5}\oplus\su{2}{2,2,1,1}{45}\oplus\su{1}{2,2,2}{50}\oplus\su{1}{3,1,1,1}{70}$ 			\\ \hline	
		$3$	& $q^{17/5}$& \Yvcentermath1$\su{4}{1}{5}\oplus\su{5}{2,2,1,1}{45}\oplus\su{1}{2,2,2}{50}\oplus\su{2}{3,1,1,1}{70}\oplus\su{1}{3,3,3,2}{105}\oplus\su{1}{3,2,1}{280}$			\\ \hline	
		\hline
	\end{tabular}
\end{table*}

\begin{table*}[!ht]
	\centering

	\caption{	\label{tab:su5_10}SU$(5)_1$ WZW model -- Tower of states starting from $\protect\su{0}{1,1}{10}$  (resp. $\protect\su{0}{1,1,1}{\overline{10}}$ by conjugation of all IRREPs).}
	\begin{tabular}{|c|c|l|}
		\hline \hline
		\rotatebox[origin=c]{90}{$\;\;L_0\;\;$} & \rotatebox[origin=c]{90}{Order} & Irreps / Multiplicities	\\	\hline
		$0$	& $q^{3/5}$& \Yvcentermath1$\su{1}{1,1}{10}$				\\ \hline				
		$1$ & $q^{8/5}$& \Yvcentermath1$\su{1}{1,1}{10}\oplus\su{1}{2}{15}\oplus\su{1}{2,2,2,1}{40}$			\\ \hline	
		$2$	& $q^{13/5}$& \Yvcentermath1$\su{3}{1,1}{10}\oplus\su{1}{2}{15}\oplus\su{2}{2,2,2,1}{40}\oplus\su{1}{3,2,1,1}{175}$ 			\\ \hline	
		$3$	& $q^{18/5}$& \Yvcentermath1$\su{5}{1,1}{10}\oplus\su{3}{2}{15}\oplus\su{1}{3,3,3,3}{35}\oplus\su{4}{2,2,2,1}{40}\oplus\su{3}{3,2,1,1}{175}\oplus\su{1}{3,2,2}{210}$			\\ \hline	
		\hline
	\end{tabular}
\end{table*}

\begin{table*}[!ht]
	\centering
	\label{tab:su6_1}
	\caption{SU$(6)_1$ WZW model -- Tower of states starting from $\protect\su{0}{0}{1}$.}
	\begin{tabular}{|c|c|l|}
		\hline \hline
		\rotatebox[origin=c]{90}{$\;\;L_0\;\;$} & \rotatebox[origin=c]{90}{Order} & Irreps / Multiplicities	\\	\hline
		$0$	& $q^{0}$& \Yvcentermath1$\su{1}{0}{1}$				\\ \hline				
		$1$ & $q^{1}$& \Yvcentermath1$\su{1}{2,1,1,1,1}{35}$			\\ \hline	
		$2$	& $q^{2}$& \Yvcentermath1$\su{1}{0}{1}\oplus\su{2}{2,1,1,1,1}{35}\oplus\su{1}{2,2,1,1}{189}$ 			\\ \hline	
		$3$	& $q^{3}$& \Yvcentermath1$\su{2}{0}{1}\oplus\su{4}{2,1,1,1,1}{35}\oplus\su{1}{2,2,2}{175}\oplus\su{2}{2,2,1,1}{189}\oplus\su{1}{3,1,1,1}{280}\oplus\su{1}{3,3,2,2,2}{\overline{280}}$			\\ \hline	
		\hline
	\end{tabular}
\end{table*}

\begin{table*}[!ht]
	\centering

	\caption{	\label{tab:su6_6}
		SU$(6)_1$ WZW model -- Tower of states starting from $\protect\su{0}{1}{6}$  (resp. $\protect\su{0}{1,1,1,1,1}{\overline{6}}$ by conjugation of all IRREPs.}
	\begin{tabular}{|c|c|l|}
		\hline \hline
		\rotatebox[origin=c]{90}{$\;\;L_0\;\;$} & \rotatebox[origin=c]{90}{Order} & Irreps / Multiplicities	\\	\hline
		$0$	& $q^{5/12}$& \Yvcentermath1$\su{1}{1}{6}$				\\ \hline				
		$1$ & $q^{17/12}$& \Yvcentermath1$\su{1}{1}{6}\oplus\su{1}{2,2,1,1,1}{84}$			\\ \hline	
		$2$	& $q^{29/12}$& \Yvcentermath1$\su{2}{1}{6}\oplus\su{2}{2,2,1,1,1}{84}\oplus\su{1}{3,1,1,1,1}{120}\oplus\su{1}{2,2,2,1}{210}$ 			\\ \hline	
		$3$	& $q^{41/12}$& \Yvcentermath1$\su{4}{1}{6}\oplus\su{5}{2,2,1,1,1}{84}\oplus\su{2}{3,1,1,1,1}{120}\oplus\su{2}{2,2,2,1}{210}\oplus\su{1}{3,3,3,2,2}{336}\oplus\su{1}{3,2,1,1}{840}$			\\ \hline	
		\hline
	\end{tabular}
\end{table*}

\begin{table*}[!ht]
	\centering

	\caption{	\label{tab:su6_15}
		SU$(6)_1$ WZW model -- Tower of states starting from $\protect\su{0}{1,1}{15}$  (resp. $\protect\su{0}{1,1,1,1}{\overline{15}}$ by conjugation of all IRREPs).}
	\begin{tabular}{|c|c|l|}
		\hline \hline
		\rotatebox[origin=c]{90}{$\;\;L_0\;\;$} & \rotatebox[origin=c]{90}{Order} & Irreps / Multiplicities	\\	\hline
		$0$	& $q^{2/3}$& \Yvcentermath1$\su{1}{1,1}{15}$				\\ \hline				
		$1$ & $q^{5/3}$& \Yvcentermath1$\su{1}{1,1}{15}\oplus\su{1}{2}{21}\oplus\su{1}{2,2,2,1,1}{105}$			\\ \hline	
		$2$	& $q^{8/3}$& \Yvcentermath1$\su{3}{1,1}{15}\oplus\su{1}{2}{21}\oplus\su{1}{2,2,2,2}{105'}\oplus\su{2}{2,2,2,1,1}{105}\oplus\su{1}{3,2,1,1,1}{384}$ 			\\ \hline	
		$3$	& $q^{11/3}$& \Yvcentermath1$\su{5}{1,1}{15}\oplus\su{3}{2}{21}\oplus\su{1}{2,2,2,2}{105'}\oplus\su{5}{2,2,2,1,1}{105}\oplus\su{1}{3,3,3,3,2}{210'}\oplus\su{3}{3,2,1,1,1}{384}\oplus\su{1}{3,2,2,1}{1050}$			\\ \hline	
		\hline
	\end{tabular}
\end{table*}

\begin{table*}[!ht]
	\centering

	\caption{	\label{tab:su7_1}
		SU$(7)_1$ WZW model -- Tower of states starting from $\protect\su{0}{0}{1}$.}
	\begin{tabular}{|c|c|l|}
		\hline \hline
		\rotatebox[origin=c]{90}{$\;\;L_0\;\;$} & \rotatebox[origin=c]{90}{Order} & Irreps / Multiplicities	\\	\hline
		$0$	& $q^{0}$& \Yvcentermath1$\su{1}{0}{1}$				\\ \hline				
		$1$ & $q^{1}$& \Yvcentermath1$\su{1}{2,1,1,1,1,1}{48}$			\\ \hline	
		$2$	& $q^{2}$& \Yvcentermath1$\su{1}{0}{1}\oplus\su{2}{2,1,1,1,1,1}{48}\oplus\su{1}{2,2,1,1,1}{392}$ 			\\ \hline	
		$3$	& $q^{3}$& \Yvcentermath1$\su{2}{0}{1}\oplus\su{4}{2,1,1,1,1,1}{48}\oplus\su{2}{2,2,1,1,1}{392}\oplus\su{1}{3,1,1,1,1}{540}\oplus\su{1}{3,3,2,2,2,2}{\overline{540}}\oplus\su{1}{2,2,2,1}{784}$			\\ \hline	
		\hline
	\end{tabular}
\end{table*}

\begin{table*}[!ht]
	\centering
	\label{tab:su7_7}
	\caption{SU$(7)_1$ WZW model -- Tower of states starting from $\protect\su{0}{1}{7}$  (resp. $\protect\su{0}{1,1,1,1,1,1}{\overline{7}}$ by conjugation of all IRREPs).}
	\begin{tabular}{|c|c|l|}
		\hline \hline
		\rotatebox[origin=c]{90}{$\;\;L_0\;\;$} & \rotatebox[origin=c]{90}{Order} & Irreps / Multiplicities	\\	\hline
		$0$	& $q^{3/7}$& \Yvcentermath1$\su{1}{1}{7}$				\\ \hline				
		$1$ & $q^{10/7}$& \Yvcentermath1$\su{1}{1}{7}\oplus\su{1}{2,2,1,1,1,1}{140}$			\\ \hline	
		$2$	& $q^{17/7}$& \Yvcentermath1$\su{2}{1}{7}\oplus\su{2}{2,2,1,1,1,1}{140}\oplus\su{1}{3,1,1,1,1,1}{189}\oplus\su{1}{2,2,2,1,1}{588}$ 			\\ \hline	
		$3$	& $q^{24/7}$& \Yvcentermath1$\su{4}{1}{7}\oplus\su{5}{2,2,1,1,1,1}{140}\oplus\su{2}{3,1,1,1,1,1}{189}\oplus\su{1}{2,2,2,2}{490'}\oplus\su{2}{2,2,2,1,1}{588}\oplus\su{1}{3,3,3,2,2,2}{840}\oplus\su{1}{3,2,1,1,1}{2016}$			\\ \hline	
		\hline
	\end{tabular}
\end{table*}

\begin{table*}[!ht]
	\centering
	\label{tab:su7_21}
	\caption{SU$(7)_1$ WZW model -- Tower of states starting from $\protect\su{0}{1,1}{21}$  (resp. $\protect\su{0}{1,1,1,1,1}{\overline{21}}$ by conjugation of all IRREPs).}
	\begin{tabular}{|c|c|l|}
		\hline \hline
		\rotatebox[origin=c]{90}{$\;\;L_0\;\;$} & \rotatebox[origin=c]{90}{Order} & Irreps / Multiplicities	\\	\hline
		$0$	& $q^{5/7}$& \Yvcentermath1	$\su{1}{1,1}{21}$	\\ \hline				
		$1$ & $q^{12/7}$& \Yvcentermath1	$\su{1}{1,1}{21}\oplus\su{1}{2}{28}\oplus\su{1}{2,2,2,1,1,1}{224}$		\\ \hline	
		$2$	& $q^{19/7}$& \Yvcentermath1	$\su{3}{1,1}{21}\oplus\su{1}{2}{28}\oplus\su{2}{2,2,2,1,1,1}{224}\oplus\su{1}{2,2,2,2,1}{490}\oplus\su{1}{3,2,1,1,1,1}{735}$		\\ \hline	
		$3$	& $q^{26/7}$& \Yvcentermath1	$\su{5}{1,1}{21}\oplus\su{3}{2}{28}\oplus\su{5}{2,2,2,1,1,1}{224}\oplus\su{2}{2,2,2,2,1}{490}\oplus\su{3}{3,2,1,1,1,1}{735}\oplus\su{1}{3,3,3,3,2,2}{756}\oplus\su{1}{3,2,2,1,1}{3402}$		\\ \hline	
		\hline
	\end{tabular}
\end{table*}

\begin{table*}[!ht]
	\centering
	\label{tab:su7_35}
	\caption{SU$(7)_1$ WZW model -- Tower of states starting from $\protect\su{0}{1,1,1}{35}$  (resp. $\protect\su{0}{1,1,1,1}{\overline{35}}$ by conjugation of all IRREPs).}
	\begin{tabular}{|c|c|l|}
		\hline \hline
		\rotatebox[origin=c]{90}{$\;\;L_0\;\;$} & \rotatebox[origin=c]{90}{Order} & Irreps / Multiplicities	\\	\hline
		$0$	& $q^{}$& \Yvcentermath1$\su{1}{1,1,1}{35}$				\\ \hline				
		$1$ & $q^{}$& \Yvcentermath1$\su{1}{1,1,1}{35}\oplus\su{1}{2,1}{112}\oplus\su{1}{2,2,2,2,1,1}{210}$			\\ \hline	
		$2$	& $q^{}$& \Yvcentermath1$\su{3}{1,1,1}{35}\oplus\su{2}{2,1}{112}\oplus\su{1}{2,2,2,2,2}{196}\oplus\su{2}{2,2,2,2,1,1}{210}\oplus\su{1}{3,2,2,1,1,1}{1323}$ 			\\ \hline	
		$3$	& $q^{}$& \Yvcentermath1$\su{6}{1,1,1}{35}\oplus\su{1}{3}{84}\oplus\su{4}{2,1}{112}\oplus\su{1}{2,2,2,2,2}{196}\oplus\su{5}{2,2,2,2,1,1}{210}\oplus\su{1}{3,3,3,3,3,2}{378}\oplus\su{1}{3,3,1,1,1,1}{1260}\oplus\su{3}{3,2,2,1,1,1}{1323}\oplus\su{1}{3,2,2,2,1}{3024}$			\\ \hline	
		\hline
	\end{tabular}
\end{table*}

\begin{table*}[!ht]
	\centering

	\caption{	\label{tab:su8_1}SU$(8)_1$ WZW model -- Tower of states starting from $\protect\su{0}{0}{1}$.}
	\begin{tabular}{|c|c|l|}
		\hline \hline
		\rotatebox[origin=c]{90}{$\;\;L_0\;\;$} & \rotatebox[origin=c]{90}{Order} & Irreps / Multiplicities	\\	\hline
		$0$	& $q^{0}$& \Yvcentermath1$\su{1}{0}{1}$				\\ \hline				
		$1$ & $q^{1}$& \Yvcentermath1$\su{1}{2,1,1,1,1,1,1}{63}$			\\ \hline	
		$2$	& $q^{2}$& \Yvcentermath1$\su{1}{0}{1}\oplus\su{2}{2,1,1,1,1,1,1}{63}\oplus\su{1}{2,2,1,1,1,1}{720}$ 			\\ \hline	
		$3$	& $q^{3}$& \Yvcentermath1$\su{2}{0}{1}\oplus\su{4}{2,1,1,1,1,1,1}{63}\oplus\su{2}{2,2,1,1,1,1}{720}\oplus\su{1}{3,1,1,1,1,1}{945}\oplus\su{1}{3,3,2,2,2,2,2}{\overline{945}}\oplus\su{1}{2,2,2,1,1}{2352}$			\\ \hline	
		\hline
	\end{tabular}
\end{table*}

\begin{table*}[!ht]
	\centering
	\label{tab:su8_8}
	\caption{SU$(8)_1$ WZW model -- Tower of states starting from $\protect\su{0}{1}{8}$  (resp. $\protect\su{0}{1,1,1,1,1,1,1}{\bar{8}}$ by conjugation of all IRREPs).}
	\begin{tabular}{|c|c|l|}
		\hline \hline
		\rotatebox[origin=c]{90}{$\;\;L_0\;\;$} & \rotatebox[origin=c]{90}{Order} & Irreps / Multiplicities	\\	\hline
		$0$	& $q^{7/16}$& \Yvcentermath1$\su{1}{1}{8}$				\\ \hline				
		$1$ & $q^{23/16}$& \Yvcentermath1$\su{1}{1}{8}\oplus\su{1}{2,2,1,1,1,1,1}{216}$			\\ \hline	
		$2$	& $q^{39/16}$& \Yvcentermath1$\su{2}{1}{8}\oplus\su{2}{2,2,1,1,1,1,1}{216}\oplus\su{1}{3,1,1,1,1,1,1}{280}\oplus\su{1}{2,2,2,1,1,1}{1344}$ 			\\ \hline	
		\hline
	\end{tabular}
\end{table*}

\begin{table*}[!ht]
	\centering
	\label{tab:su8_28}
	\caption{SU$(8)_1$ WZW model -- Tower of states starting from $\protect\su{0}{1,1}{28}$  (resp. $\protect\su{0}{1,1,1,1,1,1}{\overline{28}}$ by conjugation of all IRREPs).}
	\begin{tabular}{|c|c|l|}
		\hline \hline
		\rotatebox[origin=c]{90}{$\;\;L_0\;\;$} & \rotatebox[origin=c]{90}{Order} & Irreps / Multiplicities	\\	\hline
		$0$	& $q^{3/4}$& \Yvcentermath1$\su{1}{1,1}{28}$				\\ \hline				
		$1$ & $q^{7/4}$& \Yvcentermath1$\su{1}{1,1}{28}\oplus\su{1}{2}{36}\oplus\su{1}{2,2,2,1,1,1,1}{420}$			\\ \hline	
		$2$	& $q^{11/4}$& \Yvcentermath1$\su{3}{1,1}{28}\oplus\su{1}{2}{36}\oplus\su{2}{2,2,2,1,1,1,1}{420}\oplus\su{1}{3,2,1,1,1,1,1}{1280}\oplus\su{1}{2,2,2,2,1,1}{1512}$ 			\\ \hline	
		\hline
	\end{tabular}
\end{table*}

\begin{table*}[!ht]
	\centering
	\label{tab:su8_56}
	\caption{SU$(8)_1$ WZW model -- Tower of states starting from $\protect\su{0}{1,1,1}{56}$ (resp. $\protect\su{0}{1,1,1,1,1}{\overline{56}}$ by conjugation of all IRREPs).}
	\begin{tabular}{|c|c|l|}
		\hline \hline
		\rotatebox[origin=c]{90}{$\;\;L_0\;\;$} & \rotatebox[origin=c]{90}{Order} & Irreps / Multiplicities	\\	\hline
		$0$	& $q^{15/16}$& \Yvcentermath1$\su{1}{1,1,1}{56}$				\\ \hline				
		$1$ & $q^{31/16}$& \Yvcentermath1$\su{1}{1,1,1}{56}\oplus\su{1}{2,1}{168}\oplus\su{1}{2,2,2,2,1,1,1}{504}$			\\ \hline	
		$2$	& $q^{47/16}$& \Yvcentermath1$\su{3}{1,1,1}{56}\oplus\su{2}{2,1}{168}\oplus\su{2}{2,2,2,2,1,1,1}{504}\oplus\su{1}{2,2,2,2,2,1}{1008}\oplus\su{1}{3,2,2,1,1,1,1}{2800}$ 			\\ \hline	
		\hline
	\end{tabular}
\end{table*}

\begin{table*}[!ht]
	\centering
	\label{tab:su8_70}
	\caption{SU$(8)_1$ WZW model -- Tower of states starting from $\protect\su{0}{1,1,1,1}{70}$.}
	\begin{tabular}{|c|c|l|}
		\hline \hline
		\rotatebox[origin=c]{90}{$\;\;L_0\;\;$} & \rotatebox[origin=c]{90}{Order} & Irreps / Multiplicities	\\	\hline
		$0$	& $q^{1}$& \Yvcentermath1$\su{1}{1,1,1,1}{70}$				\\ \hline				
		$1$ & $q^{2}$& \Yvcentermath1$\su{1}{1,1,1,1}{70}\oplus\su{1}{2,1,1}{378}\oplus\su{1}{2,2,2,2,2,1,1}{\overline{378}}$			\\ \hline	
		$2$	& $q^{3}$& \Yvcentermath1$\su{3}{1,1,1,1}{70}\oplus\su{1}{2,2,2,2,2,2}{\overline{336}}\oplus\su{1}{2,2}{336}\oplus\su{2}{2,1,1}{378}\oplus\su{2}{2,2,2,2,2,1,1}{\overline{378}}\oplus\su{1}{3,2,2,2,1,1,1}{3584}$ 			\\ \hline	
		\hline
	\end{tabular}
\end{table*}

\end{widetext}

\clearpage

\section{Notes on finite size effects in ED of periodic clusters}
\label{app:FSS}

\subsection{Antisymmetric vs completely symmetric IRREPS}

\begin{figure}
	\centering
\includegraphics[width=0.9\linewidth]{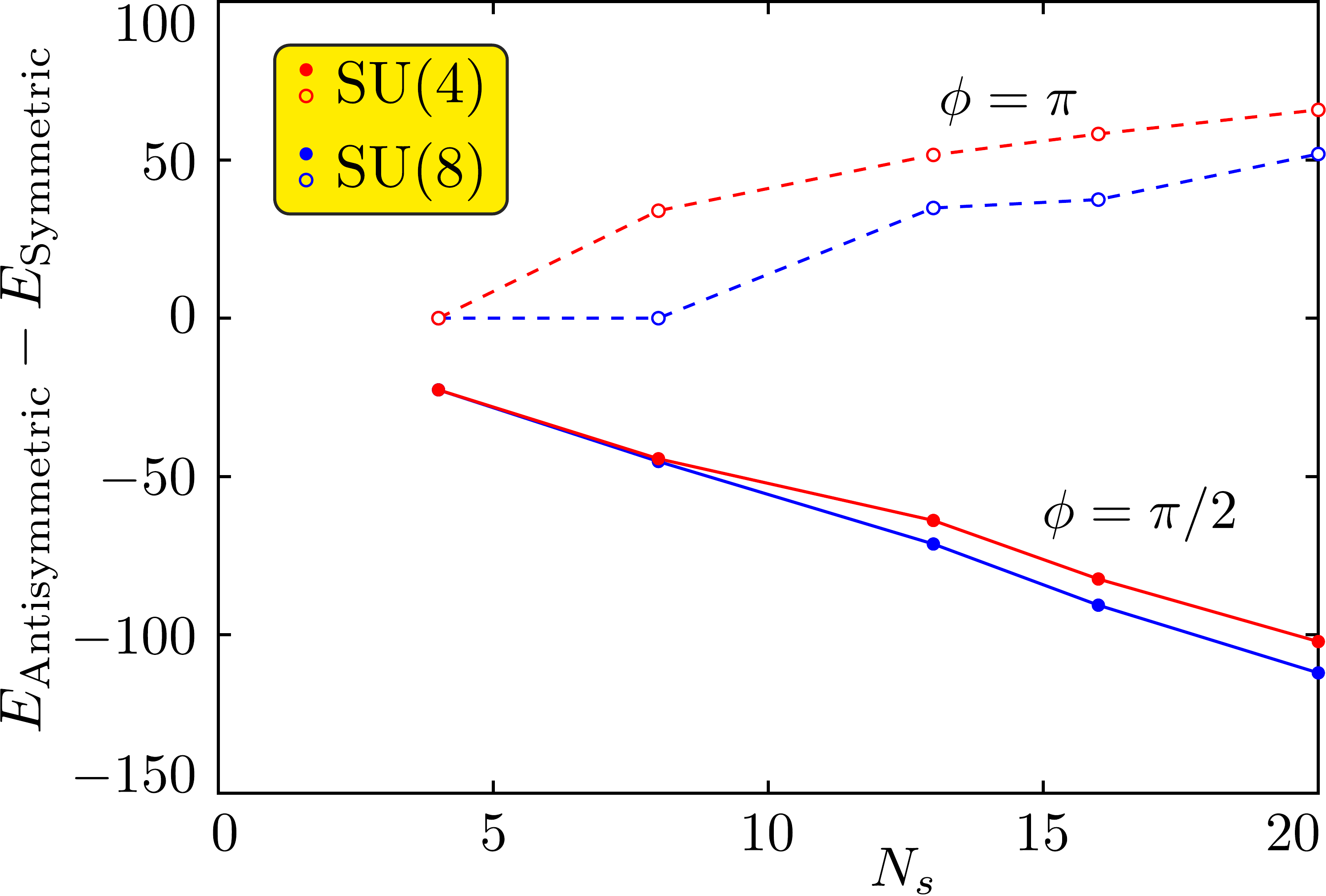}
\caption{Energy difference between the ground state of the antisymmetric IRREP $\text{aIR}_N(r_0)$ and the completely symmetric (ferromagnetic) state for $\theta=\pi/4$, $N=4$ (red) and $N=8$ (blue), $\phi=\pi/2$ (filled symbols) and $\phi=\pi$ (open symbols). In all cases, the energy difference scales approximately linearly with $N_s$, revealing a macroscopic energy difference. 	}
	\label{fig:ener_diff}
\end{figure}

In the range $\phi\in [ 0,\pi  ]$ both $J_1$ and $J_2$ couplings are antiferromagnetic but the amplitude $J_R$ of the (real) 3-site permutation changes sign, from positive to negative, at $\phi=\pi/2$.
Although a  negative $J_R$ equally favors both, the completely
symmetric multiplet (ferromagnetic) as well as the {\it completely } antisymmetric multiplet  on any triangle (see \App{app:plaquette}), on finite (periodic) clusters (with $N_s>N$), it strongly favors the ferromagnetic state with respect to the antisymmetric (antiferromagnetic) states of  $aIR_N(r_0)$.
In fact, a 3-site permutation on a triangle with $J_R<0$ cannot accommodate the complicated sign structure of antiferromagnetic states.
Note also that the energy difference is {\it macroscopic}, in the sense that it scales with the number of sites $N_s$. At $\phi=\pi/2$ where $J_R$ vanishes and the antiferromagnetic couplings $J_1$ and $J_2$ are finite, we observe the reverse, namely a macroscopic energy penalty for the ferromagnetic state with respect to the antiferromagnetic states. This is clearly evidenced in Fig.~\ref{fig:ener_diff}, showing the energy difference $E_a(N_s) - E_F(N_s)$ vs $N_s$, for $\theta=\pi/4$, and $N=4$ and $N=8$. Then, one can argue that a transition from a spin liquid phase (or several spin liquid phases) and the ferromagnetic phase should occur between $\phi=\pi/2$ and $\phi=\pi$.

\subsection{Finite size effects in low-energy spectra}

As seen in App. \App{app:plaquette}, for a given system size $N_s$ (multiple of $N$), the spectrum of the SU($N$) model includes all SU($N'$) spectra, $N'<N$. In the frustrated antiferromagnetic regime where a SU($N$) chiral spin liquid (or a singlet cluster state) is expected, SU($M$) singlets (forming a higher quadratic Casimir SU($N$) IRREP), $M< N$ also divider of $N_s$, may compete with the expected SU($N$) singlet GS of the SU($N$) model. We have observed this effect  in Fig.~\ref{fig:energies_vsPhi} for $N=8, 9, 10$ (with $N_s=16,18,20$ and $M=4,6,5$, respectively) for $\theta=\pi/4$ and small $\phi$. For instance, for $N_s=16$ and $N=8$, the high Casimir IRREP $[44440000]$ has energy given by Fig.~\ref{fig:energies_vsPhi} {\bf (c)} which is smaller at $\phi=0$ than the one of the SU($8$) singlet subspace in Fig.~\ref{fig:energies_vsPhi} {\bf (g)}.

\begin{figure}
	\centering
	\includegraphics[width=\linewidth]{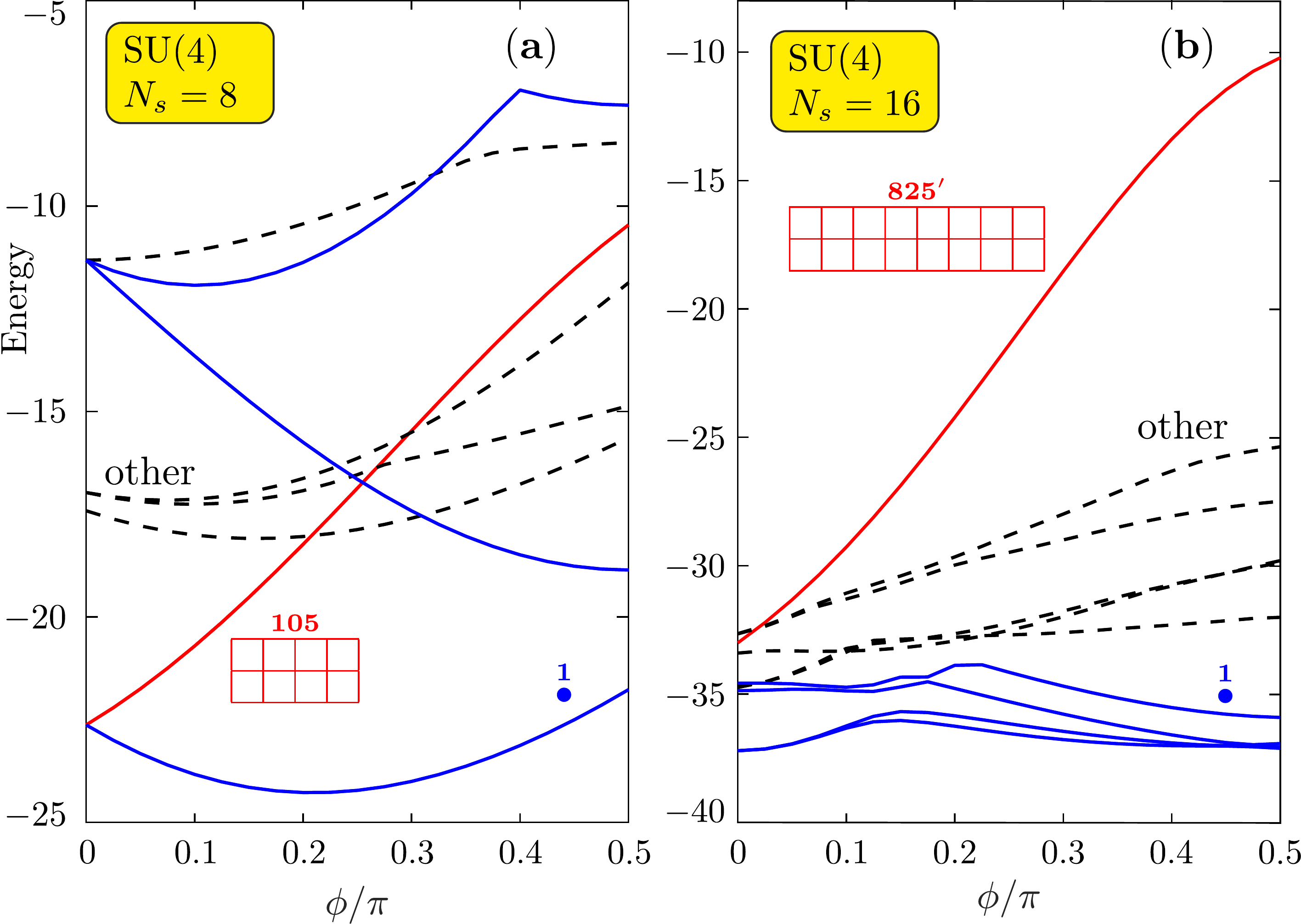}
	\caption{Low-energy spectra  of the SU($4$) model computed on $8$-site {\bf (a)} and $16$-site {\bf (b)} periodic clusters at $\theta=\pi/4$, plotted vs $\phi$.  A few of the  lowest energies of the  SU(4) singlet subspace are shown (in blue) on both panels as well as the lowest energy (in red) within the subspace of  the higher Casimir $[4400]$ (or {\bf 105}) {\bf (a)} and $[8800]$ (or {\bf 825'}) {\bf (b)} IRREPs, which can also be viewed as SU(2) singlets.	Other lowest-energy excitations are also shown for completeness. }
	\label{fig:fse}
\end{figure}

Here we argue that such a behavior is in fact a finite size effect occuring when $N_s<N^2$. To illustrate it we compare in Fig.~\ref{fig:fse} the low-energy  spectra of the $N=4$ model at $\theta=\pi/4$, versus $\phi$, on $8$-site and $16$-site clusters. For $N_s=8$, we observe that the lowest energies of the SU(4) singlets and those of  the higher Casimir IRREP $[4400]$ (also SU($2$) singlets)  are comparable. In contrast, for $N_s=16$, a clear energy separation is seen between the lowest energy states of the higher Casimir $[8800]$ IRREP (also SU($2$) singlets) and the lowest SU(4) singlets.

\section{Details on MPO--MPS implementation}
\begin{figure}[h]
	\centering
	\includegraphics[width=0.75\linewidth]{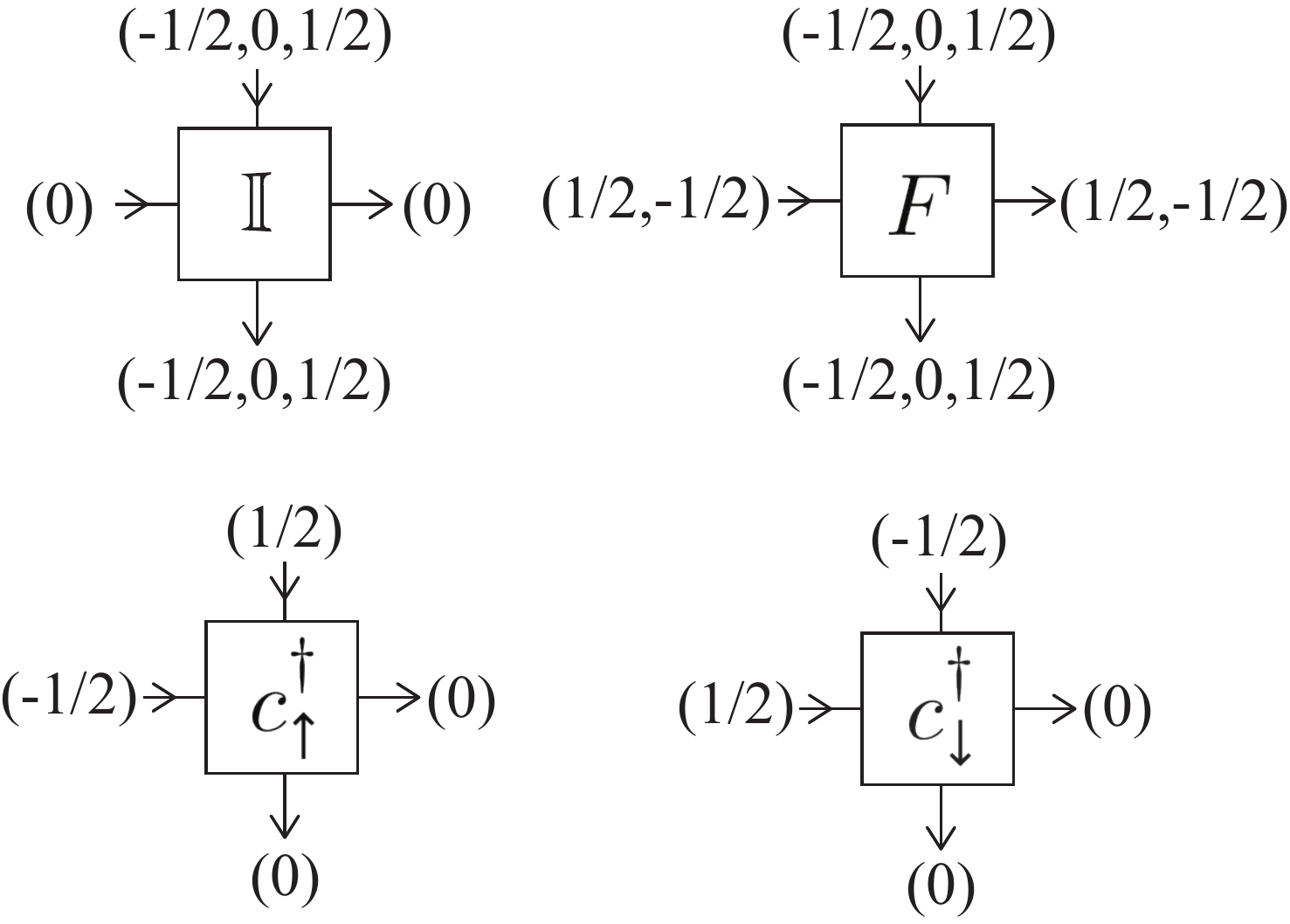}
	\caption{Graphical representation of MPO matrix elements with U$(1)$ spin symmetry for spin-$1/2$ fermions.
		Numbers in brackets indicate the possible values of $S_z$ quantum numbers, $0,-1/2$ and $1/2$ representing the $\ket{}$, $\ket{\downarrow}$ and $\ket{\uparrow}$ at each physical site, respectively. Double occupancy, $\ket{\uparrow\downarrow}$, is excluded.}
	\label{fig:U1spin}
\end{figure}

\begin{figure}[h]
	\centering
	\includegraphics[width=0.9\linewidth]{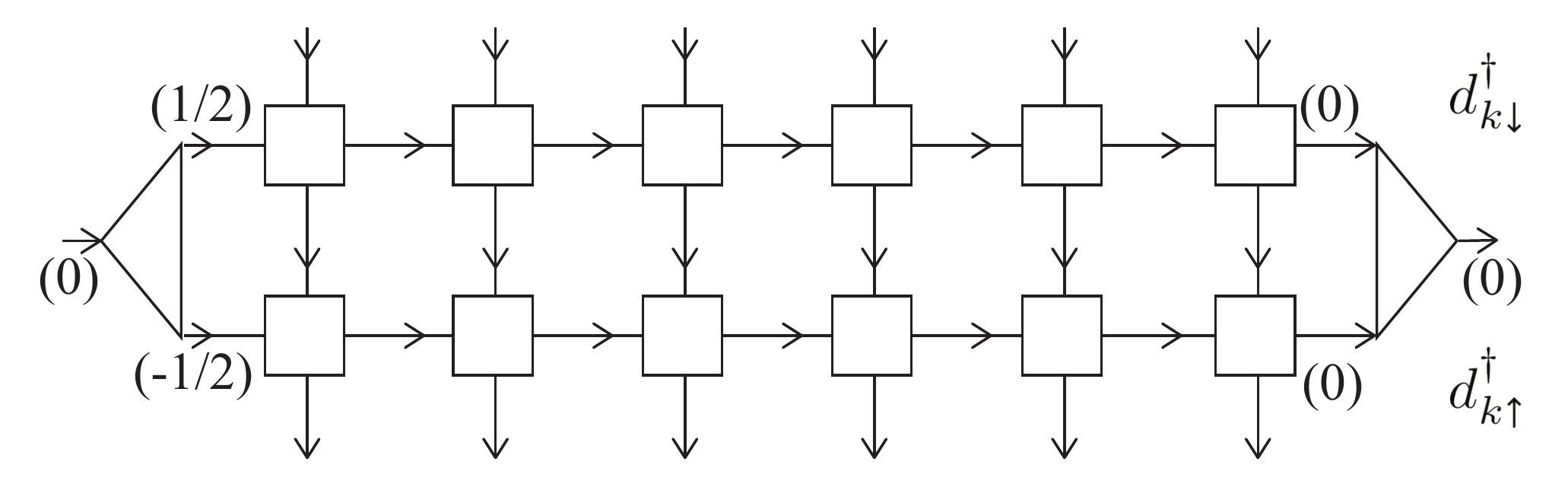}
	\caption{Graphical representation of fusing edge virtual indices of $2$ MPOs, $d^{\dagger}_{k\uparrow}$ and $d^{\dagger}_{k\downarrow}$.}
	\label{fig:U1mpo}
\end{figure}

This section describes how to cast a Slater determinant, $\ket{\Psi}=\prod_{k,\sigma}d_{k\sigma}^{\dagger}\ket{0}$, into an MPS with conserved spin symmetry.
We elaborate our implementation for $N=2$; the generalization to larger $N$ is straightforward.
For spin-$1/2$ fermions, the standard approach to express a single-particle operator $d_{k\sigma}^{\dagger}$ is to map the $L$-site spinful fermions onto a $2L$-site pseudospin-$1/2$ chain using the Jordan--Wigner transformation \cite{WYH2020,JHK2020a,Petrica2020}, namely,
\begin{equation}
	\begin{split}
		c_{\ell,\uparrow}^{\dagger} &\rightarrow \sigma^{z}_{1}\cdots\sigma^{z}_{2\ell-2}\sigma^{+}_{2\ell-1} \\
		c_{\ell,\downarrow}^{\dagger} &\rightarrow \sigma^{z}_{1}\cdots\sigma^{z}_{2\ell-2}\sigma^{z}_{2\ell-1}\sigma^{+}_{2\ell}.
	\end{split}
	\label{eq:JW}
\end{equation}
\begin{widetext}
And, $d_{k\sigma}^{\dagger} = \sum_{m,n} A_{m,n}(k) c_{m,n,\sigma}^{\dagger} = \sum_{\ell} \tilde{A}_{k\sigma,\ell}c_{\ell\sigma}^{\dagger}$  can be read as an MPO acting on the spin-$1/2$ chain
\begin{equation}
	d^{\dag}_{k\sigma} =
	\begin{pmatrix} 0 & 1 \end{pmatrix}
	\left[\prod^{2L}_{\ell=1} \begin{pmatrix}
		\mathbbm{1}_{\ell} & 0 \\
		\tilde{A}_{k\sigma,\ell} \sigma^{+}_{\ell} & \sigma^{z}_{\ell}
	\end{pmatrix}\right]
	\begin{pmatrix} 1\\0 \end{pmatrix}.
\end{equation}

For our purpose, we would like to block $2\ell-1$ and $2\ell$ sites together, which leads to
\begin{equation}
	d^{\dag}_{k\sigma} =
	\begin{pmatrix} 0 & 1 \end{pmatrix}
	\left[\prod^{L}_{j=1} \begin{pmatrix}
		\mathbbm{1}_{2j-1}\otimes\mathbbm{1}_{2j} & 0 \\
		\tilde{A}_{k\sigma,2j-1}\sigma^{+}_{2j-1}\otimes\mathbbm{1}_{2j}+\tilde{A}_{k\sigma,2j}\sigma^{z}_{2j-1}\otimes\sigma^{+}_{2j} & \sigma^{z}_{2j-1}\otimes\sigma^{z}_{2j}
	\end{pmatrix}\right]
	\begin{pmatrix} 1\\0 \end{pmatrix}.
\end{equation}
\end{widetext}
We can identify $\sigma^{+}_{2j-1}\otimes\mathbbm{1}_{2j}$ with $c^{\dagger}_{j,\uparrow}$, $\sigma^{z}_{2j-1}\otimes\sigma^{+}_{2j}$ with $c^{\dagger}_{j,\downarrow}$, and $F_j=\sigma^{z}_{2j-1}\otimes\sigma^{z}_{2j}$ with the parity operator to account for anticommutation of different sites.
In fact, we can always write the MPO in this spinful fermion basis, regardless of the number of fermion species, i.e.,
\begin{equation}
	d^{\dag}_{k\sigma} =
	\begin{pmatrix} 0 & 1 \end{pmatrix}
	\left[\prod^{L}_{j=1} \begin{pmatrix}
		\mathbb{I} & 0 \\
		\tilde{A}_{k\sigma,j}c^{\dagger}_{\sigma} & F
	\end{pmatrix}\right]
	\begin{pmatrix} 1\\0 \end{pmatrix}.
\end{equation}
This facilitates working with U$(1)$ or SU$(N)$ spin symmetry as each tensor index can be associated with a specific quantum number (see Fig.~\ref{fig:U1spin}).
With U$(1)$ spin symmetry, one can fuse the virtual indices at boundaries of each pair of MPOs to be $S_z=0$ (see Fig.~\ref{fig:U1mpo}), the resulting MPS $\ket{\Psi}$ also has $S_z=0$.
In the same way, one can easily impose SU$(2)$ spin symmetry to target spin-singlet states, provided an efficient tensor network implementation to handle Clebsch-Gordan coefficients \cite{WA2012,SS2012,HC2018,SP2020}.
We use QSpace for this purpose \cite{WA2012,WA2020}.

\begin{figure}[h]
	\includegraphics[width=0.96\linewidth]{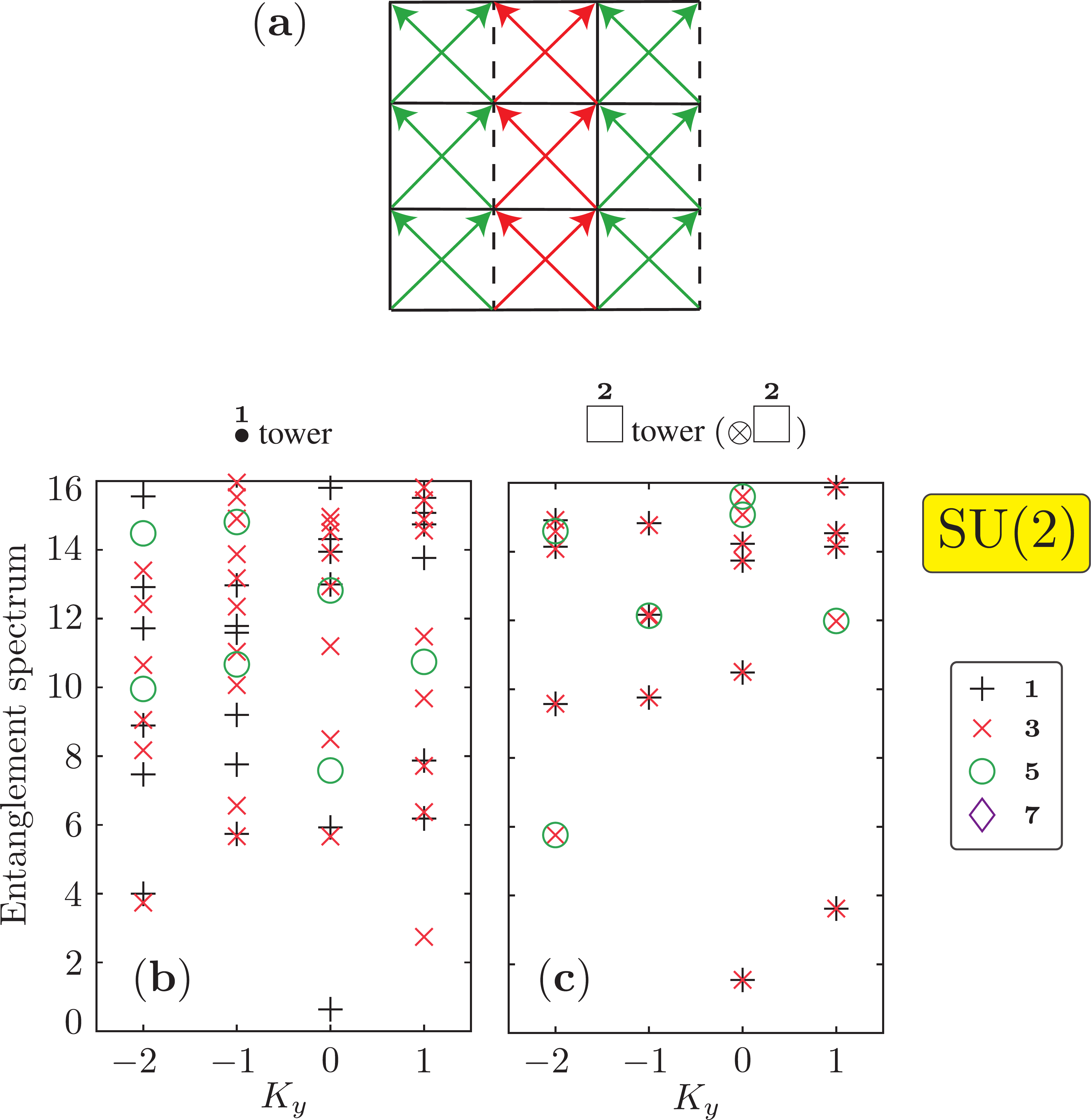}
	\caption{{\bf (a)} Illustration of the parton Hamiltonian of the SU($2$) CSL. The phase of nearest neighbor hopping is $0$ $(\pi)$ along the solid (dashed) edges.
		The phase of next-nearest-neighbor hopping is $\pi/2$ $(-\pi/2)$ along the green (red) arrows.
		{\bf (b,c)} The entanglement spectra on a $4\times12$ cylinder for the parton wave function.}
	\label{fig:part2}
\end{figure}

In Fig.~\ref{fig:part2} {\bf (b,c)}, we plot the ESs obtained from the parton construction on a $4\times12$ cylinder.
This demonstrates the efficacy of our parton approach, as we are able to prepare trial states in distinct topological sectors for iDMRG using a relatively small size cylinder.
Additionally, imposing SU$(2)$ symmetry constraint leads to an intriguing consequence: if the state is in the topologically nontrivial sector, there are multiple degenerate branches in the ES (see Fig.~\ref{fig:part2}{\bf (c)}).
This has also been observed in the SU$(2)$ iPEPS simulations previously \cite{Poilblanc2015,Poilblanc2016,Hackenbroich2018}, and was attributed to the so-called ``dressed mirror symmetry" within the virtual degrees of freedom \cite{Hackenbroich2018}.
The parton approach offers a more direct understanding --- the degeneracy equals to the number of parton states required to form a singlet superposition state.

\clearpage
\section{Modified WZW  \SUNone chiral towers of states}
\label{app:WZW2}

We list here, for $N=3$ and $4$,  the predicted ToS corresponding to the \SUN DMRG cylinders investigated and discussed in the main text.

\begin{table*}[!h]
	\centering
	\caption{SU$(3)_1$ WZW model -- The direct product of the conformal tower of the $\protect\su{0}{1}{3}$ primary (left - see Table~\protect\ref{tab:su3_3} in  ~\App{app:WZW}) with   $\protect\su{0}{1,1}{\bar{3}}$    gives a new tower (right) with a tripling of the number of states in each Virasoro level indexed by $L_0$.}
	{\scriptsize 	\begin{tabular}{|c|l|c|l|}
			\cline{1-2}\cline{4-4}
			$L_0$ & \multicolumn{1}{c|}{$\protect\su{0}{1}{3}$ tower} && \multicolumn{1}{c|}{$\protect\su{0}{1}{3}\text{ tower}\otimes \protect\su{0}{1,1}{\bar{3}}$} \\ \cline{1-2}\cline{4-4}
			$0$     & \Yvcentermath1$\su{1}{1}{3}$ && \Yvcentermath1$\su{1}{0}{1}\oplus\su{1}{2,1}{8}$ \\  \cline{1-2}\cline{4-4}
			$1$     & \Yvcentermath1$\su{1}{1}{3}\oplus\su{1}{2,2}{\bar{6}}$ &\multirow{3}{*}{$\rightarrow$} & \Yvcentermath1$\su{1}{0}{1}\oplus\su{2}{2,1}{8}\oplus\su{1}{3,3}{\overline{10}}$\\ \cline{1-2}\cline{4-4}
			$2$     & \Yvcentermath1$\su{2}{1}{3}\oplus\su{1}{2,2}{\bar{6}}\oplus\su{1}{3,1}{15}$ && \Yvcentermath1$\su{2}{0}{1}\oplus\su{4}{2,1}{8}\oplus\su{1}{3}{10}\oplus\su{1}{3,3}{\overline{10}}\oplus\su{1}{4,2}{27}$ \\ \cline{1-2}\cline{4-4}
			$3$     & \Yvcentermath1$\su{3}{1}{3}\oplus\su{3}{2,2}{\bar{6}}\oplus\su{2}{3,1}{15}$ && \Yvcentermath1$\su{3}{0}{1}\oplus\su{8}{2,1}{8}\oplus\su{2}{3}{10}\oplus\su{3}{3,3}{\overline{10}}\oplus\su{2}{4,2}{27}$\\ \cline{1-2}\cline{4-4}
	\end{tabular} }
	\label{tab:su3_copies}
\end{table*}

\begin{table*}[!h]
	\centering
	\caption{SU$(4)_1$ WZW model -- The direct product of the conformal tower of the $\protect\su{0}{1}{4}$ primary (left - see Table~\protect\ref{tab:su4_4} in  ~\App{app:WZW}) with   $\protect\su{0}{1,1,1}{\bar{4}}$    gives a new tower (right) with a quadrupling of the number of states in each Virasoro level indexed by $L_0$.}
	{\scriptsize	\begin{tabular}{|c|l|c|l|}
			\cline{1-2}\cline{4-4}
			$L_0$ & \multicolumn{1}{c|}{$\protect\su{0}{1}{4}$ tower} && \multicolumn{1}{c|}{$\protect\su{0}{1}{4}\text{ tower}\otimes \protect\su{0}{1,1,1}{\bar{4}}$} \\ \cline{1-2}\cline{4-4}
			$0$     & \Yvcentermath1$\su{1}{1}{4}$ && \Yvcentermath1$\su{1}{0}{1}\oplus\su{1}{2,1,1}{15}$ \\  \cline{1-2}\cline{4-4}
			$1$     & \Yvcentermath1$\su{1}{1}{4}\oplus\su{1}{2,2,1}{20}$ &\multirow{3}{*}{$\rightarrow$} & \Yvcentermath1$\su{1}{0}{1}\oplus\su{2}{2,1,1}{15}\oplus\su{1}{2,2}{20'}\oplus\su{1}{3,3,2}{\overline{45}}$\\ \cline{1-2}\cline{4-4}
			$2$     & \Yvcentermath1$\su{2}{1}{4}\oplus\su{2}{2,2,1}{20}\oplus\su{1}{3,1,1}{36}$ && \Yvcentermath1$\su{2}{0}{1}\oplus\su{5}{2,1,1}{15} \oplus\su{2}{2,2}{20'}\oplus\su{1}{3,1}{45}\oplus\su{2}{3,3,2}{\overline{45}}\oplus\su{1}{4,2,2}{84}$ \\ \cline{1-2}\cline{4-4}
			$3$    & \Yvcentermath1$\su{4}{1}{4}\oplus\su{1}{3,3,3}{20''}\oplus\su{4}{2,2,1}{20}\oplus\su{2}{3,1,1}{36}$&& \Yvcentermath1$\su{4}{0}{1}\oplus\su{10}{2,1,1}{15}\oplus\su{5}{2,2}{20'}\oplus\su{1}{4,4,4}{\overline{35}}\oplus\su{3}{3,1}{45}\oplus\su{5}{3,3,2}{\overline{45}}$ \\
			& \Yvcentermath1$\oplus\su{1}{3,2}{60}$ && \Yvcentermath1$\oplus\su{2}{4,2,2}{84}\oplus\su{1}{4,3,1}{175}$ \\ \cline{1-2}\cline{4-4}
	\end{tabular}}
	\label{tab:su4_copies1}
\end{table*}

\begin{table*}[!h]
	\centering
	\caption{SU$(4)_1$ WZW model -- The direct product of the conformal tower of the $\protect\su{0}{1,1}{6}$ primary (left - see Table~\protect\ref{tab:su4_6} in  ~\App{app:WZW}) with   $\protect\su{0}{1,1}{6}$    gives a new tower (right) with a multiplicative factor 6  of the number of states in each Virasoro level indexed by $m$.}
	{\scriptsize	\begin{tabular}{|c|l|c|l|}
			\cline{1-2}\cline{4-4}
			$L_0$ & \multicolumn{1}{c|}{$\protect\su{0}{1,1}{6}$ tower} && \multicolumn{1}{c|}{$\protect\su{0}{1,1}{6}\text{ tower}\otimes \protect\su{0}{1,1}{6}$} \\ \cline{1-2}\cline{4-4}
			$0$     & \Yvcentermath1$\su{1}{1,1}{6}$ && \Yvcentermath1$\su{1}{0}{1}\oplus\su{1}{2,1,1}{15}\oplus\su{1}{2,2}{20'}$ \\  \cline{1-2}\cline{4-4}
			$1$      & \Yvcentermath1$\su{1}{1,1}{6}\oplus\su{1}{2}{10}\oplus\su{1}{2,2,2}{\overline{10}}$ &\multirow{3}{*}{$\rightarrow$} & \Yvcentermath1$\su{1}{0}{1}\oplus\su{3}{2,1,1}{15}\oplus\su{1}{2,2}{20'}\oplus\su{1}{3,1}{45}\oplus\su{1}{3,3,2}{\overline{45}}$ \\  \cline{1-2}\cline{4-4}
			$2$     & \Yvcentermath1$\su{3}{1,1}{6}\oplus\su{1}{2}{10}\oplus\su{1}{2,2,2}{\overline{10}}\oplus\su{1}{3,2,1}{64}$ && \Yvcentermath1$\su{3}{0}{1}\oplus\su{6}{2,1,1}{15}\oplus\su{4}{2,2}{20'}\oplus\su{2}{3,1}{45}\oplus\su{2}{3,3,2}{\overline{45}}\oplus\su{1}{4,2,2}{84}\oplus\su{1}{4,3,1}{175}$ \\  \cline{1-2}\cline{4-4}
			$3$     & \Yvcentermath1$\su{4}{1,1}{6}\oplus\su{3}{2}{10}\oplus\su{3}{2,2,2}{\overline{10}}\oplus\su{3}{3,2,1}{64}$ && \Yvcentermath1$\su{4}{0}{1}\oplus\su{13}{2,1,1}{15}\oplus\su{7}{2,2}{20'}\oplus\su{6}{3,1}{45}\oplus\su{6}{3,3,2}{\overline{45}}\oplus\su{3}{4,2,2}{84}\oplus\su{3}{4,3,1}{175}$ \\ \cline{1-2}\cline{4-4}
	\end{tabular}}
	\label{tab:su4_copies2}
\end{table*}

\bibliography{bibliography}

\begin{thebibliography}{108}%
\makeatletter
\providecommand \@ifxundefined [1]{%
 \@ifx{#1\undefined}
}%
\providecommand \@ifnum [1]{%
 \ifnum #1\expandafter \@firstoftwo
 \else \expandafter \@secondoftwo
 \fi
}%
\providecommand \@ifx [1]{%
 \ifx #1\expandafter \@firstoftwo
 \else \expandafter \@secondoftwo
 \fi
}%
\providecommand \natexlab [1]{#1}%
\providecommand \enquote  [1]{``#1''}%
\providecommand \bibnamefont  [1]{#1}%
\providecommand \bibfnamefont [1]{#1}%
\providecommand \citenamefont [1]{#1}%
\providecommand \href@noop [0]{\@secondoftwo}%
\providecommand \href [0]{\begingroup \@sanitize@url \@href}%
\providecommand \@href[1]{\@@startlink{#1}\@@href}%
\providecommand \@@href[1]{\endgroup#1\@@endlink}%
\providecommand \@sanitize@url [0]{\catcode `\\12\catcode `\$12\catcode
  `\&12\catcode `\#12\catcode `\^12\catcode `\_12\catcode `\%12\relax}%
\providecommand \@@startlink[1]{}%
\providecommand \@@endlink[0]{}%
\providecommand \url  [0]{\begingroup\@sanitize@url \@url }%
\providecommand \@url [1]{\endgroup\@href {#1}{\urlprefix }}%
\providecommand \urlprefix  [0]{URL }%
\providecommand \Eprint [0]{\href }%
\providecommand \doibase [0]{http://dx.doi.org/}%
\providecommand \selectlanguage [0]{\@gobble}%
\providecommand \bibinfo  [0]{\@secondoftwo}%
\providecommand \bibfield  [0]{\@secondoftwo}%
\providecommand \translation [1]{[#1]}%
\providecommand \BibitemOpen [0]{}%
\providecommand \bibitemStop [0]{}%
\providecommand \bibitemNoStop [0]{.\EOS\space}%
\providecommand \EOS [0]{\spacefactor3000\relax}%
\providecommand \BibitemShut  [1]{\csname bibitem#1\endcsname}%
\let\auto@bib@innerbib\@empty
\bibitem [{\citenamefont {Misguich}\ and\ \citenamefont
  {Lhuillier}(2005)}]{Misguich2005}%
  \BibitemOpen
  \bibfield  {author} {\bibinfo {author} {\bibfnamefont {Gr\'egoire}\
  \bibnamefont {Misguich}}\ and\ \bibinfo {author} {\bibfnamefont {Claire}\
  \bibnamefont {Lhuillier}},\ }\enquote {\bibinfo {title} {Two-dimensional
  quantum antiferromagnets},}\ in\ \href {\doibase 10.1142/9789812567819}
  {\emph {\bibinfo {booktitle} {Frustrated Spin Systems}}},\ \bibinfo {editor}
  {edited by\ \bibinfo {editor} {\bibfnamefont {H.~T.}\ \bibnamefont {Diep}}}\
  (\bibinfo  {publisher} {World Scientific},\ \bibinfo {year} {2005})\ pp.\
  \bibinfo {pages} {229--306}\BibitemShut {NoStop}%
\bibitem [{\citenamefont {Savary}\ and\ \citenamefont
  {Balents}(2016)}]{Savary2016}%
  \BibitemOpen
  \bibfield  {author} {\bibinfo {author} {\bibfnamefont {Lucile}\ \bibnamefont
  {Savary}}\ and\ \bibinfo {author} {\bibfnamefont {Leon}\ \bibnamefont
  {Balents}},\ }\bibfield  {title} {\enquote {\bibinfo {title} {Quantum spin
  liquids: a review},}\ }\href {\doibase 10.1088/0034-4885/80/1/016502}
  {\bibfield  {journal} {\bibinfo  {journal} {Reports on Progress in Physics}\
  }\textbf {\bibinfo {volume} {80}},\ \bibinfo {pages} {016502} (\bibinfo
  {year} {2016})}\BibitemShut {NoStop}%
\bibitem [{\citenamefont {Zhou}\ \emph {et~al.}(2017)\citenamefont {Zhou},
  \citenamefont {Kanoda},\ and\ \citenamefont {Ng}}]{Zhou2017}%
  \BibitemOpen
  \bibfield  {author} {\bibinfo {author} {\bibfnamefont {Yi}~\bibnamefont
  {Zhou}}, \bibinfo {author} {\bibfnamefont {Kazushi}\ \bibnamefont {Kanoda}},
  \ and\ \bibinfo {author} {\bibfnamefont {Tai-Kai}\ \bibnamefont {Ng}},\
  }\bibfield  {title} {\enquote {\bibinfo {title} {Quantum spin liquid
  states},}\ }\href {\doibase 10.1103/RevModPhys.89.025003} {\bibfield
  {journal} {\bibinfo  {journal} {Rev. Mod. Phys.}\ }\textbf {\bibinfo {volume}
  {89}},\ \bibinfo {pages} {025003} (\bibinfo {year} {2017})}\BibitemShut
  {NoStop}%
\bibitem [{\citenamefont {Wen}(1990)}]{Wen1990}%
  \BibitemOpen
  \bibfield  {author} {\bibinfo {author} {\bibfnamefont {X.~G.}\ \bibnamefont
  {Wen}},\ }\bibfield  {title} {\enquote {\bibinfo {title} {Topological orders
  in rigid states},}\ }\href {\doibase 10.1142/S0217979290000139} {\bibfield
  {journal} {\bibinfo  {journal} {International Journal of Modern Physics B}\
  }\textbf {\bibinfo {volume} {04}},\ \bibinfo {pages} {239--271} (\bibinfo
  {year} {1990})}\BibitemShut {NoStop}%
\bibitem [{\citenamefont {Poilblanc}\ \emph {et~al.}(2012)\citenamefont
  {Poilblanc}, \citenamefont {Schuch}, \citenamefont {P\'erez-Garc\'{\i}a},\
  and\ \citenamefont {Cirac}}]{Poilblanc2012}%
  \BibitemOpen
  \bibfield  {author} {\bibinfo {author} {\bibfnamefont {Didier}\ \bibnamefont
  {Poilblanc}}, \bibinfo {author} {\bibfnamefont {Norbert}\ \bibnamefont
  {Schuch}}, \bibinfo {author} {\bibfnamefont {David}\ \bibnamefont
  {P\'erez-Garc\'{\i}a}}, \ and\ \bibinfo {author} {\bibfnamefont {J.~Ignacio}\
  \bibnamefont {Cirac}},\ }\bibfield  {title} {\enquote {\bibinfo {title}
  {Topological and entanglement properties of resonating valence bond wave
  functions},}\ }\href {\doibase 10.1103/PhysRevB.86.014404} {\bibfield
  {journal} {\bibinfo  {journal} {Phys. Rev. B}\ }\textbf {\bibinfo {volume}
  {86}},\ \bibinfo {pages} {014404} (\bibinfo {year} {2012})}\BibitemShut
  {NoStop}%
\bibitem [{\citenamefont {Kalmeyer}\ and\ \citenamefont
  {Laughlin}(1987)}]{Kalmeyer1987}%
  \BibitemOpen
  \bibfield  {author} {\bibinfo {author} {\bibfnamefont {V.}~\bibnamefont
  {Kalmeyer}}\ and\ \bibinfo {author} {\bibfnamefont {R.~B.}\ \bibnamefont
  {Laughlin}},\ }\bibfield  {title} {\enquote {\bibinfo {title} {Equivalence of
  the resonating-valence-bond and fractional quantum {H}all states},}\ }\href
  {\doibase 10.1103/PhysRevLett.59.2095} {\bibfield  {journal} {\bibinfo
  {journal} {Phys. Rev. Lett.}\ }\textbf {\bibinfo {volume} {59}},\ \bibinfo
  {pages} {2095--2098} (\bibinfo {year} {1987})}\BibitemShut {NoStop}%
\bibitem [{\citenamefont {Kalmeyer}\ and\ \citenamefont
  {Laughlin}(1989)}]{Kalmeyer1989}%
  \BibitemOpen
  \bibfield  {author} {\bibinfo {author} {\bibfnamefont {Vadim}\ \bibnamefont
  {Kalmeyer}}\ and\ \bibinfo {author} {\bibfnamefont {R.~B.}\ \bibnamefont
  {Laughlin}},\ }\bibfield  {title} {\enquote {\bibinfo {title} {Theory of the
  spin liquid state of the heisenberg antiferromagnet},}\ }\href {\doibase
  10.1103/PhysRevB.39.11879} {\bibfield  {journal} {\bibinfo  {journal} {Phys.
  Rev. B}\ }\textbf {\bibinfo {volume} {39}},\ \bibinfo {pages} {11879--11899}
  (\bibinfo {year} {1989})}\BibitemShut {NoStop}%
\bibitem [{\citenamefont {Laughlin}(1989)}]{laughlin1989}%
  \BibitemOpen
  \bibfield  {author} {\bibinfo {author} {\bibfnamefont {R.~B.}\ \bibnamefont
  {Laughlin}},\ }\bibfield  {title} {\enquote {\bibinfo {title} {Spin
  hamiltonian for which quantum hall wavefunction is exact},}\ }\href {\doibase
  10.1016/0003-4916(89)90339-4} {\bibfield  {journal} {\bibinfo  {journal}
  {Annals of Physics}\ }\textbf {\bibinfo {volume} {191}},\ \bibinfo {pages}
  {163--202} (\bibinfo {year} {1989})}\BibitemShut {NoStop}%
\bibitem [{\citenamefont {Wen}\ \emph {et~al.}(1989)\citenamefont {Wen},
  \citenamefont {Wilczek},\ and\ \citenamefont {Zee}}]{WWZ1989}%
  \BibitemOpen
  \bibfield  {author} {\bibinfo {author} {\bibfnamefont {X.~G.}\ \bibnamefont
  {Wen}}, \bibinfo {author} {\bibfnamefont {Frank}\ \bibnamefont {Wilczek}}, \
  and\ \bibinfo {author} {\bibfnamefont {A.}~\bibnamefont {Zee}},\ }\bibfield
  {title} {\enquote {\bibinfo {title} {Chiral spin states and
  superconductivity},}\ }\href {\doibase 10.1103/PhysRevB.39.11413} {\bibfield
  {journal} {\bibinfo  {journal} {Phys. Rev. B}\ }\textbf {\bibinfo {volume}
  {39}},\ \bibinfo {pages} {11413--11423} (\bibinfo {year} {1989})}\BibitemShut
  {NoStop}%
\bibitem [{\citenamefont {Laughlin}\ and\ \citenamefont
  {Zou}(1990)}]{Laughlin1990}%
  \BibitemOpen
  \bibfield  {author} {\bibinfo {author} {\bibfnamefont {R.~B.}\ \bibnamefont
  {Laughlin}}\ and\ \bibinfo {author} {\bibfnamefont {Z.}~\bibnamefont {Zou}},\
  }\bibfield  {title} {\enquote {\bibinfo {title} {Properties of the
  chiral-spin-liquid state},}\ }\href {\doibase 10.1103/PhysRevB.41.664}
  {\bibfield  {journal} {\bibinfo  {journal} {Phys. Rev. B}\ }\textbf {\bibinfo
  {volume} {41}},\ \bibinfo {pages} {664--687} (\bibinfo {year}
  {1990})}\BibitemShut {NoStop}%
\bibitem [{\citenamefont {Wen}(2002)}]{Wen2002}%
  \BibitemOpen
  \bibfield  {author} {\bibinfo {author} {\bibfnamefont {Xiao-Gang}\
  \bibnamefont {Wen}},\ }\bibfield  {title} {\enquote {\bibinfo {title}
  {Quantum orders and symmetric spin liquids},}\ }\href {\doibase
  10.1103/PhysRevB.65.165113} {\bibfield  {journal} {\bibinfo  {journal} {Phys.
  Rev. B}\ }\textbf {\bibinfo {volume} {65}},\ \bibinfo {pages} {165113}
  (\bibinfo {year} {2002})}\BibitemShut {NoStop}%
\bibitem [{\citenamefont {Tsui}\ \emph {et~al.}(1982)\citenamefont {Tsui},
  \citenamefont {Stormer},\ and\ \citenamefont {Gossard}}]{Tsui1982}%
  \BibitemOpen
  \bibfield  {author} {\bibinfo {author} {\bibfnamefont {D.~C.}\ \bibnamefont
  {Tsui}}, \bibinfo {author} {\bibfnamefont {H.~L.}\ \bibnamefont {Stormer}}, \
  and\ \bibinfo {author} {\bibfnamefont {A.~C.}\ \bibnamefont {Gossard}},\
  }\bibfield  {title} {\enquote {\bibinfo {title} {Two-dimensional
  magnetotransport in the extreme quantum limit},}\ }\href {\doibase
  10.1103/PhysRevLett.48.1559} {\bibfield  {journal} {\bibinfo  {journal}
  {Phys. Rev. Lett.}\ }\textbf {\bibinfo {volume} {48}},\ \bibinfo {pages}
  {1559--1562} (\bibinfo {year} {1982})}\BibitemShut {NoStop}%
\bibitem [{\citenamefont {Halperin}(1984)}]{Halperin1984}%
  \BibitemOpen
  \bibfield  {author} {\bibinfo {author} {\bibfnamefont {B.~I.}\ \bibnamefont
  {Halperin}},\ }\bibfield  {title} {\enquote {\bibinfo {title} {Statistics of
  quasiparticles and the hierarchy of fractional quantized {H}all states},}\
  }\href {\doibase 10.1103/PhysRevLett.52.1583} {\bibfield  {journal} {\bibinfo
   {journal} {Phys. Rev. Lett.}\ }\textbf {\bibinfo {volume} {52}},\ \bibinfo
  {pages} {1583--1586} (\bibinfo {year} {1984})}\BibitemShut {NoStop}%
\bibitem [{\citenamefont {Wen}(1991{\natexlab{a}})}]{Wen1991a}%
  \BibitemOpen
  \bibfield  {author} {\bibinfo {author} {\bibfnamefont {X.~G.}\ \bibnamefont
  {Wen}},\ }\bibfield  {title} {\enquote {\bibinfo {title} {Gapless boundary
  excitations in the quantum {H}all states and in the chiral spin states},}\
  }\href {\doibase 10.1103/PhysRevB.43.11025} {\bibfield  {journal} {\bibinfo
  {journal} {Phys. Rev. B}\ }\textbf {\bibinfo {volume} {43}},\ \bibinfo
  {pages} {11025--11036} (\bibinfo {year} {1991}{\natexlab{a}})}\BibitemShut
  {NoStop}%
\bibitem [{\citenamefont {Schroeter}\ \emph {et~al.}(2007)\citenamefont
  {Schroeter}, \citenamefont {Kapit}, \citenamefont {Thomale},\ and\
  \citenamefont {Greiter}}]{Schroeter2007}%
  \BibitemOpen
  \bibfield  {author} {\bibinfo {author} {\bibfnamefont {Darrell~F.}\
  \bibnamefont {Schroeter}}, \bibinfo {author} {\bibfnamefont {Eliot}\
  \bibnamefont {Kapit}}, \bibinfo {author} {\bibfnamefont {Ronny}\ \bibnamefont
  {Thomale}}, \ and\ \bibinfo {author} {\bibfnamefont {Martin}\ \bibnamefont
  {Greiter}},\ }\bibfield  {title} {\enquote {\bibinfo {title} {Spin
  hamiltonian for which the chiral spin liquid is the exact ground state},}\
  }\href {\doibase 10.1103/PhysRevLett.99.097202} {\bibfield  {journal}
  {\bibinfo  {journal} {Phys. Rev. Lett.}\ }\textbf {\bibinfo {volume} {99}},\
  \bibinfo {pages} {097202} (\bibinfo {year} {2007})}\BibitemShut {NoStop}%
\bibitem [{\citenamefont {Thomale}\ \emph {et~al.}(2009)\citenamefont
  {Thomale}, \citenamefont {Kapit}, \citenamefont {Schroeter},\ and\
  \citenamefont {Greiter}}]{Thomale2009}%
  \BibitemOpen
  \bibfield  {author} {\bibinfo {author} {\bibfnamefont {Ronny}\ \bibnamefont
  {Thomale}}, \bibinfo {author} {\bibfnamefont {Eliot}\ \bibnamefont {Kapit}},
  \bibinfo {author} {\bibfnamefont {Darrell~F.}\ \bibnamefont {Schroeter}}, \
  and\ \bibinfo {author} {\bibfnamefont {Martin}\ \bibnamefont {Greiter}},\
  }\bibfield  {title} {\enquote {\bibinfo {title} {Parent hamiltonian for the
  chiral spin liquid},}\ }\href {\doibase 10.1103/PhysRevB.80.104406}
  {\bibfield  {journal} {\bibinfo  {journal} {Phys. Rev. B}\ }\textbf {\bibinfo
  {volume} {80}},\ \bibinfo {pages} {104406} (\bibinfo {year}
  {2009})}\BibitemShut {NoStop}%
\bibitem [{\citenamefont {Nielsen}\ \emph {et~al.}(2012)\citenamefont
  {Nielsen}, \citenamefont {Cirac},\ and\ \citenamefont
  {Sierra}}]{Nielsen2012}%
  \BibitemOpen
  \bibfield  {author} {\bibinfo {author} {\bibfnamefont {Anne E.~B.}\
  \bibnamefont {Nielsen}}, \bibinfo {author} {\bibfnamefont {J.~Ignacio}\
  \bibnamefont {Cirac}}, \ and\ \bibinfo {author} {\bibfnamefont {Germ\'an}\
  \bibnamefont {Sierra}},\ }\bibfield  {title} {\enquote {\bibinfo {title}
  {Laughlin spin-liquid states on lattices obtained from conformal field
  theory},}\ }\href {\doibase 10.1103/PhysRevLett.108.257206} {\bibfield
  {journal} {\bibinfo  {journal} {Phys. Rev. Lett.}\ }\textbf {\bibinfo
  {volume} {108}},\ \bibinfo {pages} {257206} (\bibinfo {year}
  {2012})}\BibitemShut {NoStop}%
\bibitem [{\citenamefont {Greiter}\ \emph {et~al.}(2014)\citenamefont
  {Greiter}, \citenamefont {Schroeter},\ and\ \citenamefont
  {Thomale}}]{Greiter2014}%
  \BibitemOpen
  \bibfield  {author} {\bibinfo {author} {\bibfnamefont {Martin}\ \bibnamefont
  {Greiter}}, \bibinfo {author} {\bibfnamefont {Darrell~F.}\ \bibnamefont
  {Schroeter}}, \ and\ \bibinfo {author} {\bibfnamefont {Ronny}\ \bibnamefont
  {Thomale}},\ }\bibfield  {title} {\enquote {\bibinfo {title} {Parent
  {Hamiltonian} for the non-{Abelian} chiral spin liquid},}\ }\href {\doibase
  10.1103/PhysRevB.89.165125} {\bibfield  {journal} {\bibinfo  {journal} {Phys.
  Rev. B}\ }\textbf {\bibinfo {volume} {89}},\ \bibinfo {pages} {165125}
  (\bibinfo {year} {2014})}\BibitemShut {NoStop}%
\bibitem [{\citenamefont {Bauer}\ \emph {et~al.}(2014)\citenamefont {Bauer},
  \citenamefont {Cincio}, \citenamefont {Keller}, \citenamefont {Dolfi},
  \citenamefont {Vidal}, \citenamefont {Trebst},\ and\ \citenamefont
  {Ludwig}}]{Bauer2014}%
  \BibitemOpen
  \bibfield  {author} {\bibinfo {author} {\bibfnamefont {B.}~\bibnamefont
  {Bauer}}, \bibinfo {author} {\bibfnamefont {L.}~\bibnamefont {Cincio}},
  \bibinfo {author} {\bibfnamefont {B.P.}\ \bibnamefont {Keller}}, \bibinfo
  {author} {\bibfnamefont {M.}~\bibnamefont {Dolfi}}, \bibinfo {author}
  {\bibfnamefont {G.}~\bibnamefont {Vidal}}, \bibinfo {author} {\bibfnamefont
  {S.}~\bibnamefont {Trebst}}, \ and\ \bibinfo {author} {\bibfnamefont
  {A.~W.~W.}\ \bibnamefont {Ludwig}},\ }\bibfield  {title} {\enquote {\bibinfo
  {title} {Chiral spin liquid and emergent anyons in a kagome lattice {M}ott
  insulator},}\ }\href {\doibase 10.1038/ncomms6137} {\bibfield  {journal}
  {\bibinfo  {journal} {Nature Communications}\ }\textbf {\bibinfo {volume}
  {5}},\ \bibinfo {pages} {5137} (\bibinfo {year} {2014})}\BibitemShut
  {NoStop}%
\bibitem [{\citenamefont {E.~B.~Nielsen}\ \emph {et~al.}(2013)\citenamefont
  {E.~B.~Nielsen}, \citenamefont {Sierra},\ and\ \citenamefont
  {Cirac}}]{Nielsen2013}%
  \BibitemOpen
  \bibfield  {author} {\bibinfo {author} {\bibfnamefont {Anne}\ \bibnamefont
  {E.~B.~Nielsen}}, \bibinfo {author} {\bibfnamefont {Germ\'an}\ \bibnamefont
  {Sierra}}, \ and\ \bibinfo {author} {\bibfnamefont {J.~Ignacio}\ \bibnamefont
  {Cirac}},\ }\bibfield  {title} {\enquote {\bibinfo {title} {Local models of
  fractional quantum {H}all states in lattices and physical implementation},}\
  }\href {\doibase 10.1038/ncomms3864} {\bibfield  {journal} {\bibinfo
  {journal} {Nature Communications}\ }\textbf {\bibinfo {volume} {4}},\
  \bibinfo {pages} {2864} (\bibinfo {year} {2013})}\BibitemShut {NoStop}%
\bibitem [{\citenamefont {Wietek}\ and\ \citenamefont
  {L\"auchli}(2017)}]{Wietek2017}%
  \BibitemOpen
  \bibfield  {author} {\bibinfo {author} {\bibfnamefont {Alexander}\
  \bibnamefont {Wietek}}\ and\ \bibinfo {author} {\bibfnamefont {Andreas~M.}\
  \bibnamefont {L\"auchli}},\ }\bibfield  {title} {\enquote {\bibinfo {title}
  {Chiral spin liquid and quantum criticality in extended $s=\frac{1}{2}$
  {Heisenberg} models on the triangular lattice},}\ }\href {\doibase
  10.1103/PhysRevB.95.035141} {\bibfield  {journal} {\bibinfo  {journal} {Phys.
  Rev. B}\ }\textbf {\bibinfo {volume} {95}},\ \bibinfo {pages} {035141}
  (\bibinfo {year} {2017})}\BibitemShut {NoStop}%
\bibitem [{\citenamefont {Gong}\ \emph {et~al.}(2017)\citenamefont {Gong},
  \citenamefont {Zhu}, \citenamefont {Zhu}, \citenamefont {Sheng},\ and\
  \citenamefont {Yang}}]{Gong2017}%
  \BibitemOpen
  \bibfield  {author} {\bibinfo {author} {\bibfnamefont {Shou-Shu}\
  \bibnamefont {Gong}}, \bibinfo {author} {\bibfnamefont {W.}~\bibnamefont
  {Zhu}}, \bibinfo {author} {\bibfnamefont {J.-X.}\ \bibnamefont {Zhu}},
  \bibinfo {author} {\bibfnamefont {D.~N.}\ \bibnamefont {Sheng}}, \ and\
  \bibinfo {author} {\bibfnamefont {Kun}\ \bibnamefont {Yang}},\ }\bibfield
  {title} {\enquote {\bibinfo {title} {Global phase diagram and quantum spin
  liquids in a spin-$\frac{1}{2}$ triangular antiferromagnet},}\ }\href
  {\doibase 10.1103/PhysRevB.96.075116} {\bibfield  {journal} {\bibinfo
  {journal} {Phys. Rev. B}\ }\textbf {\bibinfo {volume} {96}},\ \bibinfo
  {pages} {075116} (\bibinfo {year} {2017})}\BibitemShut {NoStop}%
\bibitem [{\citenamefont {Kitaev}(2003{\natexlab{a}})}]{Kitaev2003}%
  \BibitemOpen
  \bibfield  {author} {\bibinfo {author} {\bibfnamefont {A.Yu.}\ \bibnamefont
  {Kitaev}},\ }\bibfield  {title} {\enquote {\bibinfo {title} {Fault-tolerant
  quantum computation by anyons},}\ }\href {\doibase
  http://dx.doi.org/10.1016/S0003-4916(02)00018-0} {\bibfield  {journal}
  {\bibinfo  {journal} {Annals of Physics}\ }\textbf {\bibinfo {volume}
  {303}},\ \bibinfo {pages} {2 -- 30} (\bibinfo {year}
  {2003}{\natexlab{a}})}\BibitemShut {NoStop}%
\bibitem [{\citenamefont {Kitaev}(2006)}]{Kitaev2006}%
  \BibitemOpen
  \bibfield  {author} {\bibinfo {author} {\bibfnamefont {Alexei}\ \bibnamefont
  {Kitaev}},\ }\bibfield  {title} {\enquote {\bibinfo {title} {Anyons in an
  exactly solved model and beyond},}\ }\href {\doibase
  https://doi.org/10.1016/j.aop.2005.10.005} {\bibfield  {journal} {\bibinfo
  {journal} {Annals of Physics}\ }\textbf {\bibinfo {volume} {321}},\ \bibinfo
  {pages} {2 -- 111} (\bibinfo {year} {2006})}\BibitemShut {NoStop}%
\bibitem [{\citenamefont {Yao}\ and\ \citenamefont {Kivelson}(2007)}]{Yao2007}%
  \BibitemOpen
  \bibfield  {author} {\bibinfo {author} {\bibfnamefont {Hong}\ \bibnamefont
  {Yao}}\ and\ \bibinfo {author} {\bibfnamefont {Steven~A.}\ \bibnamefont
  {Kivelson}},\ }\bibfield  {title} {\enquote {\bibinfo {title} {Exact chiral
  spin liquid with non-abelian anyons},}\ }\href {\doibase
  10.1103/PhysRevLett.99.247203} {\bibfield  {journal} {\bibinfo  {journal}
  {Phys. Rev. Lett.}\ }\textbf {\bibinfo {volume} {99}},\ \bibinfo {pages}
  {247203} (\bibinfo {year} {2007})}\BibitemShut {NoStop}%
\bibitem [{\citenamefont {Greiter}\ and\ \citenamefont
  {Thomale}(2009)}]{Greiter2009}%
  \BibitemOpen
  \bibfield  {author} {\bibinfo {author} {\bibfnamefont {Martin}\ \bibnamefont
  {Greiter}}\ and\ \bibinfo {author} {\bibfnamefont {Ronny}\ \bibnamefont
  {Thomale}},\ }\bibfield  {title} {\enquote {\bibinfo {title} {Non-abelian
  statistics in a quantum antiferromagnet},}\ }\href {\doibase
  10.1103/PhysRevLett.102.207203} {\bibfield  {journal} {\bibinfo  {journal}
  {Phys. Rev. Lett.}\ }\textbf {\bibinfo {volume} {102}},\ \bibinfo {pages}
  {207203} (\bibinfo {year} {2009})}\BibitemShut {NoStop}%
\bibitem [{\citenamefont {{Gorshkov}}\ \emph {et~al.}(2010)\citenamefont
  {{Gorshkov}}, \citenamefont {{Hermele}}, \citenamefont {{Gurarie}},
  \citenamefont {{Xu}}, \citenamefont {{Julienne}}, \citenamefont {{Ye}},
  \citenamefont {{Zoller}}, \citenamefont {{Demler}}, \citenamefont {{Lukin}},\
  and\ \citenamefont {{Rey}}}]{Gorshkov2010}%
  \BibitemOpen
  \bibfield  {author} {\bibinfo {author} {\bibfnamefont {A.~V.}\ \bibnamefont
  {{Gorshkov}}}, \bibinfo {author} {\bibfnamefont {M.}~\bibnamefont
  {{Hermele}}}, \bibinfo {author} {\bibfnamefont {V.}~\bibnamefont
  {{Gurarie}}}, \bibinfo {author} {\bibfnamefont {C.}~\bibnamefont {{Xu}}},
  \bibinfo {author} {\bibfnamefont {P.~S.}\ \bibnamefont {{Julienne}}},
  \bibinfo {author} {\bibfnamefont {J.}~\bibnamefont {{Ye}}}, \bibinfo {author}
  {\bibfnamefont {P.}~\bibnamefont {{Zoller}}}, \bibinfo {author}
  {\bibfnamefont {E.}~\bibnamefont {{Demler}}}, \bibinfo {author}
  {\bibfnamefont {M.~D.}\ \bibnamefont {{Lukin}}}, \ and\ \bibinfo {author}
  {\bibfnamefont {A.~M.}\ \bibnamefont {{Rey}}},\ }\bibfield  {title} {\enquote
  {\bibinfo {title} {{Two-orbital SU(N) magnetism with ultracold alkaline-earth
  atoms}},}\ }\href {\doibase 10.1038/nphys1535} {\bibfield  {journal}
  {\bibinfo  {journal} {Nature Physics}\ }\textbf {\bibinfo {volume} {6}},\
  \bibinfo {pages} {289--295} (\bibinfo {year} {2010})}\BibitemShut {NoStop}%
\bibitem [{\citenamefont {Hermele}\ \emph {et~al.}(2009)\citenamefont
  {Hermele}, \citenamefont {Gurarie},\ and\ \citenamefont {Rey}}]{Hermele2009}%
  \BibitemOpen
  \bibfield  {author} {\bibinfo {author} {\bibfnamefont {Michael}\ \bibnamefont
  {Hermele}}, \bibinfo {author} {\bibfnamefont {Victor}\ \bibnamefont
  {Gurarie}}, \ and\ \bibinfo {author} {\bibfnamefont {Ana~Maria}\ \bibnamefont
  {Rey}},\ }\bibfield  {title} {\enquote {\bibinfo {title} {Mott insulators of
  ultracold fermionic alkaline earth atoms: Underconstrained magnetism and
  chiral spin liquid},}\ }\href {\doibase 10.1103/PhysRevLett.103.135301}
  {\bibfield  {journal} {\bibinfo  {journal} {Phys. Rev. Lett.}\ }\textbf
  {\bibinfo {volume} {103}},\ \bibinfo {pages} {135301} (\bibinfo {year}
  {2009})}\BibitemShut {NoStop}%
\bibitem [{\citenamefont {Nataf}\ \emph {et~al.}(2016)\citenamefont {Nataf},
  \citenamefont {Lajk\'o}, \citenamefont {Wietek}, \citenamefont {Penc},
  \citenamefont {Mila},\ and\ \citenamefont {L\"auchli}}]{Nataf2016}%
  \BibitemOpen
  \bibfield  {author} {\bibinfo {author} {\bibfnamefont {Pierre}\ \bibnamefont
  {Nataf}}, \bibinfo {author} {\bibfnamefont {Mikl\'os}\ \bibnamefont
  {Lajk\'o}}, \bibinfo {author} {\bibfnamefont {Alexander}\ \bibnamefont
  {Wietek}}, \bibinfo {author} {\bibfnamefont {Karlo}\ \bibnamefont {Penc}},
  \bibinfo {author} {\bibfnamefont {Fr\'ed\'eric}\ \bibnamefont {Mila}}, \ and\
  \bibinfo {author} {\bibfnamefont {Andreas~M.}\ \bibnamefont {L\"auchli}},\
  }\bibfield  {title} {\enquote {\bibinfo {title} {Chiral spin liquids in
  triangular-lattice {SU}({N}) fermionic mott insulators with artificial gauge
  fields},}\ }\href {\doibase 10.1103/PhysRevLett.117.167202} {\bibfield
  {journal} {\bibinfo  {journal} {Phys. Rev. Lett.}\ }\textbf {\bibinfo
  {volume} {117}},\ \bibinfo {pages} {167202} (\bibinfo {year}
  {2016})}\BibitemShut {NoStop}%
\bibitem [{\citenamefont {Chen}\ \emph {et~al.}(2016)\citenamefont {Chen},
  \citenamefont {Hazzard}, \citenamefont {Rey},\ and\ \citenamefont
  {Hermele}}]{Chen2016}%
  \BibitemOpen
  \bibfield  {author} {\bibinfo {author} {\bibfnamefont {Gang}\ \bibnamefont
  {Chen}}, \bibinfo {author} {\bibfnamefont {Kaden R.~A.}\ \bibnamefont
  {Hazzard}}, \bibinfo {author} {\bibfnamefont {Ana~Maria}\ \bibnamefont
  {Rey}}, \ and\ \bibinfo {author} {\bibfnamefont {Michael}\ \bibnamefont
  {Hermele}},\ }\bibfield  {title} {\enquote {\bibinfo {title}
  {Synthetic-gauge-field stabilization of the chiral-spin-liquid phase},}\
  }\href {\doibase 10.1103/PhysRevA.93.061601} {\bibfield  {journal} {\bibinfo
  {journal} {Phys. Rev. A}\ }\textbf {\bibinfo {volume} {93}},\ \bibinfo
  {pages} {061601} (\bibinfo {year} {2016})}\BibitemShut {NoStop}%
\bibitem [{\citenamefont {He}\ \emph {et~al.}(2014)\citenamefont {He},
  \citenamefont {Sheng},\ and\ \citenamefont {Chen}}]{He2014a}%
  \BibitemOpen
  \bibfield  {author} {\bibinfo {author} {\bibfnamefont {Yin-Chen}\
  \bibnamefont {He}}, \bibinfo {author} {\bibfnamefont {D.~N.}\ \bibnamefont
  {Sheng}}, \ and\ \bibinfo {author} {\bibfnamefont {Yan}\ \bibnamefont
  {Chen}},\ }\bibfield  {title} {\enquote {\bibinfo {title} {Chiral spin liquid
  in a frustrated anisotropic {K}agome {H}eisenberg model},}\ }\href {\doibase
  10.1103/PhysRevLett.112.137202} {\bibfield  {journal} {\bibinfo  {journal}
  {Phys. Rev. Lett.}\ }\textbf {\bibinfo {volume} {112}},\ \bibinfo {pages}
  {137202} (\bibinfo {year} {2014})}\BibitemShut {NoStop}%
\bibitem [{\citenamefont {Gong}\ \emph {et~al.}(2014)\citenamefont {Gong},
  \citenamefont {Zhu},\ and\ \citenamefont {Sheng}}]{Gong2014a}%
  \BibitemOpen
  \bibfield  {author} {\bibinfo {author} {\bibfnamefont {Shou-Shu}\
  \bibnamefont {Gong}}, \bibinfo {author} {\bibfnamefont {W.}~\bibnamefont
  {Zhu}}, \ and\ \bibinfo {author} {\bibfnamefont {D.~N.}\ \bibnamefont
  {Sheng}},\ }\bibfield  {title} {\enquote {\bibinfo {title} {Emergent chiral
  spin liquid: Fractional quantum {H}all effect in a kagome {Heisenberg}
  model},}\ }\href {\doibase 10.1038/srep06317} {\bibfield  {journal} {\bibinfo
   {journal} {Scientific Reports}\ }\textbf {\bibinfo {volume} {4}},\ \bibinfo
  {pages} {6317} (\bibinfo {year} {2014})}\BibitemShut {NoStop}%
\bibitem [{\citenamefont {Wietek}\ \emph {et~al.}(2015)\citenamefont {Wietek},
  \citenamefont {Sterdyniak},\ and\ \citenamefont {L\"auchli}}]{Wietek2015}%
  \BibitemOpen
  \bibfield  {author} {\bibinfo {author} {\bibfnamefont {Alexander}\
  \bibnamefont {Wietek}}, \bibinfo {author} {\bibfnamefont {Antoine}\
  \bibnamefont {Sterdyniak}}, \ and\ \bibinfo {author} {\bibfnamefont
  {Andreas~M.}\ \bibnamefont {L\"auchli}},\ }\bibfield  {title} {\enquote
  {\bibinfo {title} {Nature of chiral spin liquids on the kagome lattice},}\
  }\href {\doibase 10.1103/PhysRevB.92.125122} {\bibfield  {journal} {\bibinfo
  {journal} {Phys. Rev. B}\ }\textbf {\bibinfo {volume} {92}},\ \bibinfo
  {pages} {125122} (\bibinfo {year} {2015})}\BibitemShut {NoStop}%
\bibitem [{\citenamefont {Boos}\ \emph {et~al.}(2020)\citenamefont {Boos},
  \citenamefont {Ganahl}, \citenamefont {Lajk\'o}, \citenamefont {Nataf},
  \citenamefont {L\"auchli}, \citenamefont {Penc}, \citenamefont {Schmidt},\
  and\ \citenamefont {Mila}}]{Boos2020}%
  \BibitemOpen
  \bibfield  {author} {\bibinfo {author} {\bibfnamefont {C.}~\bibnamefont
  {Boos}}, \bibinfo {author} {\bibfnamefont {C.~J.}\ \bibnamefont {Ganahl}},
  \bibinfo {author} {\bibfnamefont {M.}~\bibnamefont {Lajk\'o}}, \bibinfo
  {author} {\bibfnamefont {P.}~\bibnamefont {Nataf}}, \bibinfo {author}
  {\bibfnamefont {A.~M.}\ \bibnamefont {L\"auchli}}, \bibinfo {author}
  {\bibfnamefont {K.}~\bibnamefont {Penc}}, \bibinfo {author} {\bibfnamefont
  {K.~P.}\ \bibnamefont {Schmidt}}, \ and\ \bibinfo {author} {\bibfnamefont
  {F.}~\bibnamefont {Mila}},\ }\bibfield  {title} {\enquote {\bibinfo {title}
  {Time-reversal symmetry breaking abelian chiral spin liquid in mott phases of
  three-component fermions on the triangular lattice},}\ }\href {\doibase
  10.1103/PhysRevResearch.2.023098} {\bibfield  {journal} {\bibinfo  {journal}
  {Phys. Rev. Research}\ }\textbf {\bibinfo {volume} {2}},\ \bibinfo {pages}
  {023098} (\bibinfo {year} {2020})}\BibitemShut {NoStop}%
\bibitem [{\citenamefont {Pagano}\ \emph {et~al.}(2014)\citenamefont {Pagano},
  \citenamefont {Mancini}, \citenamefont {Cappellini},\ and\ \citenamefont
  {et~al.}}]{Pagano2014}%
  \BibitemOpen
  \bibfield  {author} {\bibinfo {author} {\bibfnamefont {G.}~\bibnamefont
  {Pagano}}, \bibinfo {author} {\bibfnamefont {M.}~\bibnamefont {Mancini}},
  \bibinfo {author} {\bibfnamefont {G.}~\bibnamefont {Cappellini}}, \ and\
  \bibinfo {author} {\bibnamefont {et~al.}},\ }\bibfield  {title} {\enquote
  {\bibinfo {title} {A one-dimensional liquid of fermions with tunable spin},}\
  }\href {\doibase 10.1038/nphys2878} {\bibfield  {journal} {\bibinfo
  {journal} {Nature Physics}\ }\textbf {\bibinfo {volume} {10}},\ \bibinfo
  {pages} {198--201} (\bibinfo {year} {2014})}\BibitemShut {NoStop}%
\bibitem [{\citenamefont {Zhang}\ \emph {et~al.}(2021)\citenamefont {Zhang},
  \citenamefont {Sheng},\ and\ \citenamefont {Vishwanath}}]{Zhang2021}%
  \BibitemOpen
  \bibfield  {author} {\bibinfo {author} {\bibfnamefont {Ya-Hui}\ \bibnamefont
  {Zhang}}, \bibinfo {author} {\bibfnamefont {D.~N.}\ \bibnamefont {Sheng}}, \
  and\ \bibinfo {author} {\bibfnamefont {Ashvin}\ \bibnamefont {Vishwanath}},\
  }\href@noop {} {\enquote {\bibinfo {title} {An $su(4)$ chiral spin liquid and
  quantized dipole hall effect in moiré bilayers},}\ } (\bibinfo {year}
  {2021}),\ \Eprint {http://arxiv.org/abs/2103.09825} {arXiv:2103.09825
  [cond-mat.str-el]} \BibitemShut {NoStop}%
\bibitem [{\citenamefont {{Verstraete}}\ and\ \citenamefont
  {{Cirac}}(2004)}]{Verstraete2004b}%
  \BibitemOpen
  \bibfield  {author} {\bibinfo {author} {\bibfnamefont {F.}~\bibnamefont
  {{Verstraete}}}\ and\ \bibinfo {author} {\bibfnamefont {J.~I.}\ \bibnamefont
  {{Cirac}}},\ }\bibfield  {title} {\enquote {\bibinfo {title}
  {{Renormalization algorithms for Quantum-Many Body Systems in two and higher
  dimensions}},}\ }\href@noop {} {\bibfield  {journal} {\bibinfo  {journal}
  {arXiv e-prints}\ ,\ \bibinfo {eid} {cond-mat/0407066}} (\bibinfo {year}
  {2004})},\ \Eprint {http://arxiv.org/abs/cond-mat/0407066}
  {arXiv:cond-mat/0407066 [cond-mat.str-el]} \BibitemShut {NoStop}%
\bibitem [{\citenamefont {Liao}\ \emph {et~al.}(2017)\citenamefont {Liao},
  \citenamefont {Xie}, \citenamefont {Chen}, \citenamefont {Liu}, \citenamefont
  {Xie}, \citenamefont {Huang}, \citenamefont {Normand},\ and\ \citenamefont
  {Xiang}}]{Liao2017}%
  \BibitemOpen
  \bibfield  {author} {\bibinfo {author} {\bibfnamefont {H.~J.}\ \bibnamefont
  {Liao}}, \bibinfo {author} {\bibfnamefont {Z.~Y.}\ \bibnamefont {Xie}},
  \bibinfo {author} {\bibfnamefont {J.}~\bibnamefont {Chen}}, \bibinfo {author}
  {\bibfnamefont {Z.~Y.}\ \bibnamefont {Liu}}, \bibinfo {author} {\bibfnamefont
  {H.~D.}\ \bibnamefont {Xie}}, \bibinfo {author} {\bibfnamefont {R.~Z.}\
  \bibnamefont {Huang}}, \bibinfo {author} {\bibfnamefont {B.}~\bibnamefont
  {Normand}}, \ and\ \bibinfo {author} {\bibfnamefont {T.}~\bibnamefont
  {Xiang}},\ }\bibfield  {title} {\enquote {\bibinfo {title} {Gapless
  spin-liquid ground state in the $s=1/2$ {K}agome antiferromagnet},}\ }\href
  {\doibase 10.1103/PhysRevLett.118.137202} {\bibfield  {journal} {\bibinfo
  {journal} {Phys. Rev. Lett.}\ }\textbf {\bibinfo {volume} {118}},\ \bibinfo
  {pages} {137202} (\bibinfo {year} {2017})}\BibitemShut {NoStop}%
\bibitem [{\citenamefont {Lee}\ \emph {et~al.}(2019)\citenamefont {Lee},
  \citenamefont {Kaneko}, \citenamefont {Okubo},\ and\ \citenamefont
  {Kawashima}}]{Lee2019}%
  \BibitemOpen
  \bibfield  {author} {\bibinfo {author} {\bibfnamefont {Hyun-Yong}\
  \bibnamefont {Lee}}, \bibinfo {author} {\bibfnamefont {Ryui}\ \bibnamefont
  {Kaneko}}, \bibinfo {author} {\bibfnamefont {Tsuyoshi}\ \bibnamefont
  {Okubo}}, \ and\ \bibinfo {author} {\bibfnamefont {Naoki}\ \bibnamefont
  {Kawashima}},\ }\bibfield  {title} {\enquote {\bibinfo {title} {Gapless
  {Kitaev} spin liquid to classical string gas through tensor networks},}\
  }\href {\doibase 10.1103/PhysRevLett.123.087203} {\bibfield  {journal}
  {\bibinfo  {journal} {Phys. Rev. Lett.}\ }\textbf {\bibinfo {volume} {123}},\
  \bibinfo {pages} {087203} (\bibinfo {year} {2019})}\BibitemShut {NoStop}%
\bibitem [{\citenamefont {Liu}\ \emph {et~al.}(2020)\citenamefont {Liu},
  \citenamefont {Gong}, \citenamefont {Li}, \citenamefont {Poilblanc},
  \citenamefont {Chen},\ and\ \citenamefont {Gu}}]{Liu2020}%
  \BibitemOpen
  \bibfield  {author} {\bibinfo {author} {\bibfnamefont {Wen-Yuan}\
  \bibnamefont {Liu}}, \bibinfo {author} {\bibfnamefont {Shou-Shu}\
  \bibnamefont {Gong}}, \bibinfo {author} {\bibfnamefont {Yu-Bin}\ \bibnamefont
  {Li}}, \bibinfo {author} {\bibfnamefont {Didier}\ \bibnamefont {Poilblanc}},
  \bibinfo {author} {\bibfnamefont {Wei-Qiang}\ \bibnamefont {Chen}}, \ and\
  \bibinfo {author} {\bibfnamefont {Zheng-Cheng}\ \bibnamefont {Gu}},\
  }\href@noop {} {\enquote {\bibinfo {title} {Gapless quantum spin liquid and
  global phase diagram of the spin-1/2 {$J_1$-$J_2$} square antiferromagnetic
  {Heisenberg} model},}\ } (\bibinfo {year} {2020}),\ \Eprint
  {http://arxiv.org/abs/2009.01821} {arXiv:2009.01821 [cond-mat.str-el]}
  \BibitemShut {NoStop}%
\bibitem [{\citenamefont {Schuch}\ \emph {et~al.}(2010)\citenamefont {Schuch},
  \citenamefont {Cirac},\ and\ \citenamefont
  {P{\'{e}}rez-Garc{\'{\i}}a}}]{Schuch2010a}%
  \BibitemOpen
  \bibfield  {author} {\bibinfo {author} {\bibfnamefont {Norbert}\ \bibnamefont
  {Schuch}}, \bibinfo {author} {\bibfnamefont {J.~Ignacio}\ \bibnamefont
  {Cirac}}, \ and\ \bibinfo {author} {\bibfnamefont {David}\ \bibnamefont
  {P{\'{e}}rez-Garc{\'{\i}}a}},\ }\bibfield  {title} {\enquote {\bibinfo
  {title} {{PEPS as ground states: Degeneracy and topology}},}\ }\href
  {\doibase 10.1016/j.aop.2010.05.008} {\bibfield  {journal} {\bibinfo
  {journal} {Annals of Physics}\ }\textbf {\bibinfo {volume} {325}},\ \bibinfo
  {pages} {2153--2192} (\bibinfo {year} {2010})}\BibitemShut {NoStop}%
\bibitem [{\citenamefont {Schuch}\ \emph {et~al.}(2012)\citenamefont {Schuch},
  \citenamefont {Poilblanc}, \citenamefont {Cirac},\ and\ \citenamefont
  {P{\'{e}}rez-Garc{\'{\i}}a}}]{Schuch2012}%
  \BibitemOpen
  \bibfield  {author} {\bibinfo {author} {\bibfnamefont {Norbert}\ \bibnamefont
  {Schuch}}, \bibinfo {author} {\bibfnamefont {Didier}\ \bibnamefont
  {Poilblanc}}, \bibinfo {author} {\bibfnamefont {J.~Ignacio}\ \bibnamefont
  {Cirac}}, \ and\ \bibinfo {author} {\bibfnamefont {David}\ \bibnamefont
  {P{\'{e}}rez-Garc{\'{\i}}a}},\ }\bibfield  {title} {\enquote {\bibinfo
  {title} {{Resonating valence bond states in the PEPS formalism}},}\ }\href
  {\doibase 10.1103/PhysRevB.86.115108} {\bibfield  {journal} {\bibinfo
  {journal} {Physical Review B}\ }\textbf {\bibinfo {volume} {86}},\ \bibinfo
  {pages} {115108} (\bibinfo {year} {2012})}\BibitemShut {NoStop}%
\bibitem [{\citenamefont {Chen}\ and\ \citenamefont
  {Poilblanc}(2018)}]{Chen2018a}%
  \BibitemOpen
  \bibfield  {author} {\bibinfo {author} {\bibfnamefont {Ji-Yao}\ \bibnamefont
  {Chen}}\ and\ \bibinfo {author} {\bibfnamefont {Didier}\ \bibnamefont
  {Poilblanc}},\ }\bibfield  {title} {\enquote {\bibinfo {title} {Topological
  $\mathbb{Z}_{2}$ resonating-valence-bond spin liquid on the square
  lattice},}\ }\href {\doibase 10.1103/PhysRevB.97.161107} {\bibfield
  {journal} {\bibinfo  {journal} {Phys. Rev. B}\ }\textbf {\bibinfo {volume}
  {97}},\ \bibinfo {pages} {161107} (\bibinfo {year} {2018})}\BibitemShut
  {NoStop}%
\bibitem [{\citenamefont {Poilblanc}\ \emph {et~al.}(2015)\citenamefont
  {Poilblanc}, \citenamefont {Cirac},\ and\ \citenamefont
  {Schuch}}]{Poilblanc2015}%
  \BibitemOpen
  \bibfield  {author} {\bibinfo {author} {\bibfnamefont {Didier}\ \bibnamefont
  {Poilblanc}}, \bibinfo {author} {\bibfnamefont {J.~Ignacio}\ \bibnamefont
  {Cirac}}, \ and\ \bibinfo {author} {\bibfnamefont {Norbert}\ \bibnamefont
  {Schuch}},\ }\bibfield  {title} {\enquote {\bibinfo {title} {Chiral
  topological spin liquids with projected entangled pair states},}\ }\href
  {\doibase 10.1103/PhysRevB.91.224431} {\bibfield  {journal} {\bibinfo
  {journal} {Phys. Rev. B}\ }\textbf {\bibinfo {volume} {91}},\ \bibinfo
  {pages} {224431} (\bibinfo {year} {2015})}\BibitemShut {NoStop}%
\bibitem [{\citenamefont {Chen}\ \emph
  {et~al.}(2018{\natexlab{a}})\citenamefont {Chen}, \citenamefont
  {Vanderstraeten}, \citenamefont {Capponi},\ and\ \citenamefont
  {Poilblanc}}]{Chen2018b}%
  \BibitemOpen
  \bibfield  {author} {\bibinfo {author} {\bibfnamefont {Ji-Yao}\ \bibnamefont
  {Chen}}, \bibinfo {author} {\bibfnamefont {Laurens}\ \bibnamefont
  {Vanderstraeten}}, \bibinfo {author} {\bibfnamefont {Sylvain}\ \bibnamefont
  {Capponi}}, \ and\ \bibinfo {author} {\bibfnamefont {Didier}\ \bibnamefont
  {Poilblanc}},\ }\bibfield  {title} {\enquote {\bibinfo {title} {Non-abelian
  chiral spin liquid in a quantum antiferromagnet revealed by an i{P}{E}{P}{S}
  study},}\ }\href {\doibase 10.1103/PhysRevB.98.184409} {\bibfield  {journal}
  {\bibinfo  {journal} {Phys. Rev. B}\ }\textbf {\bibinfo {volume} {98}},\
  \bibinfo {pages} {184409} (\bibinfo {year} {2018}{\natexlab{a}})}\BibitemShut
  {NoStop}%
\bibitem [{\citenamefont {Francesco}\ \emph {et~al.}(1997)\citenamefont
  {Francesco}, \citenamefont {Mathieu},\ and\ \citenamefont
  {S\'en\'echal}}]{Franscesco1997}%
  \BibitemOpen
  \bibfield  {author} {\bibinfo {author} {\bibfnamefont {Philippe}\
  \bibnamefont {Francesco}}, \bibinfo {author} {\bibfnamefont {Pierre}\
  \bibnamefont {Mathieu}}, \ and\ \bibinfo {author} {\bibfnamefont {David}\
  \bibnamefont {S\'en\'echal}},\ }\href {\doibase 10.1007/978-1-4612-2256-9}
  {\emph {\bibinfo {title} {Conformal Field Theory}}}\ (\bibinfo  {publisher}
  {Springer-Verlag New York},\ \bibinfo {year} {1997})\BibitemShut {NoStop}%
\bibitem [{\citenamefont {Chen}\ \emph {et~al.}(2020)\citenamefont {Chen},
  \citenamefont {Capponi}, \citenamefont {Wietek}, \citenamefont {Mambrini},
  \citenamefont {Schuch},\ and\ \citenamefont {Poilblanc}}]{Chen2020}%
  \BibitemOpen
  \bibfield  {author} {\bibinfo {author} {\bibfnamefont {Ji-Yao}\ \bibnamefont
  {Chen}}, \bibinfo {author} {\bibfnamefont {Sylvain}\ \bibnamefont {Capponi}},
  \bibinfo {author} {\bibfnamefont {Alexander}\ \bibnamefont {Wietek}},
  \bibinfo {author} {\bibfnamefont {Matthieu}\ \bibnamefont {Mambrini}},
  \bibinfo {author} {\bibfnamefont {Norbert}\ \bibnamefont {Schuch}}, \ and\
  \bibinfo {author} {\bibfnamefont {Didier}\ \bibnamefont {Poilblanc}},\
  }\bibfield  {title} {\enquote {\bibinfo {title} {$\mathrm{SU}(3{)}_{1}$
  chiral spin liquid on the square lattice: A view from symmetric projected
  entangled pair states},}\ }\href {\doibase 10.1103/PhysRevLett.125.017201}
  {\bibfield  {journal} {\bibinfo  {journal} {Phys. Rev. Lett.}\ }\textbf
  {\bibinfo {volume} {125}},\ \bibinfo {pages} {017201} (\bibinfo {year}
  {2020})}\BibitemShut {NoStop}%
\bibitem [{\citenamefont {Poilblanc}(2017)}]{Poilblanc2017b}%
  \BibitemOpen
  \bibfield  {author} {\bibinfo {author} {\bibfnamefont {Didier}\ \bibnamefont
  {Poilblanc}},\ }\bibfield  {title} {\enquote {\bibinfo {title} {Investigation
  of the chiral antiferromagnetic {H}eisenberg model using projected entangled
  pair states},}\ }\href {\doibase 10.1103/PhysRevB.96.121118} {\bibfield
  {journal} {\bibinfo  {journal} {Phys. Rev. B}\ }\textbf {\bibinfo {volume}
  {96}},\ \bibinfo {pages} {121118} (\bibinfo {year} {2017})}\BibitemShut
  {NoStop}%
\bibitem [{Note1()}]{Note1}%
  \BibitemOpen
  \bibinfo {note} {The chiral spin liquid phase should also exist away from
  $J_2=J_1/2$, due to its gapped nature.}\BibitemShut {Stop}%
\bibitem [{Note2()}]{Note2}%
  \BibitemOpen
  \bibinfo {note} {This can be extended to all fundamental IRREPs of SU($N$):
  $P_{ij}={\protect \bf J}_i\cdot {\protect \bf J}_j +\protect \frac {1}{N}$,
  where $J^\alpha $ are the generators defined in Eq.~(\ref {eq:liecom}) of
  ~\protect \App {app:WZW}. Note, the usual SU($2$) spin operators are given by
  ${\protect \bf S}=(1/\protect \sqrt {2}) {\protect \bf J}$.}\BibitemShut
  {Stop}%
\bibitem [{Note3()}]{Note3}%
  \BibitemOpen
  \bibinfo {note} {This decomposition holds only for $N=2$ (in the fundamental
  representation)}\BibitemShut {NoStop}%
\bibitem [{\citenamefont {Poilblanc}\ \emph {et~al.}(2016)\citenamefont
  {Poilblanc}, \citenamefont {Schuch},\ and\ \citenamefont
  {Affleck}}]{Poilblanc2016}%
  \BibitemOpen
  \bibfield  {author} {\bibinfo {author} {\bibfnamefont {Didier}\ \bibnamefont
  {Poilblanc}}, \bibinfo {author} {\bibfnamefont {Norbert}\ \bibnamefont
  {Schuch}}, \ and\ \bibinfo {author} {\bibfnamefont {Ian}\ \bibnamefont
  {Affleck}},\ }\bibfield  {title} {\enquote {\bibinfo {title}
  {$\mathrm{SU}(2{)}_{1}$ chiral edge modes of a critical spin liquid},}\
  }\href {\doibase 10.1103/PhysRevB.93.174414} {\bibfield  {journal} {\bibinfo
  {journal} {Phys. Rev. B}\ }\textbf {\bibinfo {volume} {93}},\ \bibinfo
  {pages} {174414} (\bibinfo {year} {2016})}\BibitemShut {NoStop}%
\bibitem [{\citenamefont {Nataf}\ and\ \citenamefont {Mila}(2014)}]{Nataf2014}%
  \BibitemOpen
  \bibfield  {author} {\bibinfo {author} {\bibfnamefont {Pierre}\ \bibnamefont
  {Nataf}}\ and\ \bibinfo {author} {\bibfnamefont {Fr\'ed\'eric}\ \bibnamefont
  {Mila}},\ }\bibfield  {title} {\enquote {\bibinfo {title} {Exact
  diagonalization of {H}eisenberg $\mathrm{SU}(n)$ models},}\ }\href {\doibase
  10.1103/PhysRevLett.113.127204} {\bibfield  {journal} {\bibinfo  {journal}
  {Phys. Rev. Lett.}\ }\textbf {\bibinfo {volume} {113}},\ \bibinfo {pages}
  {127204} (\bibinfo {year} {2014})}\BibitemShut {NoStop}%
\bibitem [{\citenamefont {Wan}\ \emph {et~al.}(2017)\citenamefont {Wan},
  \citenamefont {Nataf},\ and\ \citenamefont {Mila}}]{Wan2017}%
  \BibitemOpen
  \bibfield  {author} {\bibinfo {author} {\bibfnamefont {Kianna}\ \bibnamefont
  {Wan}}, \bibinfo {author} {\bibfnamefont {Pierre}\ \bibnamefont {Nataf}}, \
  and\ \bibinfo {author} {\bibfnamefont {Fr\'ed\'eric}\ \bibnamefont {Mila}},\
  }\bibfield  {title} {\enquote {\bibinfo {title} {Exact diagonalization of
  {SU($N$)} {H}eisenberg and {A}ffleck-{K}ennedy-{L}ieb-{T}asaki chains using
  the full {SU($N$)} symmetry},}\ }\href {\doibase 10.1103/PhysRevB.96.115159}
  {\bibfield  {journal} {\bibinfo  {journal} {Phys. Rev. B}\ }\textbf {\bibinfo
  {volume} {96}},\ \bibinfo {pages} {115159} (\bibinfo {year}
  {2017})}\BibitemShut {NoStop}%
\bibitem [{\citenamefont {Alex}\ \emph {et~al.}(2011)\citenamefont {Alex},
  \citenamefont {Kalus}, \citenamefont {Huckleberry},\ and\ \citenamefont {von
  Delft}}]{Alex2011}%
  \BibitemOpen
  \bibfield  {author} {\bibinfo {author} {\bibfnamefont {Arne}\ \bibnamefont
  {Alex}}, \bibinfo {author} {\bibfnamefont {Matthias}\ \bibnamefont {Kalus}},
  \bibinfo {author} {\bibfnamefont {Alan}\ \bibnamefont {Huckleberry}}, \ and\
  \bibinfo {author} {\bibfnamefont {Jan}\ \bibnamefont {von Delft}},\
  }\bibfield  {title} {\enquote {\bibinfo {title} {A numerical algorithm for
  the explicit calculation of {SU($N$)} and $\mbox{SL}(n,\mathbb {C})$sl(n,c)
  {C}lebsch–{G}ordan coefficients},}\ }\href {\doibase 10.1063/1.3521562}
  {\bibfield  {journal} {\bibinfo  {journal} {Journal of Mathematical Physics}\
  }\textbf {\bibinfo {volume} {52}},\ \bibinfo {pages} {023507} (\bibinfo
  {year} {2011})}\BibitemShut {NoStop}%
\bibitem [{Note4()}]{Note4}%
  \BibitemOpen
  \bibinfo {note} {Both states are translationally invariant and have different
  $\pm 1$ characters under $\pi /2$-rotation, for $C_4$-symmetric
  clusters.}\BibitemShut {Stop}%
\bibitem [{\citenamefont {Haghshenas}\ \emph {et~al.}(2018)\citenamefont
  {Haghshenas}, \citenamefont {Lan}, \citenamefont {Gong},\ and\ \citenamefont
  {Sheng}}]{Haghshenas2018}%
  \BibitemOpen
  \bibfield  {author} {\bibinfo {author} {\bibfnamefont {R.}~\bibnamefont
  {Haghshenas}}, \bibinfo {author} {\bibfnamefont {Wang-Wei}\ \bibnamefont
  {Lan}}, \bibinfo {author} {\bibfnamefont {Shou-Shu}\ \bibnamefont {Gong}}, \
  and\ \bibinfo {author} {\bibfnamefont {D.~N.}\ \bibnamefont {Sheng}},\
  }\bibfield  {title} {\enquote {\bibinfo {title} {Quantum phase diagram of
  spin-1 ${J}_{1}\text{\ensuremath{-}}{J}_{2}$ {H}eisenberg model on the square
  lattice: An infinite projected entangled-pair state and density matrix
  renormalization group study},}\ }\href {\doibase 10.1103/PhysRevB.97.184436}
  {\bibfield  {journal} {\bibinfo  {journal} {Phys. Rev. B}\ }\textbf {\bibinfo
  {volume} {97}},\ \bibinfo {pages} {184436} (\bibinfo {year}
  {2018})}\BibitemShut {NoStop}%
\bibitem [{\citenamefont {Chen}\ \emph
  {et~al.}(2018{\natexlab{b}})\citenamefont {Chen}, \citenamefont {Capponi},\
  and\ \citenamefont {Poilblanc}}]{Chen2018c}%
  \BibitemOpen
  \bibfield  {author} {\bibinfo {author} {\bibfnamefont {Ji-Yao}\ \bibnamefont
  {Chen}}, \bibinfo {author} {\bibfnamefont {Sylvain}\ \bibnamefont {Capponi}},
  \ and\ \bibinfo {author} {\bibfnamefont {Didier}\ \bibnamefont {Poilblanc}},\
  }\bibfield  {title} {\enquote {\bibinfo {title} {Discrete lattice symmetry
  breaking in a two-dimensional frustrated spin-1 {H}eisenberg model},}\ }\href
  {\doibase 10.1103/PhysRevB.98.045106} {\bibfield  {journal} {\bibinfo
  {journal} {Phys. Rev. B}\ }\textbf {\bibinfo {volume} {98}},\ \bibinfo
  {pages} {045106} (\bibinfo {year} {2018}{\natexlab{b}})}\BibitemShut
  {NoStop}%
\bibitem [{\citenamefont {Estienne}\ and\ \citenamefont
  {Bernevig}(2012)}]{Estienne2012}%
  \BibitemOpen
  \bibfield  {author} {\bibinfo {author} {\bibfnamefont {Benoit}\ \bibnamefont
  {Estienne}}\ and\ \bibinfo {author} {\bibfnamefont {B.~Andrei}\ \bibnamefont
  {Bernevig}},\ }\bibfield  {title} {\enquote {\bibinfo {title} {Spin-singlet
  quantum hall states and jack polynomials with a prescribed symmetry},}\
  }\href {\doibase https://doi.org/10.1016/j.nuclphysb.2011.12.007} {\bibfield
  {journal} {\bibinfo  {journal} {Nucl. Phys. B}\ }\textbf {\bibinfo {volume}
  {857}},\ \bibinfo {pages} {185 -- 206} (\bibinfo {year} {2012})}\BibitemShut
  {NoStop}%
\bibitem [{\citenamefont {Sterdyniak}\ \emph {et~al.}(2013)\citenamefont
  {Sterdyniak}, \citenamefont {Repellin}, \citenamefont {Bernevig},\ and\
  \citenamefont {Regnault}}]{Sterdyniak2013}%
  \BibitemOpen
  \bibfield  {author} {\bibinfo {author} {\bibfnamefont {A.}~\bibnamefont
  {Sterdyniak}}, \bibinfo {author} {\bibfnamefont {C.}~\bibnamefont
  {Repellin}}, \bibinfo {author} {\bibfnamefont {B.~Andrei}\ \bibnamefont
  {Bernevig}}, \ and\ \bibinfo {author} {\bibfnamefont {N.}~\bibnamefont
  {Regnault}},\ }\bibfield  {title} {\enquote {\bibinfo {title} {Series of
  abelian and non-abelian states in $c> 1$ fractional chern insulators},}\
  }\href {\doibase 10.1103/PhysRevB.87.205137} {\bibfield  {journal} {\bibinfo
  {journal} {Phys. Rev. B}\ }\textbf {\bibinfo {volume} {87}},\ \bibinfo
  {pages} {205137} (\bibinfo {year} {2013})}\BibitemShut {NoStop}%
\bibitem [{\citenamefont {Zhang}\ \emph {et~al.}(2012)\citenamefont {Zhang},
  \citenamefont {Grover}, \citenamefont {Turner}, \citenamefont {Oshikawa},\
  and\ \citenamefont {Vishwanath}}]{ZY2012}%
  \BibitemOpen
  \bibfield  {author} {\bibinfo {author} {\bibfnamefont {Yi}~\bibnamefont
  {Zhang}}, \bibinfo {author} {\bibfnamefont {Tarun}\ \bibnamefont {Grover}},
  \bibinfo {author} {\bibfnamefont {Ari}\ \bibnamefont {Turner}}, \bibinfo
  {author} {\bibfnamefont {Masaki}\ \bibnamefont {Oshikawa}}, \ and\ \bibinfo
  {author} {\bibfnamefont {Ashvin}\ \bibnamefont {Vishwanath}},\ }\bibfield
  {title} {\enquote {\bibinfo {title} {Quasiparticle statistics and braiding
  from ground-state entanglement},}\ }\href {\doibase
  10.1103/PhysRevB.85.235151} {\bibfield  {journal} {\bibinfo  {journal} {Phys.
  Rev. B}\ }\textbf {\bibinfo {volume} {85}},\ \bibinfo {pages} {235151}
  (\bibinfo {year} {2012})}\BibitemShut {NoStop}%
\bibitem [{\citenamefont {Abrikosov}(1965)}]{AAA1965}%
  \BibitemOpen
  \bibfield  {author} {\bibinfo {author} {\bibfnamefont {A.~A.}\ \bibnamefont
  {Abrikosov}},\ }\bibfield  {title} {\enquote {\bibinfo {title} {Electron
  scattering on magnetic impurities in metals and anomalous resistivity
  effects},}\ }\href {\doibase 10.1103/PhysicsPhysiqueFizika.2.5} {\bibfield
  {journal} {\bibinfo  {journal} {Physics Physique Fizika}\ }\textbf {\bibinfo
  {volume} {2}},\ \bibinfo {pages} {5--20} (\bibinfo {year}
  {1965})}\BibitemShut {NoStop}%
\bibitem [{\citenamefont {Wen}(1991{\natexlab{b}})}]{WXG1991}%
  \BibitemOpen
  \bibfield  {author} {\bibinfo {author} {\bibfnamefont {X.~G.}\ \bibnamefont
  {Wen}},\ }\bibfield  {title} {\enquote {\bibinfo {title} {Non-abelian
  statistics in the fractional quantum hall states},}\ }\href {\doibase
  10.1103/PhysRevLett.66.802} {\bibfield  {journal} {\bibinfo  {journal} {Phys.
  Rev. Lett.}\ }\textbf {\bibinfo {volume} {66}},\ \bibinfo {pages} {802--805}
  (\bibinfo {year} {1991}{\natexlab{b}})}\BibitemShut {NoStop}%
\bibitem [{\citenamefont {Auerbach}()}]{AA1998}%
  \BibitemOpen
  \bibfield  {author} {\bibinfo {author} {\bibfnamefont {Assa}\ \bibnamefont
  {Auerbach}},\ }\bibfield  {title} {\enquote {\bibinfo {title} {Interacting
  electrons and quantum magnetism},}\ }\href@noop {} {\bibinfo  {journal}
  {(Springer, Berlin, 1998)}\ }\BibitemShut {NoStop}%
\bibitem [{\citenamefont {Tu}\ \emph {et~al.}(2013)\citenamefont {Tu},
  \citenamefont {Zhang},\ and\ \citenamefont {Qi}}]{THH2013}%
  \BibitemOpen
\bibfield  {journal} {  }\bibfield  {author} {\bibinfo {author} {\bibfnamefont
  {Hong-Hao}\ \bibnamefont {Tu}}, \bibinfo {author} {\bibfnamefont
  {Yi}~\bibnamefont {Zhang}}, \ and\ \bibinfo {author} {\bibfnamefont
  {Xiao-Liang}\ \bibnamefont {Qi}},\ }\bibfield  {title} {\enquote {\bibinfo
  {title} {Momentum polarization: An entanglement measure of topological spin
  and chiral central charge},}\ }\href {\doibase 10.1103/PhysRevB.88.195412}
  {\bibfield  {journal} {\bibinfo  {journal} {Phys. Rev. B}\ }\textbf {\bibinfo
  {volume} {88}},\ \bibinfo {pages} {195412} (\bibinfo {year}
  {2013})}\BibitemShut {NoStop}%
\bibitem [{\citenamefont {Mei}\ and\ \citenamefont {Wen}(2015)}]{MJW2015}%
  \BibitemOpen
  \bibfield  {author} {\bibinfo {author} {\bibfnamefont {Jia-Wei}\ \bibnamefont
  {Mei}}\ and\ \bibinfo {author} {\bibfnamefont {Xiao-Gang}\ \bibnamefont
  {Wen}},\ }\bibfield  {title} {\enquote {\bibinfo {title} {Modular matrices
  from universal wave-function overlaps in {G}utzwiller-projected parton wave
  functions},}\ }\href {\doibase 10.1103/PhysRevB.91.125123} {\bibfield
  {journal} {\bibinfo  {journal} {Phys. Rev. B}\ }\textbf {\bibinfo {volume}
  {91}},\ \bibinfo {pages} {125123} (\bibinfo {year} {2015})}\BibitemShut
  {NoStop}%
\bibitem [{\citenamefont {Wu}\ \emph {et~al.}(2020)\citenamefont {Wu},
  \citenamefont {Wang},\ and\ \citenamefont {Tu}}]{WYH2020}%
  \BibitemOpen
  \bibfield  {author} {\bibinfo {author} {\bibfnamefont {Ying-Hai}\
  \bibnamefont {Wu}}, \bibinfo {author} {\bibfnamefont {Lei}\ \bibnamefont
  {Wang}}, \ and\ \bibinfo {author} {\bibfnamefont {Hong-Hao}\ \bibnamefont
  {Tu}},\ }\bibfield  {title} {\enquote {\bibinfo {title} {Tensor network
  representations of parton wave functions},}\ }\href {\doibase
  10.1103/PhysRevLett.124.246401} {\bibfield  {journal} {\bibinfo  {journal}
  {Phys. Rev. Lett.}\ }\textbf {\bibinfo {volume} {124}},\ \bibinfo {pages}
  {246401} (\bibinfo {year} {2020})}\BibitemShut {NoStop}%
\bibitem [{\citenamefont {Li}\ and\ \citenamefont {Haldane}(2008)}]{LH2008}%
  \BibitemOpen
  \bibfield  {author} {\bibinfo {author} {\bibfnamefont {Hui}\ \bibnamefont
  {Li}}\ and\ \bibinfo {author} {\bibfnamefont {F.~D.~M.}\ \bibnamefont
  {Haldane}},\ }\bibfield  {title} {\enquote {\bibinfo {title} {Entanglement
  spectrum as a generalization of entanglement entropy: Identification of
  topological order in non-abelian fractional quantum hall effect states},}\
  }\href {\doibase 10.1103/PhysRevLett.101.010504} {\bibfield  {journal}
  {\bibinfo  {journal} {Phys. Rev. Lett.}\ }\textbf {\bibinfo {volume} {101}},\
  \bibinfo {pages} {010504} (\bibinfo {year} {2008})}\BibitemShut {NoStop}%
\bibitem [{\citenamefont {Jin}\ \emph {et~al.}(2021)\citenamefont {Jin},
  \citenamefont {Tu},\ and\ \citenamefont {Zhou}}]{JHK2020b}%
  \BibitemOpen
  \bibfield  {author} {\bibinfo {author} {\bibfnamefont {Hui-Ke}\ \bibnamefont
  {Jin}}, \bibinfo {author} {\bibfnamefont {Hong-Hao}\ \bibnamefont {Tu}}, \
  and\ \bibinfo {author} {\bibfnamefont {Yi}~\bibnamefont {Zhou}},\ }\bibfield
  {title} {\enquote {\bibinfo {title} {Density matrix renormalization group
  boosted by gutzwiller projected wave functions},}\ }\href {\doibase
  10.1103/PhysRevB.104.L020409} {\bibfield  {journal} {\bibinfo  {journal}
  {Phys. Rev. B}\ }\textbf {\bibinfo {volume} {104}},\ \bibinfo {pages}
  {L020409} (\bibinfo {year} {2021})}\BibitemShut {NoStop}%
\bibitem [{\citenamefont {Li}\ \emph {et~al.}(2013)\citenamefont {Li},
  \citenamefont {Weichselbaum},\ and\ \citenamefont {von Delft}}]{Li13}%
  \BibitemOpen
  \bibfield  {author} {\bibinfo {author} {\bibfnamefont {W.}~\bibnamefont
  {Li}}, \bibinfo {author} {\bibfnamefont {A.}~\bibnamefont {Weichselbaum}}, \
  and\ \bibinfo {author} {\bibfnamefont {J.}~\bibnamefont {von Delft}},\
  }\bibfield  {title} {\enquote {\bibinfo {title} {Identifying
  symmetry-protected topological order by entanglement entropy},}\ }\href
  {\doibase 10.1103/PhysRevB.88.245121} {\bibfield  {journal} {\bibinfo
  {journal} {Phys. Rev. B}\ }\textbf {\bibinfo {volume} {88}},\ \bibinfo
  {pages} {245121--245129} (\bibinfo {year} {2013})}\BibitemShut {NoStop}%
\bibitem [{\citenamefont {Kitaev}(2003{\natexlab{b}})}]{AYK2003}%
  \BibitemOpen
  \bibfield  {author} {\bibinfo {author} {\bibfnamefont {A.Yu.}\ \bibnamefont
  {Kitaev}},\ }\bibfield  {title} {\enquote {\bibinfo {title} {Fault-tolerant
  quantum computation by anyons},}\ }\href {\doibase
  https://doi.org/10.1016/S0003-4916(02)00018-0} {\bibfield  {journal}
  {\bibinfo  {journal} {Annals of Physics}\ }\textbf {\bibinfo {volume}
  {303}},\ \bibinfo {pages} {2 -- 30} (\bibinfo {year}
  {2003}{\natexlab{b}})}\BibitemShut {NoStop}%
\bibitem [{\citenamefont {Cincio}\ and\ \citenamefont {Vidal}(2013)}]{CL2013}%
  \BibitemOpen
  \bibfield  {author} {\bibinfo {author} {\bibfnamefont {L.}~\bibnamefont
  {Cincio}}\ and\ \bibinfo {author} {\bibfnamefont {G.}~\bibnamefont {Vidal}},\
  }\bibfield  {title} {\enquote {\bibinfo {title} {Characterizing topological
  order by studying the ground states on an infinite cylinder},}\ }\href
  {\doibase 10.1103/PhysRevLett.110.067208} {\bibfield  {journal} {\bibinfo
  {journal} {Phys. Rev. Lett.}\ }\textbf {\bibinfo {volume} {110}},\ \bibinfo
  {pages} {067208} (\bibinfo {year} {2013})}\BibitemShut {NoStop}%
\bibitem [{\citenamefont {Cincio}\ and\ \citenamefont {Vidal}()}]{CLarXiv}%
  \BibitemOpen
  \bibfield  {author} {\bibinfo {author} {\bibfnamefont {L.}~\bibnamefont
  {Cincio}}\ and\ \bibinfo {author} {\bibfnamefont {G.}~\bibnamefont {Vidal}},\
  }\bibfield  {title} {\enquote {\bibinfo {title} {Characterizing topological
  order by studying the ground states on an infinite cylinder},}\ }\href
  {https://arxiv.org/abs/1208.2623} {\bibinfo  {journal} {arXiv:1208.2623
  [cond-mat.str-el]}\ }\BibitemShut {NoStop}%
\bibitem [{\citenamefont {Yan}\ \emph {et~al.}(2011)\citenamefont {Yan},
  \citenamefont {Huse},\ and\ \citenamefont {White}}]{YS2011}%
  \BibitemOpen
\bibfield  {journal} {  }\bibfield  {author} {\bibinfo {author} {\bibfnamefont
  {Simeng}\ \bibnamefont {Yan}}, \bibinfo {author} {\bibfnamefont {David~A.}\
  \bibnamefont {Huse}}, \ and\ \bibinfo {author} {\bibfnamefont {Steven~R.}\
  \bibnamefont {White}},\ }\bibfield  {title} {\enquote {\bibinfo {title}
  {Spin-liquid ground state of the s = 1/2 kagome heisenberg
  antiferromagnet},}\ }\href {\doibase 10.1126/science.1201080} {\bibfield
  {journal} {\bibinfo  {journal} {Science}\ }\textbf {\bibinfo {volume}
  {332}},\ \bibinfo {pages} {1173--1176} (\bibinfo {year} {2011})}\BibitemShut
  {NoStop}%
\bibitem [{\citenamefont {Zaletel}\ \emph {et~al.}(2013)\citenamefont
  {Zaletel}, \citenamefont {Mong},\ and\ \citenamefont {Pollmann}}]{ZMP2013}%
  \BibitemOpen
  \bibfield  {author} {\bibinfo {author} {\bibfnamefont {Michael~P.}\
  \bibnamefont {Zaletel}}, \bibinfo {author} {\bibfnamefont {Roger S.~K.}\
  \bibnamefont {Mong}}, \ and\ \bibinfo {author} {\bibfnamefont {Frank}\
  \bibnamefont {Pollmann}},\ }\bibfield  {title} {\enquote {\bibinfo {title}
  {Topological characterization of fractional quantum hall ground states from
  microscopic hamiltonians},}\ }\href {\doibase 10.1103/PhysRevLett.110.236801}
  {\bibfield  {journal} {\bibinfo  {journal} {Phys. Rev. Lett.}\ }\textbf
  {\bibinfo {volume} {110}},\ \bibinfo {pages} {236801} (\bibinfo {year}
  {2013})}\BibitemShut {NoStop}%
\bibitem [{\citenamefont {Saadatmand}\ and\ \citenamefont
  {McCulloch}(2016)}]{SSN2016}%
  \BibitemOpen
  \bibfield  {author} {\bibinfo {author} {\bibfnamefont {S.~N.}\ \bibnamefont
  {Saadatmand}}\ and\ \bibinfo {author} {\bibfnamefont {I.~P.}\ \bibnamefont
  {McCulloch}},\ }\bibfield  {title} {\enquote {\bibinfo {title} {Symmetry
  fractionalization in the topological phase of the spin-$\frac{1}{2}$
  ${J}_{1}\text{\ensuremath{-}}{J}_{2}$ triangular {H}eisenberg model},}\
  }\href {\doibase 10.1103/PhysRevB.94.121111} {\bibfield  {journal} {\bibinfo
  {journal} {Phys. Rev. B}\ }\textbf {\bibinfo {volume} {94}},\ \bibinfo
  {pages} {121111} (\bibinfo {year} {2016})}\BibitemShut {NoStop}%
\bibitem [{\citenamefont {Hu}\ \emph {et~al.}(2019)\citenamefont {Hu},
  \citenamefont {Zhu}, \citenamefont {Eggert},\ and\ \citenamefont
  {He}}]{HS2019}%
  \BibitemOpen
  \bibfield  {author} {\bibinfo {author} {\bibfnamefont {Shijie}\ \bibnamefont
  {Hu}}, \bibinfo {author} {\bibfnamefont {W.}~\bibnamefont {Zhu}}, \bibinfo
  {author} {\bibfnamefont {Sebastian}\ \bibnamefont {Eggert}}, \ and\ \bibinfo
  {author} {\bibfnamefont {Yin-Chen}\ \bibnamefont {He}},\ }\bibfield  {title}
  {\enquote {\bibinfo {title} {Dirac spin liquid on the spin-$1/2$ triangular
  {H}eisenberg antiferromagnet},}\ }\href {\doibase
  10.1103/PhysRevLett.123.207203} {\bibfield  {journal} {\bibinfo  {journal}
  {Phys. Rev. Lett.}\ }\textbf {\bibinfo {volume} {123}},\ \bibinfo {pages}
  {207203} (\bibinfo {year} {2019})}\BibitemShut {NoStop}%
\bibitem [{\citenamefont {Szasz}\ \emph {et~al.}(2020)\citenamefont {Szasz},
  \citenamefont {Motruk}, \citenamefont {Zaletel},\ and\ \citenamefont
  {Moore}}]{SA2020}%
  \BibitemOpen
  \bibfield  {author} {\bibinfo {author} {\bibfnamefont {Aaron}\ \bibnamefont
  {Szasz}}, \bibinfo {author} {\bibfnamefont {Johannes}\ \bibnamefont
  {Motruk}}, \bibinfo {author} {\bibfnamefont {Michael~P.}\ \bibnamefont
  {Zaletel}}, \ and\ \bibinfo {author} {\bibfnamefont {Joel~E.}\ \bibnamefont
  {Moore}},\ }\bibfield  {title} {\enquote {\bibinfo {title} {Chiral spin
  liquid phase of the triangular lattice {H}ubbard model: A density matrix
  renormalization group study},}\ }\href {\doibase 10.1103/PhysRevX.10.021042}
  {\bibfield  {journal} {\bibinfo  {journal} {Phys. Rev. X}\ }\textbf {\bibinfo
  {volume} {10}},\ \bibinfo {pages} {021042} (\bibinfo {year}
  {2020})}\BibitemShut {NoStop}%
\bibitem [{Note5()}]{Note5}%
  \BibitemOpen
  \bibinfo {note} {For SU($N$)$_1$ CSL all topological sectors can be obtained
  in this way. However, for some topological phases, other types of anyon
  sectors can appear, such as a ``defect line'' cutting along the $x$
  direction. This is also very common and appears in, e.g., $\protect \mathbbm
  {Z}_2$~\protect \cite {Poilblanc2012} and Ising topological phases. Then,
  adopting tailored boundaries in DMRG is not sufficient to detect such
  topological sectors.}\BibitemShut {Stop}%
\bibitem [{\citenamefont {White}(1992)}]{WS1992}%
  \BibitemOpen
  \bibfield  {author} {\bibinfo {author} {\bibfnamefont {Steven~R.}\
  \bibnamefont {White}},\ }\bibfield  {title} {\enquote {\bibinfo {title}
  {Density matrix formulation for quantum renormalization groups},}\ }\href
  {\doibase 10.1103/PhysRevLett.69.2863} {\bibfield  {journal} {\bibinfo
  {journal} {Phys. Rev. Lett.}\ }\textbf {\bibinfo {volume} {69}},\ \bibinfo
  {pages} {2863--2866} (\bibinfo {year} {1992})}\BibitemShut {NoStop}%
\bibitem [{\citenamefont {McCulloch}()}]{McCulloch2008}%
  \BibitemOpen
  \bibfield  {author} {\bibinfo {author} {\bibfnamefont {I.~P.}\ \bibnamefont
  {McCulloch}},\ }\bibfield  {title} {\enquote {\bibinfo {title} {Infinite size
  density matrix renormalization group, revisited},}\ }\href
  {http://arxiv.org/abs/0804.2509} {\bibinfo  {journal} {arXiv:0804.2509
  [cond-mat.str-el]}\ }\BibitemShut {NoStop}%
\bibitem [{Note6()}]{Note6}%
  \BibitemOpen
\bibfield  {journal} {  }\bibinfo {note} {This is similar to the AKLT state
  with periodic boundary conditions, which has four-fold degeneracy in the
  entanglement spectrum rather than the two-fold degeneracy suggested by the
  $D=2$ MPS representation~\protect \cite {Li13}.}\BibitemShut {Stop}%
\bibitem [{\citenamefont {Weichselbaum}(2012)}]{WA2012}%
  \BibitemOpen
  \bibfield  {author} {\bibinfo {author} {\bibfnamefont {Andreas}\ \bibnamefont
  {Weichselbaum}},\ }\bibfield  {title} {\enquote {\bibinfo {title}
  {Non-abelian symmetries in tensor networks: A quantum symmetry space
  approach},}\ }\href {\doibase http://dx.doi.org/10.1016/j.aop.2012.07.009}
  {\bibfield  {journal} {\bibinfo  {journal} {Annals of Physics}\ }\textbf
  {\bibinfo {volume} {327}},\ \bibinfo {pages} {2972 -- 3047} (\bibinfo {year}
  {2012})}\BibitemShut {NoStop}%
\bibitem [{\citenamefont {Weichselbaum}(2020)}]{WA2020}%
  \BibitemOpen
  \bibfield  {author} {\bibinfo {author} {\bibfnamefont {Andreas}\ \bibnamefont
  {Weichselbaum}},\ }\bibfield  {title} {\enquote {\bibinfo {title} {X-symbols
  for non-abelian symmetries in tensor networks},}\ }\href {\doibase
  10.1103/PhysRevResearch.2.023385} {\bibfield  {journal} {\bibinfo  {journal}
  {Phys. Rev. Research}\ }\textbf {\bibinfo {volume} {2}},\ \bibinfo {pages}
  {023385} (\bibinfo {year} {2020})}\BibitemShut {NoStop}%
\bibitem [{Note7()}]{Note7}%
  \BibitemOpen
  \bibinfo {note} {To describe non-Abelian SU($N$)$_k$ CSL, $k>1$, we speculate
  that one should include all IRREPS in $\protect \cal V$ with up to $k$
  columns, consistently with the SU($2$)$_2$ case~\protect \cite
  {Chen2018b}.}\BibitemShut {Stop}%
\bibitem [{\citenamefont {Mambrini}\ \emph {et~al.}(2016)\citenamefont
  {Mambrini}, \citenamefont {Or\'us},\ and\ \citenamefont
  {Poilblanc}}]{Mambrini2016}%
  \BibitemOpen
  \bibfield  {author} {\bibinfo {author} {\bibfnamefont {Matthieu}\
  \bibnamefont {Mambrini}}, \bibinfo {author} {\bibfnamefont {Rom\'an}\
  \bibnamefont {Or\'us}}, \ and\ \bibinfo {author} {\bibfnamefont {Didier}\
  \bibnamefont {Poilblanc}},\ }\bibfield  {title} {\enquote {\bibinfo {title}
  {Systematic construction of spin liquids on the square lattice from tensor
  networks with $\text{SU(2)}$ symmetry},}\ }\href {\doibase
  10.1103/PhysRevB.94.205124} {\bibfield  {journal} {\bibinfo  {journal} {Phys.
  Rev. B}\ }\textbf {\bibinfo {volume} {94}},\ \bibinfo {pages} {205124}
  (\bibinfo {year} {2016})}\BibitemShut {NoStop}%
\bibitem [{\citenamefont {Nishino}\ and\ \citenamefont
  {Okunishi}(1996)}]{Nishino1996}%
  \BibitemOpen
  \bibfield  {author} {\bibinfo {author} {\bibfnamefont {Tomotoshi}\
  \bibnamefont {Nishino}}\ and\ \bibinfo {author} {\bibfnamefont {Kouichi}\
  \bibnamefont {Okunishi}},\ }\bibfield  {title} {\enquote {\bibinfo {title}
  {Corner transfer matrix renormalization group method},}\ }\href {\doibase
  10.1143/JPSJ.65.891} {\bibfield  {journal} {\bibinfo  {journal} {Journal of
  the Physical Society of Japan}\ }\textbf {\bibinfo {volume} {65}},\ \bibinfo
  {pages} {891--894} (\bibinfo {year} {1996})}\BibitemShut {NoStop}%
\bibitem [{\citenamefont {Or{\'{u}}s}\ and\ \citenamefont
  {Vidal}(2009)}]{Orus2009}%
  \BibitemOpen
  \bibfield  {author} {\bibinfo {author} {\bibfnamefont {Rom{\'{a}}n}\
  \bibnamefont {Or{\'{u}}s}}\ and\ \bibinfo {author} {\bibfnamefont
  {Guifr{\'{e}}}\ \bibnamefont {Vidal}},\ }\bibfield  {title} {\enquote
  {\bibinfo {title} {{Simulation of two-dimensional quantum systems on an
  infinite lattice revisited: Corner transfer matrix for tensor
  contraction}},}\ }\href {\doibase 10.1103/PhysRevB.80.094403} {\bibfield
  {journal} {\bibinfo  {journal} {Physical Review B}\ }\textbf {\bibinfo
  {volume} {80}},\ \bibinfo {pages} {094403} (\bibinfo {year}
  {2009})}\BibitemShut {NoStop}%
\bibitem [{\citenamefont {Poilblanc}\ and\ \citenamefont
  {Mambrini}(2017)}]{Poilblanc2017a}%
  \BibitemOpen
  \bibfield  {author} {\bibinfo {author} {\bibfnamefont {Didier}\ \bibnamefont
  {Poilblanc}}\ and\ \bibinfo {author} {\bibfnamefont {Matthieu}\ \bibnamefont
  {Mambrini}},\ }\bibfield  {title} {\enquote {\bibinfo {title} {Quantum
  critical phase with infinite projected entangled paired states},}\ }\href
  {\doibase 10.1103/PhysRevB.96.014414} {\bibfield  {journal} {\bibinfo
  {journal} {Phys. Rev. B}\ }\textbf {\bibinfo {volume} {96}},\ \bibinfo
  {pages} {014414} (\bibinfo {year} {2017})}\BibitemShut {NoStop}%
\bibitem [{\citenamefont {Cirac}\ \emph {et~al.}(2011)\citenamefont {Cirac},
  \citenamefont {Poilblanc}, \citenamefont {Schuch},\ and\ \citenamefont
  {Verstraete}}]{Cirac2011}%
  \BibitemOpen
  \bibfield  {author} {\bibinfo {author} {\bibfnamefont {J.~Ignacio}\
  \bibnamefont {Cirac}}, \bibinfo {author} {\bibfnamefont {Didier}\
  \bibnamefont {Poilblanc}}, \bibinfo {author} {\bibfnamefont {Norbert}\
  \bibnamefont {Schuch}}, \ and\ \bibinfo {author} {\bibfnamefont {Frank}\
  \bibnamefont {Verstraete}},\ }\bibfield  {title} {\enquote {\bibinfo {title}
  {{Entanglement spectrum and boundary theories with projected entangled-pair
  states}},}\ }\href {\doibase 10.1103/PhysRevB.83.245134} {\bibfield
  {journal} {\bibinfo  {journal} {Phys. Rev. B}\ }\textbf {\bibinfo {volume}
  {83}},\ \bibinfo {pages} {245134} (\bibinfo {year} {2011})}\BibitemShut
  {NoStop}%
\bibitem [{\citenamefont {Hackenbroich}\ \emph {et~al.}(2018)\citenamefont
  {Hackenbroich}, \citenamefont {Sterdyniak},\ and\ \citenamefont
  {Schuch}}]{Hackenbroich2018}%
  \BibitemOpen
  \bibfield  {author} {\bibinfo {author} {\bibfnamefont {Anna}\ \bibnamefont
  {Hackenbroich}}, \bibinfo {author} {\bibfnamefont {Antoine}\ \bibnamefont
  {Sterdyniak}}, \ and\ \bibinfo {author} {\bibfnamefont {Norbert}\
  \bibnamefont {Schuch}},\ }\bibfield  {title} {\enquote {\bibinfo {title}
  {Interplay of {SU(2)}, point group, and translational symmetry for projected
  entangled pair states: Application to a chiral spin liquid},}\ }\href
  {\doibase 10.1103/PhysRevB.98.085151} {\bibfield  {journal} {\bibinfo
  {journal} {Phys. Rev. B}\ }\textbf {\bibinfo {volume} {98}},\ \bibinfo
  {pages} {085151} (\bibinfo {year} {2018})}\BibitemShut {NoStop}%
\bibitem [{\citenamefont {Dubail}\ and\ \citenamefont
  {Read}(2015)}]{Dubail2015}%
  \BibitemOpen
  \bibfield  {author} {\bibinfo {author} {\bibfnamefont {J.}~\bibnamefont
  {Dubail}}\ and\ \bibinfo {author} {\bibfnamefont {N.}~\bibnamefont {Read}},\
  }\bibfield  {title} {\enquote {\bibinfo {title} {Tensor network trial states
  for chiral topological phases in two dimensions and a no-go theorem in any
  dimension},}\ }\href {\doibase 10.1103/PhysRevB.92.205307} {\bibfield
  {journal} {\bibinfo  {journal} {Phys. Rev. B}\ }\textbf {\bibinfo {volume}
  {92}},\ \bibinfo {pages} {205307} (\bibinfo {year} {2015})}\BibitemShut
  {NoStop}%
\bibitem [{\citenamefont {Gauth\'e}\ \emph {et~al.}(2020)\citenamefont
  {Gauth\'e}, \citenamefont {Capponi}, \citenamefont {Mambrini},\ and\
  \citenamefont {Poilblanc}}]{Gauthe2020}%
  \BibitemOpen
  \bibfield  {author} {\bibinfo {author} {\bibfnamefont {Olivier}\ \bibnamefont
  {Gauth\'e}}, \bibinfo {author} {\bibfnamefont {Sylvain}\ \bibnamefont
  {Capponi}}, \bibinfo {author} {\bibfnamefont {Matthieu}\ \bibnamefont
  {Mambrini}}, \ and\ \bibinfo {author} {\bibfnamefont {Didier}\ \bibnamefont
  {Poilblanc}},\ }\bibfield  {title} {\enquote {\bibinfo {title} {Quantum spin
  liquid phases in the bilinear-biquadratic two-{SU($4$)}-fermion hamiltonian
  on the square lattice},}\ }\href {\doibase 10.1103/PhysRevB.101.205144}
  {\bibfield  {journal} {\bibinfo  {journal} {Phys. Rev. B}\ }\textbf {\bibinfo
  {volume} {101}},\ \bibinfo {pages} {205144} (\bibinfo {year}
  {2020})}\BibitemShut {NoStop}%
\bibitem [{\citenamefont {Weichselbaum}\ \emph {et~al.}(2018)\citenamefont
  {Weichselbaum}, \citenamefont {Capponi}, \citenamefont {Lecheminant},
  \citenamefont {Tsvelik},\ and\ \citenamefont {L\"auchli}}]{WA2018_SUN}%
  \BibitemOpen
  \bibfield  {author} {\bibinfo {author} {\bibfnamefont {A.}~\bibnamefont
  {Weichselbaum}}, \bibinfo {author} {\bibfnamefont {S.}~\bibnamefont
  {Capponi}}, \bibinfo {author} {\bibfnamefont {P.}~\bibnamefont
  {Lecheminant}}, \bibinfo {author} {\bibfnamefont {A.~M.}\ \bibnamefont
  {Tsvelik}}, \ and\ \bibinfo {author} {\bibfnamefont {A.~M.}\ \bibnamefont
  {L\"auchli}},\ }\bibfield  {title} {\enquote {\bibinfo {title} {Unified phase
  diagram of antiferromagnetic su($n$) spin ladders},}\ }\href {\doibase
  10.1103/PhysRevB.98.085104} {\bibfield  {journal} {\bibinfo  {journal} {Phys.
  Rev. B}\ }\textbf {\bibinfo {volume} {98}},\ \bibinfo {pages} {085104}
  (\bibinfo {year} {2018})}\BibitemShut {NoStop}%
\bibitem [{\citenamefont {Halperin}(1983)}]{Halperin1983}%
  \BibitemOpen
  \bibfield  {author} {\bibinfo {author} {\bibfnamefont {B.~I.}\ \bibnamefont
  {Halperin}},\ }\bibfield  {title} {\enquote {\bibinfo {title} {Theory of the
  quantized hall conductance},}\ }\href {http://doi.org/10.5169/seals-115362}
  {\bibfield  {journal} {\bibinfo  {journal} {Helv. Phys. Acta}\ }\textbf
  {\bibinfo {volume} {56}},\ \bibinfo {pages} {75} (\bibinfo {year}
  {1983})}\BibitemShut {NoStop}%
\bibitem [{\citenamefont {Tu}\ \emph {et~al.}(2014)\citenamefont {Tu},
  \citenamefont {Nielsen},\ and\ \citenamefont {Sierra}}]{THH2014}%
  \BibitemOpen
  \bibfield  {author} {\bibinfo {author} {\bibfnamefont {Hong-Hao}\
  \bibnamefont {Tu}}, \bibinfo {author} {\bibfnamefont {A.~E.~B.}\ \bibnamefont
  {Nielsen}}, \ and\ \bibinfo {author} {\bibfnamefont {Germ{\'a}n}\
  \bibnamefont {Sierra}},\ }\bibfield  {title} {\enquote {\bibinfo {title}
  {Quantum spin models for the $\mathrm{SU}(n)_{1}$ {W}ess-{Z}umino-{W}itten
  model},}\ }\href {https://doi.org/10.1016/j.nuclphysb.2014.06.027} {\bibfield
   {journal} {\bibinfo  {journal} {Nucl. Phys. B}\ }\textbf {\bibinfo {volume}
  {886}},\ \bibinfo {pages} {328} (\bibinfo {year} {2014})}\BibitemShut
  {NoStop}%
\bibitem [{\citenamefont {Bondesan}\ and\ \citenamefont
  {Quella}(2014)}]{BQ2014}%
  \BibitemOpen
  \bibfield  {author} {\bibinfo {author} {\bibfnamefont {Roberto}\ \bibnamefont
  {Bondesan}}\ and\ \bibinfo {author} {\bibfnamefont {Thomas}\ \bibnamefont
  {Quella}},\ }\bibfield  {title} {\enquote {\bibinfo {title} {Infinite matrix
  product states for long-range {SU($N$)} spin models},}\ }\href
  {https://doi.org/10.1016/j.nuclphysb.2014.07.002} {\bibfield  {journal}
  {\bibinfo  {journal} {Nucl. Phys. B}\ }\textbf {\bibinfo {volume} {886}},\
  \bibinfo {pages} {483} (\bibinfo {year} {2014})}\BibitemShut {NoStop}%
\bibitem [{\citenamefont {Regnault}\ and\ \citenamefont
  {Bernevig}(2011)}]{Regnault2011}%
  \BibitemOpen
  \bibfield  {author} {\bibinfo {author} {\bibfnamefont {N.}~\bibnamefont
  {Regnault}}\ and\ \bibinfo {author} {\bibfnamefont {B.~Andrei}\ \bibnamefont
  {Bernevig}},\ }\bibfield  {title} {\enquote {\bibinfo {title} {Fractional
  {C}hern insulator},}\ }\href {\doibase 10.1103/PhysRevX.1.021014} {\bibfield
  {journal} {\bibinfo  {journal} {Phys. Rev. X}\ }\textbf {\bibinfo {volume}
  {1}},\ \bibinfo {pages} {021014} (\bibinfo {year} {2011})}\BibitemShut
  {NoStop}%
\bibitem [{\citenamefont {Bernevig}\ and\ \citenamefont
  {Regnault}(2012)}]{Bernevig2012}%
  \BibitemOpen
  \bibfield  {author} {\bibinfo {author} {\bibfnamefont {B.~Andrei}\
  \bibnamefont {Bernevig}}\ and\ \bibinfo {author} {\bibfnamefont
  {N.}~\bibnamefont {Regnault}},\ }\bibfield  {title} {\enquote {\bibinfo
  {title} {Emergent many-body translational symmetries of {Abelian} and
  {non-Abelian} fractionally filled topological insulators},}\ }\href {\doibase
  10.1103/PhysRevB.85.075128} {\bibfield  {journal} {\bibinfo  {journal} {Phys.
  Rev. B}\ }\textbf {\bibinfo {volume} {85}},\ \bibinfo {pages} {075128}
  (\bibinfo {year} {2012})}\BibitemShut {NoStop}%
\bibitem [{\citenamefont {Georgi}(1999)}]{Georgi-book-99}%
  \BibitemOpen
  \bibfield  {author} {\bibinfo {author} {\bibfnamefont {H.}~\bibnamefont
  {Georgi}},\ }\href@noop {} {\emph {\bibinfo {title} {{L}ie Algebras in
  Particle Physics}}}\ (\bibinfo  {publisher} {Perseus Books},\ \bibinfo {year}
  {1999})\BibitemShut {NoStop}%
\bibitem [{\citenamefont {Goddard}\ and\ \citenamefont
  {Olive}(1986)}]{Goddard-O-86}%
  \BibitemOpen
  \bibfield  {author} {\bibinfo {author} {\bibfnamefont {P.}~\bibnamefont
  {Goddard}}\ and\ \bibinfo {author} {\bibfnamefont {D.}~\bibnamefont
  {Olive}},\ }\bibfield  {title} {\enquote {\bibinfo {title} {{K}ac-{M}oody and
  {V}irasoro algebras in relation to quantum physics},}\ }\href
  {https://www.worldscientific.com/doi/abs/10.1142/S0217751X86000149}
  {\bibfield  {journal} {\bibinfo  {journal} {Int. J. Mod. Phys. A}\ }\textbf
  {\bibinfo {volume} {1}},\ \bibinfo {pages} {303--414} (\bibinfo {year}
  {1986})}\BibitemShut {NoStop}%
\bibitem [{\citenamefont {Knizhnik}\ and\ \citenamefont
  {Zamolodchikov}(1984)}]{Knizhnik-Z-84}%
  \BibitemOpen
  \bibfield  {author} {\bibinfo {author} {\bibfnamefont {V.~G.}\ \bibnamefont
  {Knizhnik}}\ and\ \bibinfo {author} {\bibfnamefont {A.~B.}\ \bibnamefont
  {Zamolodchikov}},\ }\bibfield  {title} {\enquote {\bibinfo {title} {Current
  algebra and {W}ess-{Z}umino model in two dimensions},}\ }\href
  {http://www.sciencedirect.com/science/article/B6TVC-4718WH6-11C/2/78488410cf4d8d21de2be2fb78354edb}
  {\bibfield  {journal} {\bibinfo  {journal} {Nucl. Phys. B}\ }\textbf
  {\bibinfo {volume} {247}},\ \bibinfo {pages} {83--103} (\bibinfo {year}
  {1984})}\BibitemShut {NoStop}%
\bibitem [{\citenamefont {Ginsparg}(1988)}]{WZWreview1988}%
  \BibitemOpen
  \bibfield  {author} {\bibinfo {author} {\bibfnamefont {P.}~\bibnamefont
  {Ginsparg}},\ }\bibfield  {title} {\enquote {\bibinfo {title} {Applied
  conformal field theory},}\ }in\ \href@noop {} {\emph {\bibinfo {booktitle}
  {Fields, Strings and Critical Phenomena}}},\ \bibinfo {editor} {edited by\
  \bibinfo {editor} {\bibfnamefont {E.}~\bibnamefont {Br\'ezin}}\ and\ \bibinfo
  {editor} {\bibfnamefont {J.}~\bibnamefont {Zinn-Justin}}},\ \bibinfo
  {organization} {Les {H}ouches summer school}\ (\bibinfo  {publisher}
  {North-Holland},\ \bibinfo {year} {1988})\BibitemShut {NoStop}%
\bibitem [{\citenamefont {Jin}\ \emph {et~al.}(2020)\citenamefont {Jin},
  \citenamefont {Tu},\ and\ \citenamefont {Zhou}}]{JHK2020a}%
  \BibitemOpen
  \bibfield  {author} {\bibinfo {author} {\bibfnamefont {Hui-Ke}\ \bibnamefont
  {Jin}}, \bibinfo {author} {\bibfnamefont {Hong-Hao}\ \bibnamefont {Tu}}, \
  and\ \bibinfo {author} {\bibfnamefont {Yi}~\bibnamefont {Zhou}},\ }\bibfield
  {title} {\enquote {\bibinfo {title} {Efficient tensor network representation
  for {G}utzwiller projected states of paired fermions},}\ }\href {\doibase
  10.1103/PhysRevB.101.165135} {\bibfield  {journal} {\bibinfo  {journal}
  {Phys. Rev. B}\ }\textbf {\bibinfo {volume} {101}},\ \bibinfo {pages}
  {165135} (\bibinfo {year} {2020})}\BibitemShut {NoStop}%
\bibitem [{\citenamefont {Petrica}\ \emph {et~al.}(2021)\citenamefont
  {Petrica}, \citenamefont {Zheng}, \citenamefont {Chan},\ and\ \citenamefont
  {Clark}}]{Petrica2020}%
  \BibitemOpen
  \bibfield  {author} {\bibinfo {author} {\bibfnamefont {Gabriel}\ \bibnamefont
  {Petrica}}, \bibinfo {author} {\bibfnamefont {Bo-Xiao}\ \bibnamefont
  {Zheng}}, \bibinfo {author} {\bibfnamefont {Garnet Kin-Lic}\ \bibnamefont
  {Chan}}, \ and\ \bibinfo {author} {\bibfnamefont {Bryan~K.}\ \bibnamefont
  {Clark}},\ }\bibfield  {title} {\enquote {\bibinfo {title} {Finite and
  infinite matrix product states for gutzwiller projected mean-field wave
  functions},}\ }\href {\doibase 10.1103/PhysRevB.103.125161} {\bibfield
  {journal} {\bibinfo  {journal} {Phys. Rev. B}\ }\textbf {\bibinfo {volume}
  {103}},\ \bibinfo {pages} {125161} (\bibinfo {year} {2021})}\BibitemShut
  {NoStop}%
\bibitem [{\citenamefont {Singh}\ and\ \citenamefont {Vidal}(2012)}]{SS2012}%
  \BibitemOpen
  \bibfield  {author} {\bibinfo {author} {\bibfnamefont {Sukhwinder}\
  \bibnamefont {Singh}}\ and\ \bibinfo {author} {\bibfnamefont {Guifre}\
  \bibnamefont {Vidal}},\ }\bibfield  {title} {\enquote {\bibinfo {title}
  {Tensor network states and algorithms in the presence of a global su(2)
  symmetry},}\ }\href {\doibase 10.1103/PhysRevB.86.195114} {\bibfield
  {journal} {\bibinfo  {journal} {Phys. Rev. B}\ }\textbf {\bibinfo {volume}
  {86}},\ \bibinfo {pages} {195114} (\bibinfo {year} {2012})}\BibitemShut
  {NoStop}%
\bibitem [{\citenamefont {Hubig}(2018)}]{HC2018}%
  \BibitemOpen
  \bibfield  {author} {\bibinfo {author} {\bibfnamefont {Claudius}\
  \bibnamefont {Hubig}},\ }\bibfield  {title} {\enquote {\bibinfo {title}
  {{Abelian and non-abelian symmetries in infinite projected entangled pair
  states}},}\ }\href {\doibase 10.21468/SciPostPhys.5.5.047} {\bibfield
  {journal} {\bibinfo  {journal} {SciPost Phys.}\ }\textbf {\bibinfo {volume}
  {5}},\ \bibinfo {pages} {47} (\bibinfo {year} {2018})}\BibitemShut {NoStop}%
\bibitem [{\citenamefont {Schmoll}\ \emph {et~al.}(2020)\citenamefont
  {Schmoll}, \citenamefont {Singh}, \citenamefont {Rizzi},\ and\ \citenamefont
  {Orús}}]{SP2020}%
  \BibitemOpen
  \bibfield  {author} {\bibinfo {author} {\bibfnamefont {Philipp}\ \bibnamefont
  {Schmoll}}, \bibinfo {author} {\bibfnamefont {Sukhbinder}\ \bibnamefont
  {Singh}}, \bibinfo {author} {\bibfnamefont {Matteo}\ \bibnamefont {Rizzi}}, \
  and\ \bibinfo {author} {\bibfnamefont {Román}\ \bibnamefont {Orús}},\
  }\bibfield  {title} {\enquote {\bibinfo {title} {A programming guide for
  tensor networks with global {SU($2$)} symmetry},}\ }\href {\doibase
  https://doi.org/10.1016/j.aop.2020.168232} {\bibfield  {journal} {\bibinfo
  {journal} {Annals of Physics}\ }\textbf {\bibinfo {volume} {419}},\ \bibinfo
  {pages} {168232} (\bibinfo {year} {2020})}\BibitemShut {NoStop}%
\end{thebibliography}%
\end{document}